\documentclass[appendixfloats]{emulateapj}
\usepackage[]{graphicx,times,url,amssymb,longtable,subfigure}

\newcommand{\Ha}{\mbox{H$\alpha $}}
\newcommand{\SII}{\mbox{[S$\rm II$]}}

\newcommand{\OIII}{\mbox{[O$\rm III$]}}
\newcommand{\Op}{\mbox{${\rm O^{+}}$}}
\newcommand{\Opp}{\mbox{${\rm O^{2+}}$}}
\newcommand{\Sp}{\mbox{${\rm S^{+}}$}}
\newcommand{\Spp}{\mbox{${\rm S^{2+}}$}}
\newcommand{\HI}{{\sc Hi}}
\newcommand{\HII}{{\sc Hii}}
\newcommand{\Msun}{\mbox{${\cal M}_\odot$}}

\newcommand{\AV}{\mbox{$A_{\rm V}$}}
\newcommand{\NHI}{$N{\rm (HI)}$}
\newcommand{\Hplus}{$\rm H^{+}$}
\newcommand{\LHa}{\mbox{$L$}}
\newcommand{\Lesc}{\mbox{$L_{\rm esc}$}}

\newcommand{\fesc}{\mbox{$f_{\rm {esc}}$}}
\newcommand{\fescgal}{\mbox{$f_{\rm {esc,gal}}$}}
\newcommand{\fescave}{\mbox{$\langle f_{\rm esc} \rangle$}}

\newcommand{\taulyc}{\mbox{$\tau_{\rm LyC}$}}
\newcommand{\teff}{\mbox{$T_{\rm eff}$}}
\newcommand{\SHa}{\mbox{$S{\rm (H\alpha)}$}}
\newcommand{\ergs}{\mbox{${\rm erg\ s^{-1}}$}}

\begin{document}
\shorttitle{Optical Depth of LMC and SMC \HII\ regions}
\shortauthors{Pellegrini et al.}

\title{The Optical Depth of \HII\ regions in the Magellanic Clouds}
\author{E.~W. Pellegrini$^1$, M.~S. Oey$^1$, P.~F. Winkler$^2$, 
S.~D. Points$^3$, R.~C. Smith$^3$, A.~E. Jaskot$^1$, and J. Zastrow$^1$}

\affil{
$^1$Department of Astronomy, University of Michigan, 500 Church
  Street, Ann Arbor, MI 48109, USA,
  $^2$Department of Physics, Middlebury College, Middlebury, VT 05753,
  USA,
  $^3$Cerro Tololo Inter-American Observatory, National Optical Astronomy Observatory, Casilla 603, La Serena, Chile }

\email{pelleger@umich.edu}
\slugcomment{}
\begin{abstract}
-THIS ELECTRONIC VERSION INCLUDES TABLES 5 AND 6 AS REVISED IN THE ERRATUM-\\
- THE ERRATUM HAS BEEN BUNDLED AS A SEPARATE FILE IN THIS ARXIV SUBMISSION -\\
  We exploit ionization-parameter mapping as a powerful tool to
  measure the optical depth of star-forming \HII\ regions.  Our
  simulations using the photoionization code {\sc Cloudy} and our
  new, {\sc SurfBright} surface brightness simulator demonstrate that
  this technique can directly diagnose most density-bounded, optically
  thin nebulae using spatially resolved emission line data.  We apply
  this method to the Large and Small Magellanic Clouds, using the data
  from the Magellanic Clouds Emission Line Survey.  We generate new
  \HII\ region catalogs based on photoionization criteria set by the
  observed ionization structure in the \SII/\OIII\ ratio and \Ha\
  surface brightness.  The luminosity functions from these catalogs
  generally agree with those from \Ha-only surveys.  We then use
  ionization-parameter mapping to crudely classify all the nebulae
  into optically thick vs optically thin categories, yielding
  fundamental new insights into Lyman continuum radiation
  transfer.  We find that in both galaxies, the frequency of optically
  thin objects correlates with \Ha\ luminosity, and that the numbers
  of these objects dominate above $\log\LHa/({\rm erg\ s^{-1}}) \geq
  37.0$.  The frequencies of optically thin objects are 40\% and 33\%
  in the LMC and SMC, respectively.  Similarly, the frequency of optically thick regions
  correlates with \HI\ column density, with optically thin objects
  dominating at the lowest \NHI.  The integrated escape luminosity of
  ionizing radiation is dominated by the largest regions, and
  corresponds to luminosity-weighted, ionizing escape fractions from
  the \HII\ region population of $\ge 0.42$ and $\ge$ 0.40 in the LMC
  and SMC, respectively.  These values correspond to global galactic
  escape fractions of 4\% and 11\%, respectively.  This is sufficient
  to power the ionization rate of the observed diffuse ionized gas in
  both galaxies.  Since our optical depth estimates tend to be
  underestimates, and also omit the contribution from field stars
  without nebulae, our results suggest the possibility of significant
  galactic escape fractions of Lyman continuum radiation.\\
-THIS ELECTRONIC VERSION INCLUDES TABLES 5 AND 6 AS REVISED IN THE ERRATUM-
\end{abstract}


\section{Introduction}
\label{sec:Introduction}

Few of the rich and complex disciplines in astrophysics affect our
understanding of the Universe as deeply as the diffusion of ionizing
radiation from stars and its interaction with surrounding matter.  Of
all the known sinks and sources of energy, the ionizing radiation
released by O stars during their short lives has great consequences
by 1) determining the structure and energy balance of the
interstellar medium (ISM) in galaxies, 2) generating diagnostics of stellar
populations and interstellar conditions, and 3) providing an important
source of the Lyman continuum radiation field during cosmic
reionization.

The luminosity and spectral energy distribution (SED) of massive stars
make them a powerful source of ionizing radiation within star forming
galaxies \citep[e.g.,][]{Abbott1982,Reynolds1984}.  Their power has
been demonstrated by studies of nearby galaxies which show that the
Lyman-continuum (LyC) radiation from O-stars embedded within
\HII\ regions, combined with those in the field, is luminous enough to
balance the incessant recombination and cooling of the diffuse, warm
ionized medium (WIM) in galaxies (e.g., \citealt{Oey1997, Hoopes2000,
  Oey2004}; for a recent review of the WIM see \citealt{Haffner2009}).
Radiative transfer calculations also demonstrate that injecting
ionizing radiation from stars into the WIM not only heats the gas, but
also acts to decrease its cooling efficiency \citep{Cantalupo2010},
preventing the catastrophic cooling of warm diffuse gas which would
lead to unregulated star formation \citep[e.g.][]{Parravano1988,
  Ostriker2010}.  In these ways, ionizing stellar radiation can
strongly influence both the ISM structure and star formation rates of
galaxies.

The H Balmer recombination lines form beacons of star
formation across the universe (e.g., \citealt{Cowie1998}).  When
ionizing photons are all absorbed by gas, H recombination lines are an
accurate diagnostic of $Q{\rm (H^0)}$, the rate at which H ionizing
radiation is produced by stars.  Multiwavelength emission-line
observations and theoretical stellar SEDs are routinely used with
observed recombination rates to infer the stellar populations of
distant galaxies \citep[e.g.,][]{Sullivan2004, Iglesias2004}.

Ionizing radiation from stars may ultimately escape into the intergalactic
medium (IGM) before being absorbed.  This radiation may be an
important source of the cosmic background UV field during the epoch of
reionization, some time between redshift $z\sim 11$
\citep{Komatsu2011} and $z\sim 6$ \citep{Fan2002}.  During this time,
star-forming galaxies are believed to contribute 10-20\% of
their total ionizing radiation budget to sustain reionization, because
the UV and X-ray field from AGN alone was likely insufficient
\citep{Sokasian2003}.  Recent detections of faint Ly$\alpha$
emitting galaxies by \citet{Dressler2011} support this view with
evidence that aggregate LyC radiation from faint
galaxies during this epoch is sufficient to sustain cosmic
reionization.  

Understanding the radiative transfer of LyC photons from massive stars
is therefore a fundamental problem, and although they are well
understood in general terms, \HII\ regions still present a
computational challenge because of their greatly varying densities,
small-scale structure, and irregular nature.  Consequently,
\citet{Paardekooper2011} identified radiative transfer of individual
nebulae as the main bottleneck which limits our ability to determine
the escape fraction of ionizing radiation from star forming galaxies
in cosmological simulations.  It is thus imperative to understand
radiation transport within \HII\ regions if we are to understand
fundamental properties of the universe.

There has been a variety of approaches to evaluate the optical depth
of \HII\ regions.  The most direct method compares the ionization rate
derived from \Ha\ luminosities $\LHa$ to that predicted from the
observed ionizing stellar population.  Using this approach,
\citet{Oey1997} found that up to half of all ionizing photons
generated by stars escape \HII\ regions to ionize the WIM, also known
as diffuse ionized gas (DIG).  However,
theoretical predictions for the LyC photon emission rate $Q{\rm (H^0)}$ have
decreased significantly (e.g., \citealt{Martins2005, Smith2002}), and are
now generally 
consistent with the observed \HII\ region luminosities
(e.g., \citealt{Voges2008, Zastrow2011a}).  Clearly, until the ionizing fluxes
and SEDs of massive stars are definitively established, comparing
predicted and observed $\LHa$ will be subject to large systematic
uncertainties.  Identifying all the ionizing stars is also difficult
in regions with significant extinction and crowding.

Other studies attempt to evaluate nebular optical depth by modeling
nebular emission lines from ions with different ionization potentials
averaged over the entire \HII\ region (e.g.,
\citealt{Relano2002, Iglesias2002, Giammanco2004, Kehrig2010}). 
However, inhomogeneous, optically-thin nebulae may
contain many optically-thick cloudlets.  Since the emission-line
volume-emissivity is proportional to the square of the electron
density, the resulting spatially integrated spectra can be
dominated by these dense clumps and resemble the spectrum of an
optically-thick, homogeneous nebula, despite small
clump-covering-factors \citep{Giammanco2004}. Typically, these
  studies do not resolve the spatial structure of the emitting
  gas. Observations either integrate all the nebular light and lose
  all spatial information, or study structure from a single long slit
  spectrum. By simplifying the line fluxes of an entire \HII\ region to a
  single value, valuable information about the true structure of the
  gas is lost.

The correlation between DIG surface brightness and proximity to \HII\
regions is another key piece of evidence for the
leakage of ionizing radiation from discrete \HII\ regions, and can be
used to estimate the optical depth.  \citet{Seon2009} used
these correlations to test a model of M51 where leaking \HII\ regions
explain the observed DIG and \Ha\ surface brightness distributions, similar to
the method used by Zurita et al.  (2002) in NGC~157.  However,
\citet{Seon2009} found that this model requires a highly
rarefied or porous ISM with an anomalously low dust abundance.  These
details are inconsistent with the known properties of M51, suggesting
the models do not fully explain the propagation of radiation in real
galaxies.

Thus, existing methods to determine nebular optical depth are subject
to large uncertainties; clearly it would be preferable to have a
diagnostic that is reliable, effective and simple.  Here, we offer
such a diagnostic, using an approach that makes it possible to
accurately characterize the optical depth of individual \HII\ regions
in the nearest galaxies.  In \S \ref{sec:method}, we describe our
method; in \S \ref{sec:IPMMCs}, we apply our technique to the Magellanic Clouds, and
use it to generate a new, physically motivated \HII\ region catalog;
and we evaluate our technique in \S \ref{sec:HIItau}.  Our results yield powerful new
insights on the radiative transfer of LyC radiation from massive stars
in these galaxies, which we present in \S \ref{sec:GlobFesc}.

\section{Ionization-Parameter Mapping}
\label{sec:method}

With the recent availability of wide-field, narrow-band imaging and
tunable filters, the potential of spatially resolved, emission-line
diagnostics as constraints on nebular models is being more fully
realized, and these techniques can now be applied to entire
populations of extragalactic nebulae.  We revisit a largely overlooked
approach, ionization-parameter mapping (IPM), which is capable of
directly assessing the optical depth of ionizing radiation in
individual \HII\ regions (e.g., \citealt{Koeppen1979}).  The technique
is based on emission-line ratio mapping, which has been previously employed
(e.g., \citealt{Heydari1981,Pogge1988a,Pogge1988b}); here, we
present a modern development, demonstration, and application. The
  current approach is driven by newly available data with unprecedented
  sensitivity, resolution, and spatial completeness.  We leverage this
  data against recent developments in the ability to predict spatially
  resolved, emission line diagnostics with photoionization models.
Our method thus balances the quantitative diagnostics of spectroscopy and
the spatial coverage of imaging, yielding a powerful method that is
both observationally efficient and straightforward enough to be
applied to entire galaxies.

\subsection{Evaluating nebular optical depth}
\label{sec:evalnebtau}

\begin{figure*}
\centering
\includegraphics[scale=0.7]{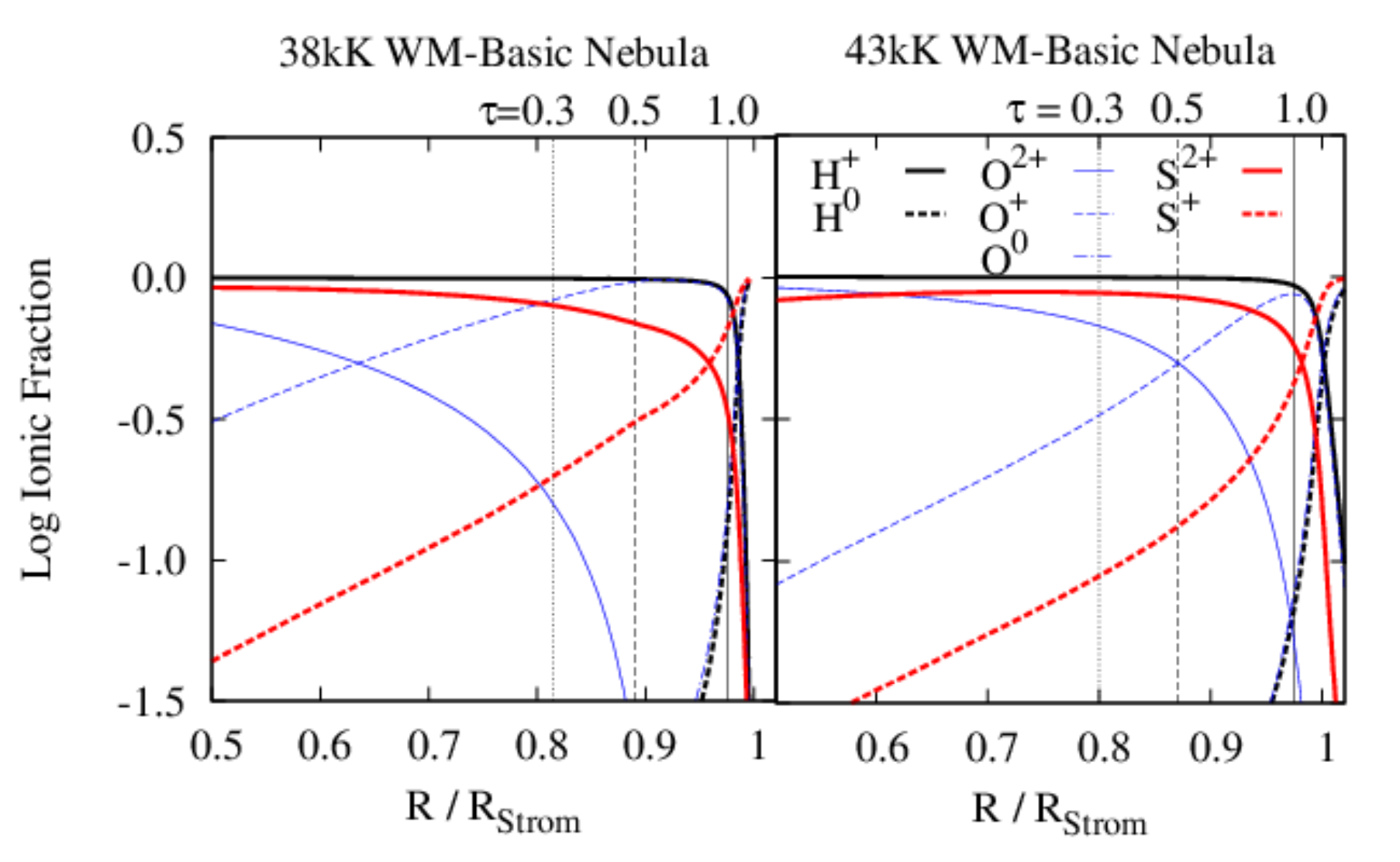}
\caption{\footnotesize The radial ionization structure for
  Str\"omgren sphere \HII\ regions photoionized by a $T_{\rm eff} =
  38$ kK star (left panel) and a 43 kK star (right panel), assuming a gas 
  density of $10\ {\rm cm^{-3}}$.  The logarithmic ionization fractions
  of H$^+$(solid, thick black line), H$^{0}$ (dashed, thick black line), O$^{2+}$ (solid,
  thin blue line), O$^{+}$ (dashed, thin blue line), O$^{0}$ (dash-dot,
  blue line), S$^{2+}$ (solid, thick red line), and S$^{+}$
  (thick, dashed red line) are plotted.  The vertical lines mark 
  radii where \taulyc = 0.3 (dotted), 0.5 (dashed) to 1.0 (solid).}
\label{fig:structure}
\end{figure*}

For classic, optically-thick \HII\ regions, there is a transition zone
between the central, highly excited region and the neutral
environment.  These transition zones are characterized by a strong
decrease in the excitation, and hence also in the gas ionization
parameter, which traces the degree of ionization and photon-to-gas
density.  Figure \ref{fig:structure}$a$ and \ref{fig:structure}$b$
show the radial ionic structure of Str\"omgren spheres generated by a
38,000 K and 43,000 K star, respectively.  These demonstrate the
transition from highly ionized inner zones dominated by \Opp\ and
\Spp\ to outer envelopes dominated by \Op\ and \Sp. The low-ionization
transition zone is thicker than the narrow H$^0$/H$^+$ ionization
front where the \SII\ volume emissivity peaks \citep{agn2006}; this
results from the sensitivity of the \SII/\OIII\ ratio to the radial
difference between the \Opp\ and \Hplus\ recombination fronts, which
are in turn determined by the LyC optical depth $\taulyc$ and stellar
effective temperature $\teff$.  This large scale gradient is a key
feature to the application of ionization parameter mapping at great
distances.  For the models in Figure~\ref{fig:structure}, the assumed
ionizing SED is a single WM-Basic stellar atmosphere \citep{Smith2002}
defined by a variable $\teff$ and fixed $Q{\rm (H^0)}=10^{49} \rm
s^{-1}$, equivalent to one O6~V star.  Calculations were performed
using the {\sc Cloudy} photoionization code, version C08.00
\citep{Ferland1998}, adopting gas-phase abundances equal to those of
the 30~Doradus star-forming region, having log(O/H) = $-3.75$
\citep{Pellegrini2011}.  Our models include dust with a gas-to-dust
ratio of $\AV/N{\rm (H)} = 1.8\times10^{-23}~{\rm cm^{-2}}$, which is
consistent with the ionized gas studied by \citet{Pellegrini2011}.  We
use dust with an LMC size-distribution described by
\citet{WeingartnerDraine2001}, although our results are not sensitive
to the dust abundance.  The initial H density $n{\rm_H}$ is equal to
$10~{\rm cm^{-3}}$, and the distance $r_{0} = 0.1~{\rm pc}$ between
the illuminated face of the cloud and the ionizing source.  Deeper in
the cloud $n_{\rm H}$ is set by a hydrostatic equation of state with
no magnetic field (${\bf B} =0$~G) described in
\citet{Pellegrini2007}.

\begin{figure}[t]
\centering
\includegraphics*[scale=0.45]{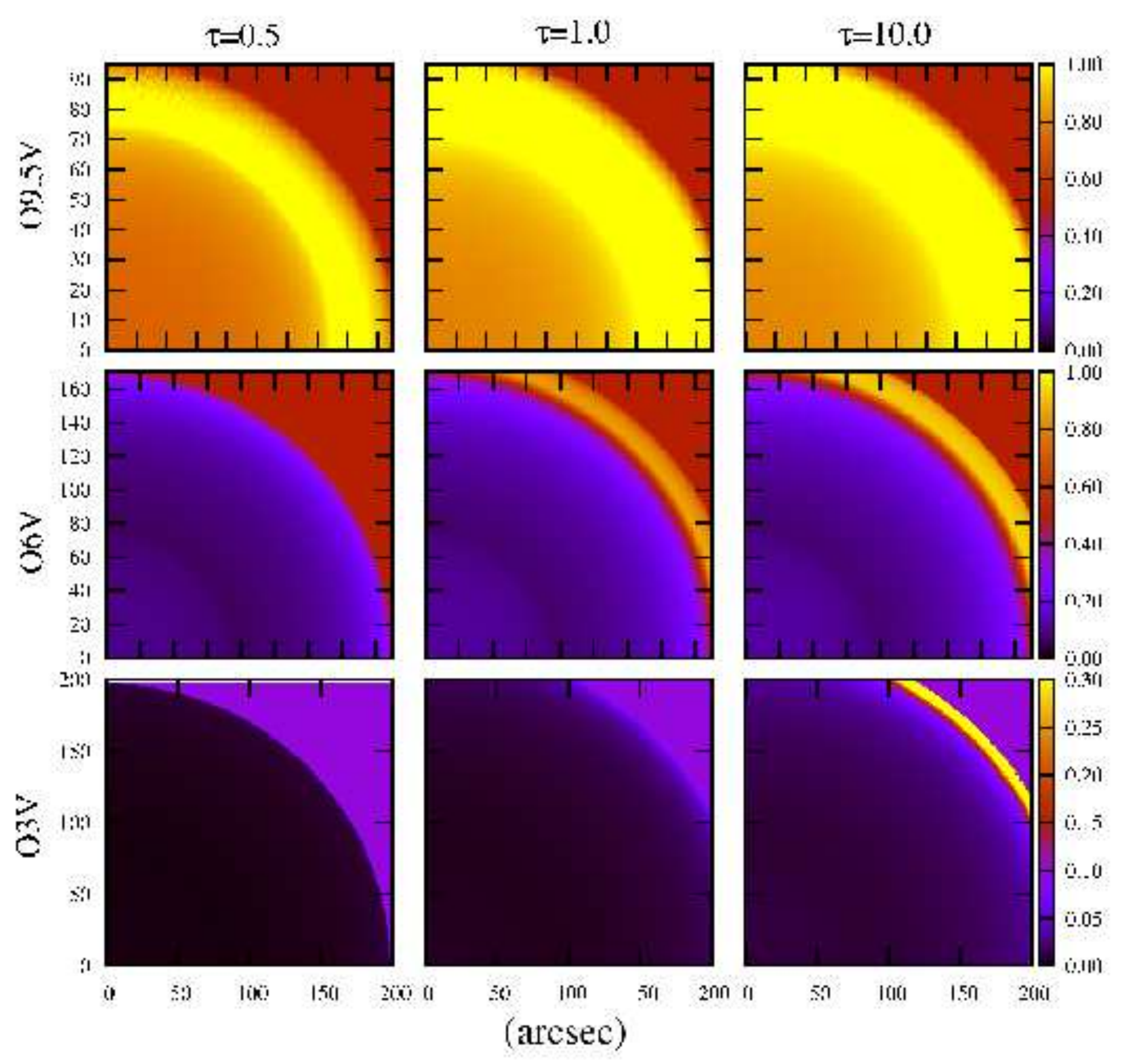} 
\caption{\footnotesize Models of the observed \SII/\OIII\
  surface brightness ratios for nebulae ionized by O3 V, O6 V, and
  O9.5 V stars, at optical depths $\tau_{\rm LyC}$ of 0.5, 1.0, and 10.
A uniform background is assumed, and the units on both axes are in arcsec,
projecting the objects at the LMC distance.}
\label{fig:LMC_Models_nH_10}
\end{figure}

\begin{figure}[ht]
\centering
\includegraphics[width=2.2in]{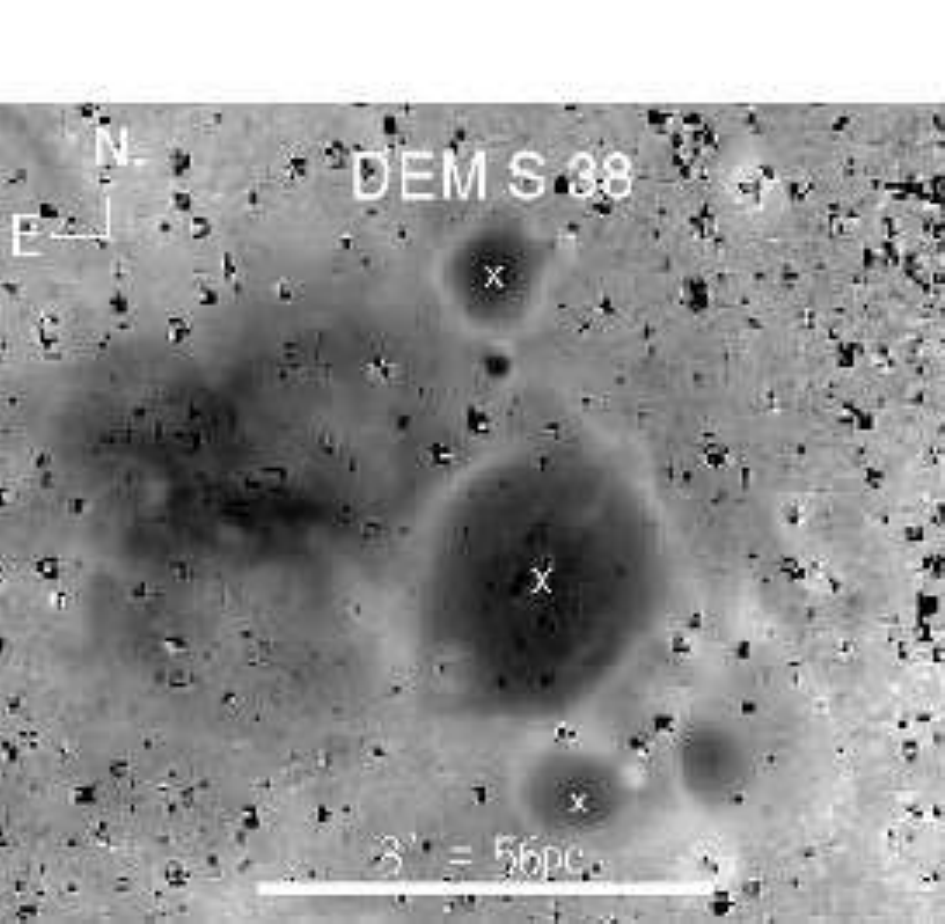}
\caption{\footnotesize A map of the \SII/\OIII\ ratio centered on HII
  region DEM S38.  The central region is a classic example of an
  optically thick nebula with highly ionized gas (low \SII/\OIII;
  dark) surrounded by an ionization transition zone with higher
  \SII/\OIII\ (lighter grayscale).  The same effect is
 seen in the regions to the north and south of DEM S38, which are
  marked with crosses.  In contrast, the irregular \HII\ region seen to
  the east (left) of DEM S38 shows no evidence such a transition zone
  between the high ionization region and the galactic background,
  indicating that it is optically thin.}
\label{fig:DEMS38_S2O3}
\end{figure}

Figure~\ref{fig:structure} demonstrates that we can estimate
$\tau_{\rm LyC}$ from the observed ion stratification within the
nebula, which depends strongly on \taulyc.  While nebulae ionized by
different \teff\ have greatly differing structure, the optical depth
is strongly constrained by the radial structure in two ions, and
essentially uniquely determined by three ions.  Figure
\ref{fig:LMC_Models_nH_10} shows models of the observed
surface-brightness ratios for the \SII$\lambda\lambda6716,6731$ and
\OIII$\lambda5007$ emission lines for a series of \HII\ regions with
LMC element abundances.  We calculate the projected 2D surface
brightness of our models according to equation 2 of
~\citet{Pellegrini2009} using the {\sc SurfBright} routine, which is
described in Appendix A. We have added a constant, noiseless
background of $1\times10^{-15}\rm{~erg~s^{-1}~cm^{-2}~arcsec^{-2}}$,
consistent with typical \HII\ region observations.  We note that
decreasing the background component will enhance the predicted
contrast, while an increase reduces contrast. The model parameters of
these simulations are similar to those of the models in
Figure~\ref{fig:structure}. The nebulae are ionized by a single
  WM-Basic \citep{Smith2002} stellar SED with an ionizing luminosity
  equivalent to a cluster of ten O6 V stars, and $\teff$ equal to
  30,500 K, 38,000 K or 44,500 K.  These $\teff$ correspond to O9.5
V, O6 V and O3 V spectral types, respectively, using the spectral type
-- $T_{\rm eff}$ calibration of ~\citet{Martins2005}. A single
  \teff\ is often used to represent the SED of ionizing clusters,
  which is a reasonable approximation since the earliest spectral type
  dominates the SED \citep[e.g.,][]{Oey2000}.
Figure~\ref{fig:LMC_Models_nH_10} shows models for $\tau_{\rm LyC} =
0.5,$ 1.0, and 10.0, at each \teff, where the cloud is truncated at
various radii to simulate the different $\tau_{\rm LyC}$.

Figure~\ref{fig:LMC_Models_nH_10} demonstrates how the optical
depth and \teff\ determine the observed ionic structure that is
rendered by ionization-parameter mapping.  In general, the $\taulyc =
0.5$ models show no low-ionization transition layer, although there is
an exception for the latest spectral type.  For early and mid-O spectral
types, the morphology in these ionization-parameter maps is an especially
strong discriminant for the optical depth.  And as shown in
Figure~\ref{fig:structure}, \taulyc\ can be fully constrained when
surface-brightness ratios are obtained for three radially varying ions
instead of two.  We further discuss the use and limitations of our method
in \S \ref{sec:limitations} below.

Figure \ref{fig:DEMS38_S2O3} shows the observed ratio map of
\SII/\OIII\ for a star-forming complex centered on the nebula DEM~S38,
from the Magellanic Clouds Emission-Line Survey (MCELS;
\citealt{Smith1998, Points2005, Smith2005, Winkler2005}).  We clearly
see an envelope of low-ionization gas surrounding a high-excitation
interior in each \HII\ region marked with an X (DEM~S38 and the two
regions to the north and south), strongly suggesting that these
objects are optically thick.  In contrast, the nebula east of DEM~S38
shows high ionization throughout, and no evidence of an internal
gradient in gas ionization state.  This indicates that the object is
optically thin.  

\begin{figure}[t]
\vspace{6mm}
\centering
\includegraphics[scale =0.5]{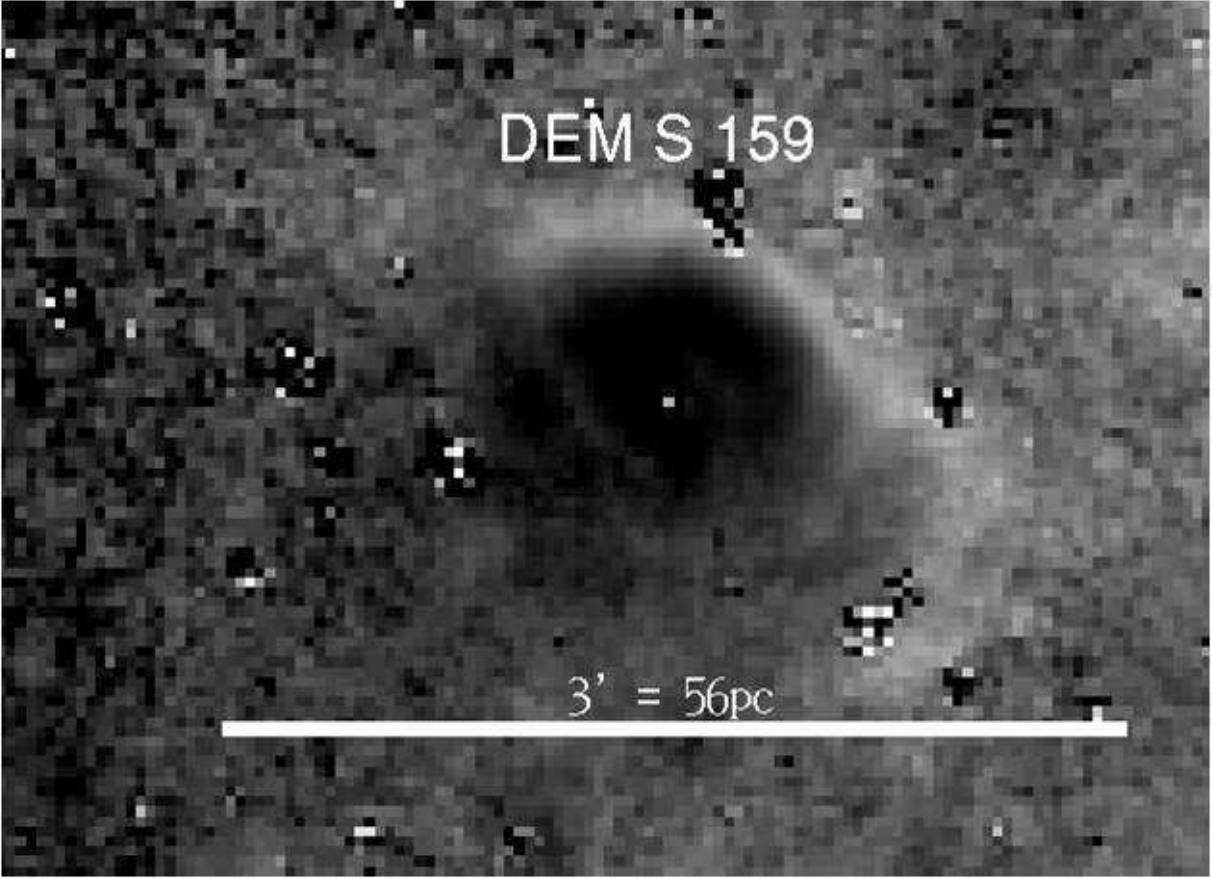}
\caption{\footnotesize DEM S159 shows blister-like features,
exhibiting nebular traits for both high and low optical depth.  As in
  Figure \ref{fig:DEMS38_S2O3}, the ratio of \SII/\OIII\ reveals the
  presence of highly ionized gas (dark) in this region.  This is
  confined to the northwest by a pronounced ionization transition zone,
  but not to the southeast.}
\label{fig:DEMS159_S2O3}
\end{figure}

We also see that the object DEM~S159 (Figure
\ref{fig:DEMS159_S2O3}) shows the intermediate morphology of a blister-like
\HII\ region.  Like DEM~S38, there is a central region of highly
ionized gas, but a transition zone of weakly ionized gas is found only
to the north, while toward the south, the nebula remains highly
ionized throughout, like our \taulyc=0.5 models in
Figure~\ref{fig:LMC_Models_nH_10}.  Since all of the nebula is ionized
by the same SED, DEM~S159 must be optically thick to the north, and
optically thin to the south.  Thus, Figures~\ref{fig:DEMS38_S2O3} and
\ref{fig:DEMS159_S2O3} vividly demonstrate the viability of
ionization-parameter mapping as a technique to evaluate \taulyc.  The
morphology of the ionization structure in these objects is
qualitatively consistent with our models, and in \S \ref{sec:HIItau} below, we also
show quantitatively that observations are consistent with predictions.

Furthermore, the contrasting gas morphology between the spherical,
optically thick nebulae and the irregular, optically thin object in
Figure~\ref{fig:DEMS38_S2O3} is not a coincidence.  In the MCELS data
for the Magellanic Clouds, most of the optically thick objects showing
low-ionization envelopes look like classical, spherical, Str\"omgren
spheres.  The opposite is true for optically thin objects, which are
more complex and irregular in morphology.  This is consistent with
recent radiation-MHD simulations by \citet{Arthur2011}, which show
that the highly ionized, density-bounded nebulae powered by the
hottest stars are subject to strong radiative feedback and gas
instabilities, generating irregular gas morphologies.  Thus, the gas
morphologies are fully consistent with the interpretation that objects
having low-ionization transition zones are generally optically thick
and radiation-bounded.

\begin{figure}[b]
\centering
\subfigure[DEM S15]{
\includegraphics[scale=0.4]{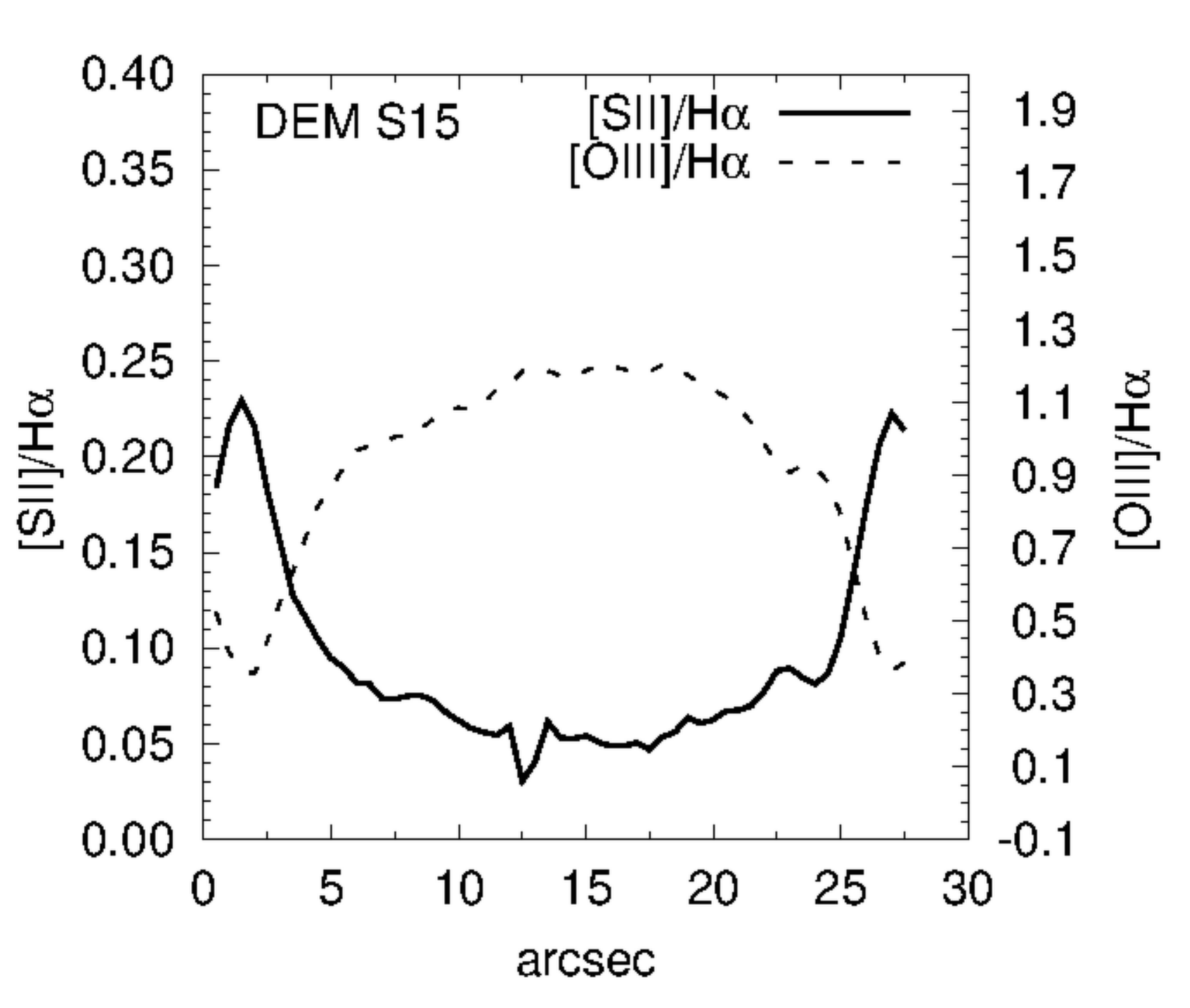}
\label{fig:DEMS15}
}
\subfigure[N59]{
\includegraphics[scale=0.4]{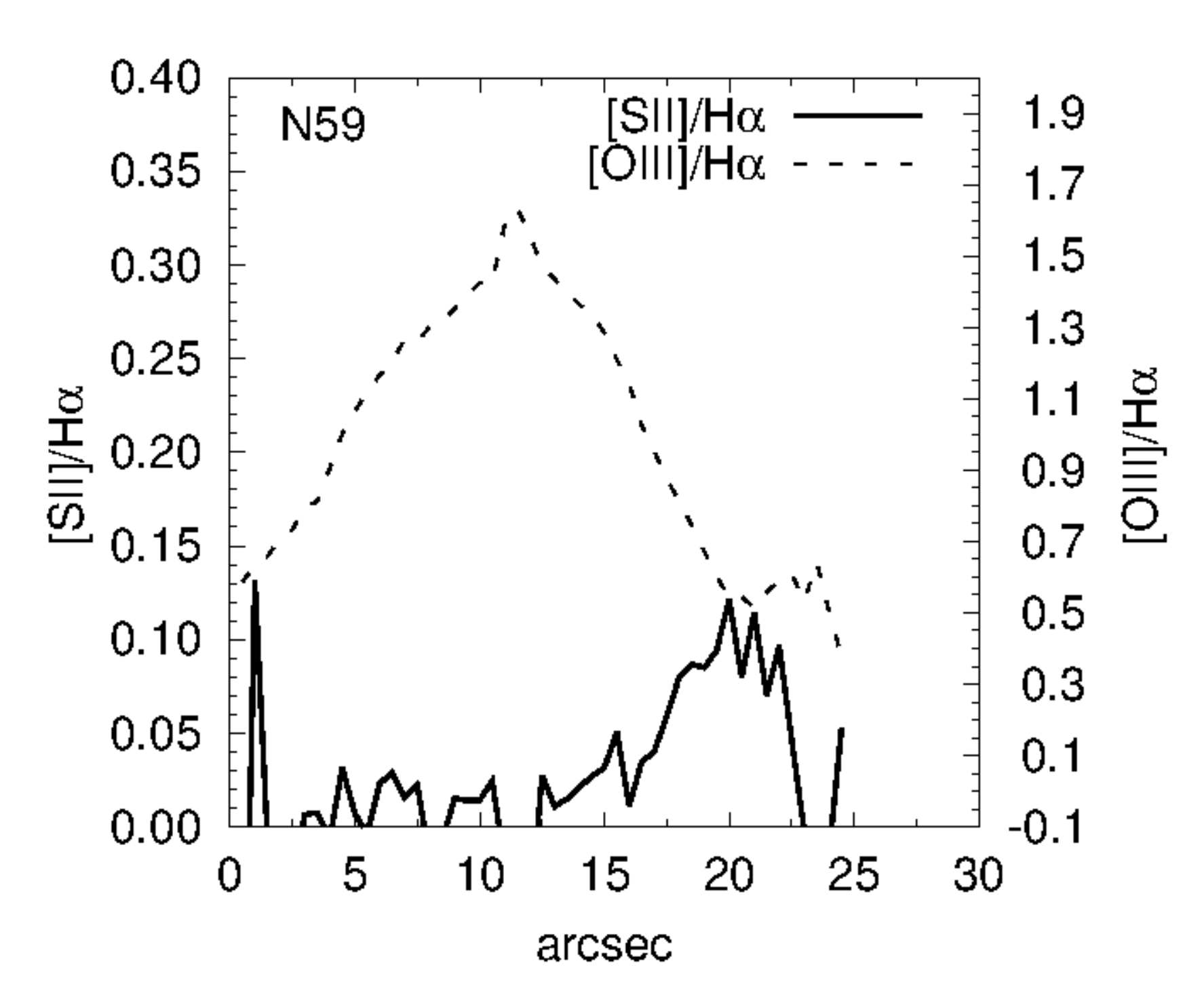}
\label{fig:N59}
}
\caption{\footnotesize The spatial profiles of \OIII/\Ha\ and \SII/\Ha\ across
the center of the optically thick object DEM~S15 (top) and optically
thin region N59 (bottom).} 
\label{fig:Profile}
\end{figure}

Finally, ionization-parameter mapping also constrains the optical
depth in the line of sight, since the low-ionization, transition zone
should also exist along these photon paths in an optically thick
nebula.  This was explored by \citet{Pellegrini2011} who found a lower
limit of \SII/\Ha\ $\sim 0.05$ for LMC nebulae that are optically thick in
the line of sight.  Lower values of \SII/\Ha\ indicate that the
low-ionization transition zone is missing or depleted, and therefore
that the region is optically thin to the Lyman continuum.  In
Figure~\ref{fig:Profile} we compare the line of sight emission line
ratios of DEM~S15 and N59, a pair of optically thick and optically
thin \HII\ regions.  DEM~S15 is a classical Str\"omgren sphere,
limb-brightened in the lower ionization species, and it shows a
central line-of-sight \SII/\Ha$= 0.05$, consistent with an optically
thick nebula of SMC metallicity.  In contrast, in N59, the
central \SII/\Ha\ ratio is essentially zero across much of the object,
and thus no transition zone is seen in those sight lines,
demonstrating that the object is optically thin.

\subsection{Limitations for 2-ion mapping}
\label{sec:limitations}

Ionization-parameter mapping is tremendously powerful, and can even be
done with only two radially varying ions.  When using only two ions,
we caution that the technique has three limitations.  Ostensibly,
the most important quantity to be derived with this technique is the
escape fraction of LyC photons from an individual \HII\ region, \fesc\,
defined as
\begin{equation}
\fesc = e^{-\taulyc} \quad .
\end{equation}
Ionization-parameter mapping based on only two ions can
provide only lower limits on \fesc, because the observed morphology
becomes degenerate at high \fesc.  This can be seen in the bottom row
of Figure \ref{fig:LMC_Models_nH_10} for $\tau_{\rm LyC} \le 1.0$,
which corresponds to $\fesc \ge 40\%$.  In these cases, background
emission masks the very faint, lower-ionization emission lines in
fully ionized gas, and the ratio ceases to directly track changes in
the \HII\ region ionization structure.  This problem worsens as
$T_{\rm eff}$ increases, and it becomes more difficult to identify the
transition to neutral gas.  However, only the hottest ionizing stars in
the local universe will have $\teff\sim 44,500$ K, and when three ions
are available, the degeneracies are resolved.

There is also a degeneracy between optically thin nebulae ionized by
cool stars ($T_{\rm eff} \lesssim 34,000K$) and optically thick
regions heated by hotter stars.  The degeneracy exists where cool
stars do not emit much radiation above 35 eV to generate O$^{2+}$, and
so these nebulae are entirely dominated by O$^{+}$.  Again, ionization
parameter mapping based on three ions, adding S$^{2+}$ for example,
can resolve the degeneracy (Figure~\ref{fig:structure}).  However, we
stress that this problem applies primarily to the lowest-luminosity
objects, and as we show below, their aggregate luminosity is
insignificant compared to the total amount of energy found to be
escaping all \HII\ regions. 

Finally, we again caution that for a population of randomly oriented
blister \HII\ regions, it is likely that the orientation of some
objects will cause a projected ionization-parameter gradient that
appears optically thick on the limb, but is optically thin in the line
of sight.  The most extreme example is of a half-sphere, blister
nebula viewed directly face-on: despite having $\fesc = 0.5$, the
projected region is circular and will show an ionization transition
zone associated with the optically thick half.  However, as discussed
in \S \ref{sec:evalnebtau}, the ionic ratios of \SII, \OIII, and \Ha\
across the central region of these nebulae should show a deficit in
lower-ionization species that is incompatible with optically thick
models (Figure~\ref{fig:N59}).  With further constraints on the
ionizing SED and a quantitative evaluation of these ratios, we can
still measure their optical depth.  Thus, there may be instances where
optically thin \HII\ regions are initially misidentified as optically
thick, but these can be identified by quantitative examination of
spatially resolved ionic ratios.  If objects are misidentified, this
again would favor underestimates of \taulyc.

\begin{deluxetable*}{lcccc}
\tabletypesize{\footnotesize}
\tablecaption{MCELS 1-$\sigma$ surface brightness detection limits}
\tablehead{\colhead{}&
\colhead{${S(\rm H\alpha)}$}&
\colhead{${\rm EM(H\alpha)}$}&
\colhead{${S(\OIII)}$}&
\colhead{${S\SII)}$}\\
\colhead{Galaxy}&
\colhead{ ${\rm \ergs\ cm^{-2} arcsec^{-2}}$}&
\colhead{${\rm pc\ cm^{-6}}$}&
\colhead{ ${\rm erg\ s^{-1}\ cm^{-2}\ arcsec^{-2}}$}&
\colhead{ ${\rm erg\ s^{-1}\ cm^{-2}\ arcsec^{-2}}$}
} 
\hline
\startdata
LMC& $7.0E-18$ & 3.5 & $1.4E-17$ & $5.2E-18$\\
SMC& $7.2E-18$ & 3.6 & $5.4E-17$ & $1.0E-17$
\enddata
\label{tab:MCELS}
\tablecomments{Values shown are the 1-$\sigma$ uncertainties in a
  single pixel for surface brightness and emission measure as shown.
  The LMC data are not continuum-subtracted and have a pixel scale of
  3 arcsec, while the SMC data are continuum subtracted and have a
  pixel scale of 2 arcsec.}
\end{deluxetable*}

\begin{figure*}[]
\centering
\includegraphics*[scale=0.45]{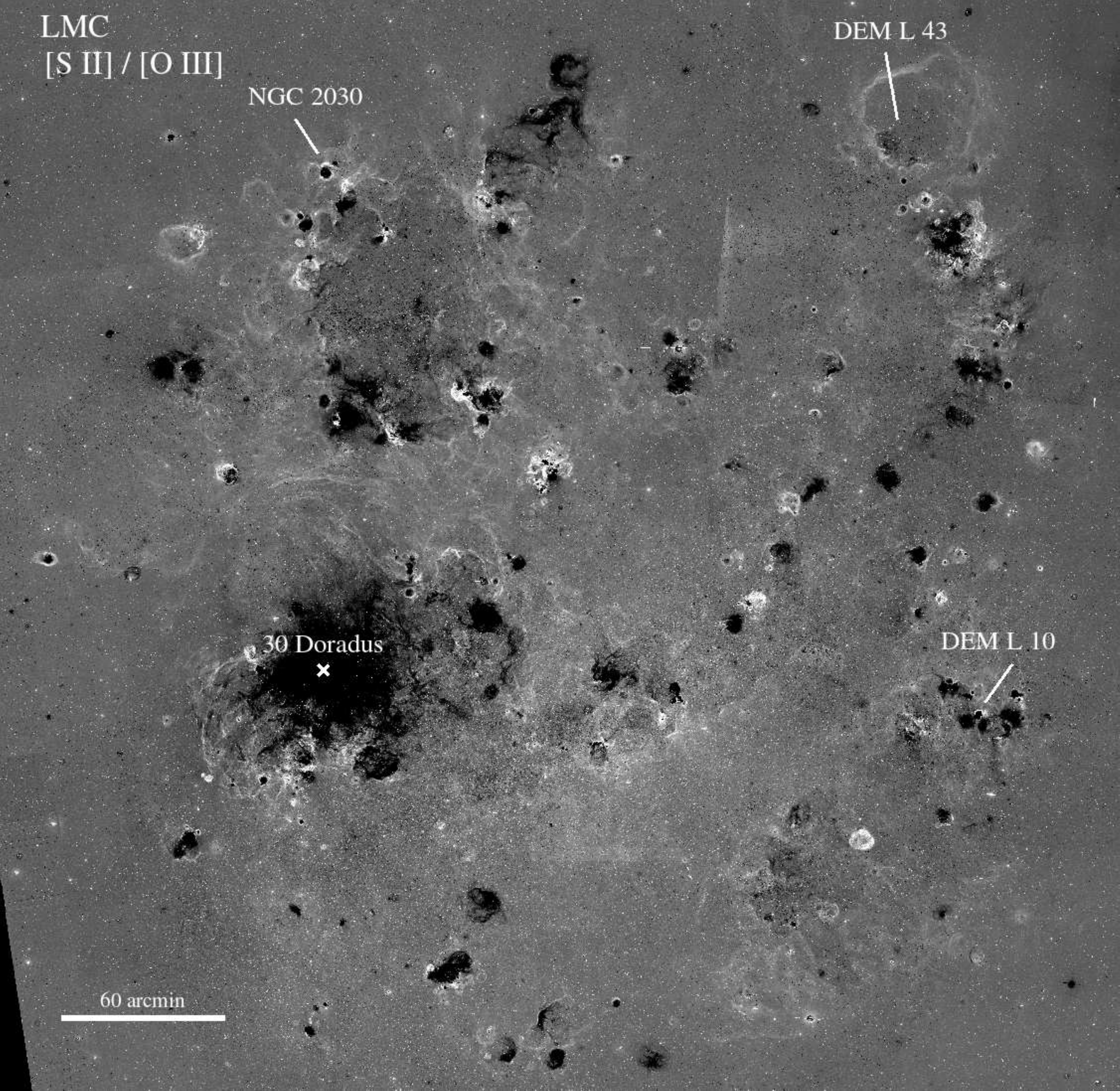}
\caption{\footnotesize The map of log [S II]/[O III] for the entire
  LMC galaxy created from the MCELS narrow-band imaging data.  Black
  corresponds to low [S II]/[O III]; this ratio map is not based on
  continuum-subtracted data.  North is up, east to the left in the
  center of the field.  }
\label{fig:LMCS2O3}
\end{figure*}

\begin{figure*}[]
\centering
\includegraphics*[scale=0.5]{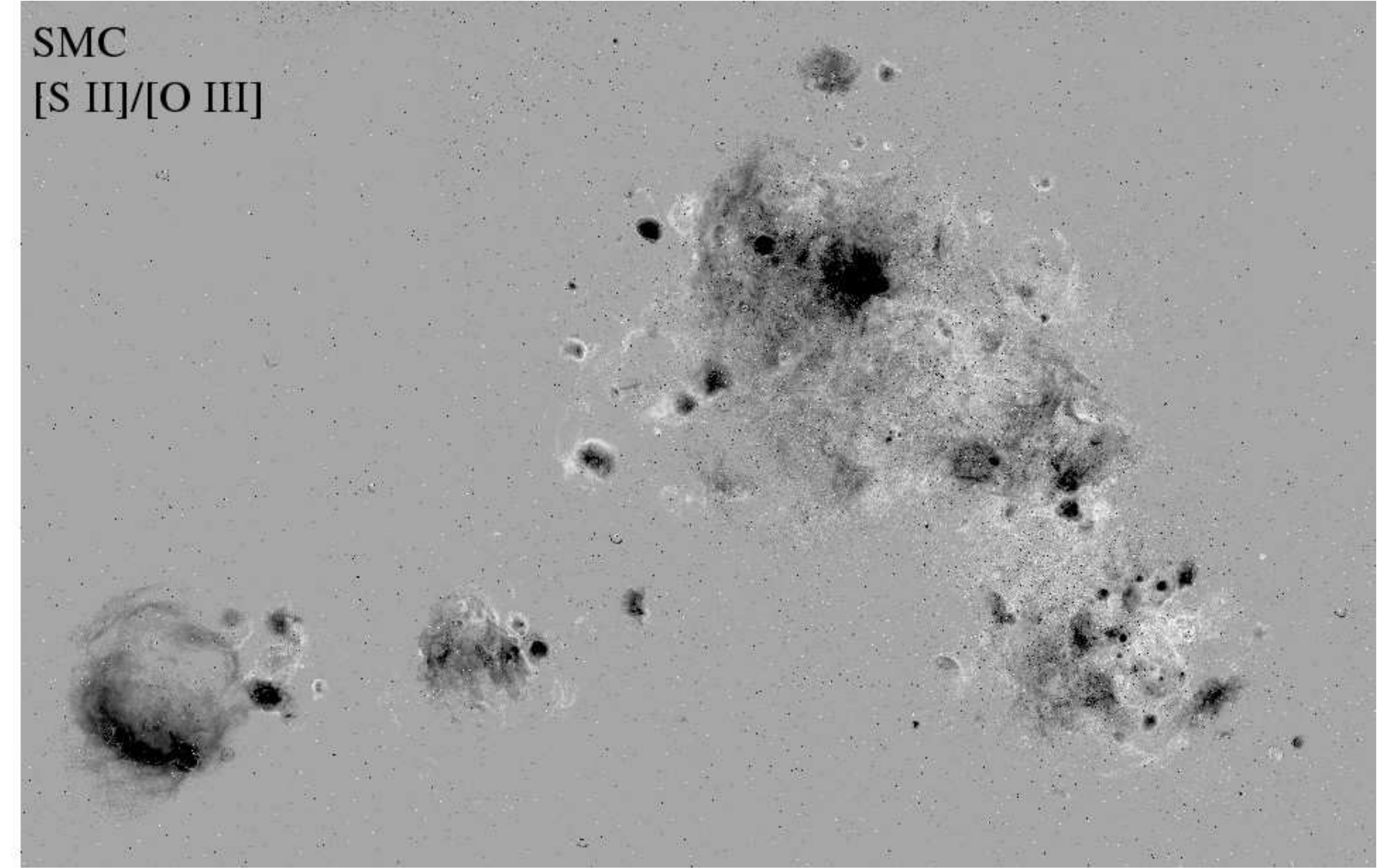}
\caption{\footnotesize Same as figure~\ref{fig:LMCS2O3} for the SMC
  galaxy.  This ratio map is based on continuum-subtracted data.}
\label{fig:SMCS2O3}
\end{figure*}

Hence, the caveats identified above can be resolved by
ionization-parameter mapping in three ions and quantitative
evaluation of the entire nebular projection.
We note that these fairly manageable issues all work to underestimate the
optical depth of \HII\ regions.  Thus, ionization-parameter mapping 
shows great promise as a powerful tool in studies of the ISM. 
Emission-line ratio maps neutralize variations in surface 
brightness, clearly revealing changes in ionization structure for
bright and faint regions alike.  The power of the
technique is that it allows us to identify optically thin \HII\
regions by the absence of the low-ionization envelope, which almost
always indicates that the nebula is density-bounded.

\section{Ionization-Parameter Mapping of the LMC and SMC}
\label{sec:IPMMCs}

We now apply our technique of ionization-parameter mapping to the
Large Magellanic Cloud (LMC) and Small Magellanic Cloud (SMC), which
have been mapped with narrow-band emission-line imaging by the MCELS
survey.  This is a spatially complete, flux-limited survey carried out
at the Cerro Tololo Inter-American Observatory (CTIO) with the
University of Michigan's Curtis 0.6/0.9m Schmidt telescope.  Over the
course of 5 years, the LMC and SMC were imaged in
\SII$\lambda\lambda6717,6731$, \OIII\ $\lambda$5007, and \Ha, with
respective filter widths of 50, 40 and 30\AA.  The
\Ha\ filter bandpass includes [N~II]$\lambda\lambda$6548,6584 at a
reduced throughput.  The final product, mosaics in both low and
high-ionization line emission, and in the \Ha\ recombination line,
traces the ionized ISM at both large and small scales.  The process of
mosaicking the images resulted in a binned pixel scale of 3.0 and 2.0
arcsec pixel$^{-1}$ for the LMC and SMC, respectively. These
correspond to a spatial scale of 0.7 pc and 0.6 pc for distances of
49 kpc ~\citep{Macri2006} and 61 kpc \citep{Hilditch2005},
respectively, with an effective resolution of $\sim 5$ arcsec.  The
1-$\sigma$ surface-brightness limit of each band is listed in
Table~\ref{tab:MCELS}.  These are the sensitivities per pixel,
expressed as surface brightness in $\rm \ergs\ cm^{-2}~arcsec^{-2}$
and \Ha\ emission measure (EM) in $\rm pc\ cm^{-6}$.  Such depth is
important to form a complete understanding of the WIM ionization,
and the dependence of \fesc\ on star-formation intensity and \HII\ region
properties.

The MCELS survey includes continuum observations centered at 5130~\AA\
and 6850~\AA, with effective bandpasses of 155~\AA\ and 95~\AA\ ,
respectively.  These were used to produce a continuum-subtracted mosaic
of the SMC \citep{Winkler2005}.  Based on spectrophotometric
observations of the SMC region NGC 346 by \citet{Tsamis2003}, we
estimate that the flux calibration of the continuum-subtracted data
has uncertainties on the order of 20\%.  At present, the LMC data are
not yet continuum-subtracted.  To flux-calibrate the LMC data, we used
spectrophotometric observations by ~\citet{Pellegrini2010}, extracting
MCELS line fluxes along the length of slit position 5 in that paper to
determine the flux constants.  We also compare against the
flux-calibrated, narrow-band data obtained on the SOAR telescope in a
30 arcsec circular aperture at the position, $\alpha =$ 05:38:56.9,
$\delta = $--69:05:21.8 (J2000) \citep{Pellegrini2010}.  We find that
the comparisons agree within approximately 20\%, which then
corresponds to the systematic uncertainty in our flux calibration.

We generated line-ratio maps of \SII/\OIII\ for both the LMC and SMC
using {\sc IRAF}\footnote{IRAF is distributed by the National Optical
  Astronomy Observatory, which is operated by the Association of
  Universities for Research in Astronomy (AURA) under cooperative
  agreement with the National Science Foundation.}, with the LMC 
maps based on non-continuum-subtracted emission-line images, and the
SMC maps based on continuum-subtracted images.  These ratio maps, which
probe the ionization parameter in the ionized gas, are
shown in Figures ~\ref{fig:LMCS2O3} and ~\ref{fig:SMCS2O3}, respectively.

\subsection{Ionization-based HII region catalogs}
\label{sec:IBHIIcat}

\begin{figure*}[t]
\vspace{6mm}
\centering
\includegraphics[scale =0.3]{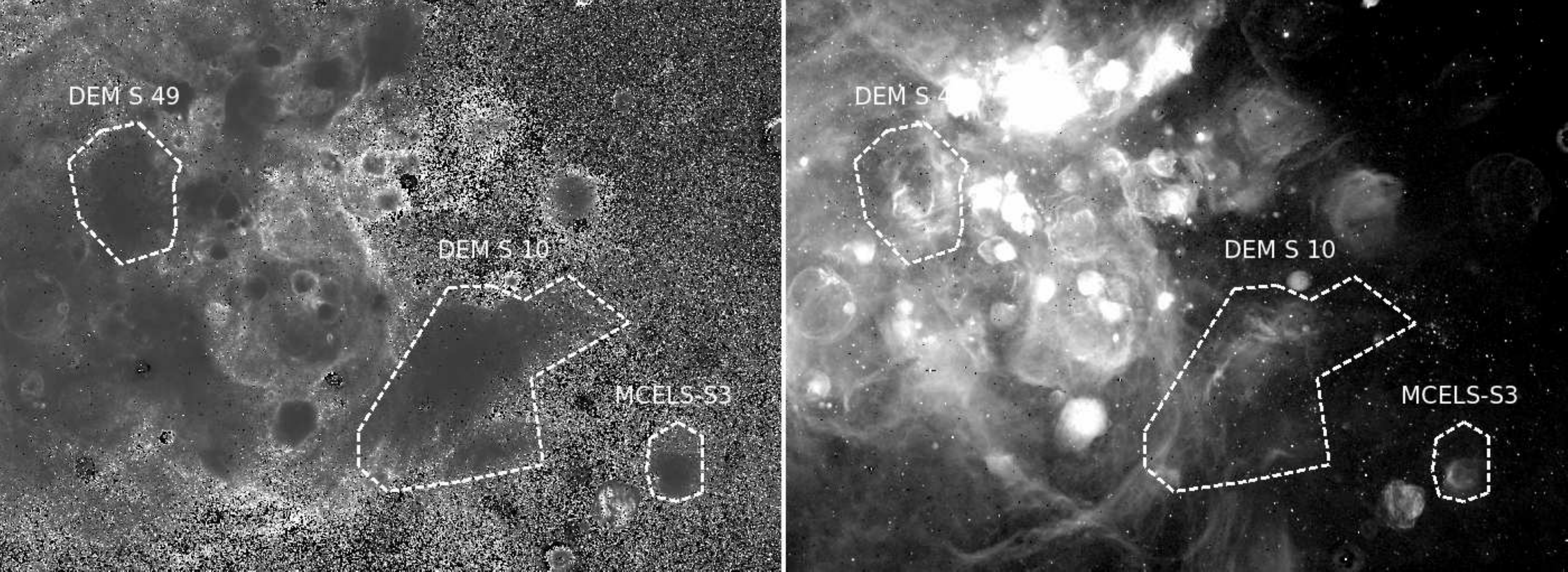}
\caption{\footnotesize \SII/\OIII\ ratio map (left) and \Ha\ (right)
  for an SMC region, demonstrating the advantage of using \SII/\OIII\
  in combination with \Ha\ to define the boundaries of extended \HII\
  regions coincident with complex background structures (e.g.,
  DEM~S49), and those with extended, faint emission (e.g., DEM~S10,
  MCELS-S3).  We indicate \HII\ region boundaries for
 DEM~S10, DEM~S49 and MCELS-S3 with dashed lines in both
  images. }
\label{fig:enhancementexample}
\end{figure*}

Ionization-parameter mapping allows us to assign physically-motivated
\HII\ region boundaries in complex, confused regions with multiple
ionizing sources.  In areas where \HII\ regions are
  overlapping, or are found in complex ionized backgrounds, the
  ionization stratification makes it possible to
  isolate individual photoionized regions, which is impossible with
  imaging in only \Ha\ or any single line.  In particular,
ionization-parameter mapping allows us to
define nebular boundaries based on both ionization structure and
\Ha\ surface brightness $S$(\Ha) morphology.  The examples in
  Figure~\ref{fig:enhancementexample} 
demonstrate how ionization-parameter mapping generates contrast
between the DIG and low surface-brightness, extended \HII\ regions that are
independently ionized entities.  In the \Ha\ image (right
panel of Figure~\ref{fig:enhancementexample}), the objects DEM~S10 and
DEM~S49 are amorphous regions that blend into the surrounding DIG with
ambiguous boundaries.  In contrast, the \SII/\OIII\ ratio map clearly
shows them as distinct regions.  The boundaries of optically thick
objects are usually unambiguous because these are characterized by a
stratified ionization structure as described above, accompanied
by a sharp decrease in surface brightness.  

Since previous nebular
catalogs for the Magellanic Clouds are based only on \Ha\ morphology
(e.g., \citealt{Henize1956, DEM}), we use these more sophisticated
criteria based on ionization-parameter mapping to compile a more
physically-based catalog of \HII\ regions in 
these galaxies.  In the case of extended,
optically thin objects that show only gradual changes in \SII/\Ha\ or
\SII/\OIII, we define the \HII\ region boundary to be the point at
which either the \SHa\ or the ionic ratio become indistinguishable
from the DIG, whichever is larger in size.  We
defined photometric apertures with polygons in SAOImage DS9,
for both target objects and local background regions; we used the
FUNCNTS routine from FUNTOOLS\footnote{\footnotesize
  https://www.cfa.harvard.edu/$\sim$john/funtools/} to measure the fluxes.

The photometry of faint objects, especially seen in \SII\
and \OIII\ filters, are at risk of being contaminated by stellar
continuum in the LMC, where our data are not continuum subtracted.  
We minimize this contamination by avoiding foreground
Galactic stars and also sampling the local density of field stars with
our background apertures.  Despite the careful creation of apertures,
the difference between stellar populations inside and outside the HII
regions may still introduce significant errors, since the most
massive, brightest stars often reside within HII regions.  Thus,
errors in the background subtraction dominate the flux uncertainties
for both galaxies, and they are largest for low surface-brightness
objects.  We therefore find that the median local background surface 
brightness for nebulae in the LMC is larger than for the SMC:
$7.0\times 10^{-16}$ and $1.3\times 10^{-16}\ \rm erg\
s^{-1}cm^{-2}arcsec^{-2}$, respectively, in \Ha.  However, we stress that high
surface-brightness emission dominates most objects, yielding 
median background uncertainties of 6\% and 8\% in the LMC and SMC,
respectively.  For the LMC, the discrete stellar contributions can
increase this uncertainty to about 18\%.  This is consistent with
a comparison of our background-subtracted fluxes of bright, isolated
LMC and SMC \HII\ regions to their fluxes reported by
\citet{Kennicutt1989}.  We find that the independent measurements
agree within 20\%, which now also includes systematic uncertainties. 


We further explored the limiting case of applying a constant background
to all objects.  For the LMC, we calculated this background from the
mean of three locations, two in the north and one in the south to
estimate the contamination from sources producing a constant
background such as the sky, large scale diffuse emission, etc.  We
find the mean, constant background $S{(\Ha)} = 5.2 \times 10^{-16}\
\rm erg\ s^{-1} cm^{-2} arcsec^{-2}$.  For the SMC, we determined the
background of the continuum-subtracted \Ha\ image to be consistent
with zero ($7\times 10^{-18} \pm 7\ \rm erg\ s^{-1} cm^{-2}
arcsec^{-2}$; cf. Table~\ref{tab:MCELS}).  For both galaxies,
subtracting the median backgrounds affects the
resulting $\LHa$ by no more than 0.2 dex, and does not
substantively change our results.

Our \HII\ region catalogs for the LMC and SMC defined with these
ionization-based criteria are presented in Appendix~B,
Tables~\ref{tab:LMCObjs} and \ref{tab:SMCObjs}, which give
luminosities and associated \HI\ column densities for 401 objects in
the LMC, and 214 in the SMC.  

\subsection{The HII region Luminosity Function}
\label{sec:HIILF}

\begin{figure}
\centering
\subfigure[LMC]{
\includegraphics[scale=0.4]{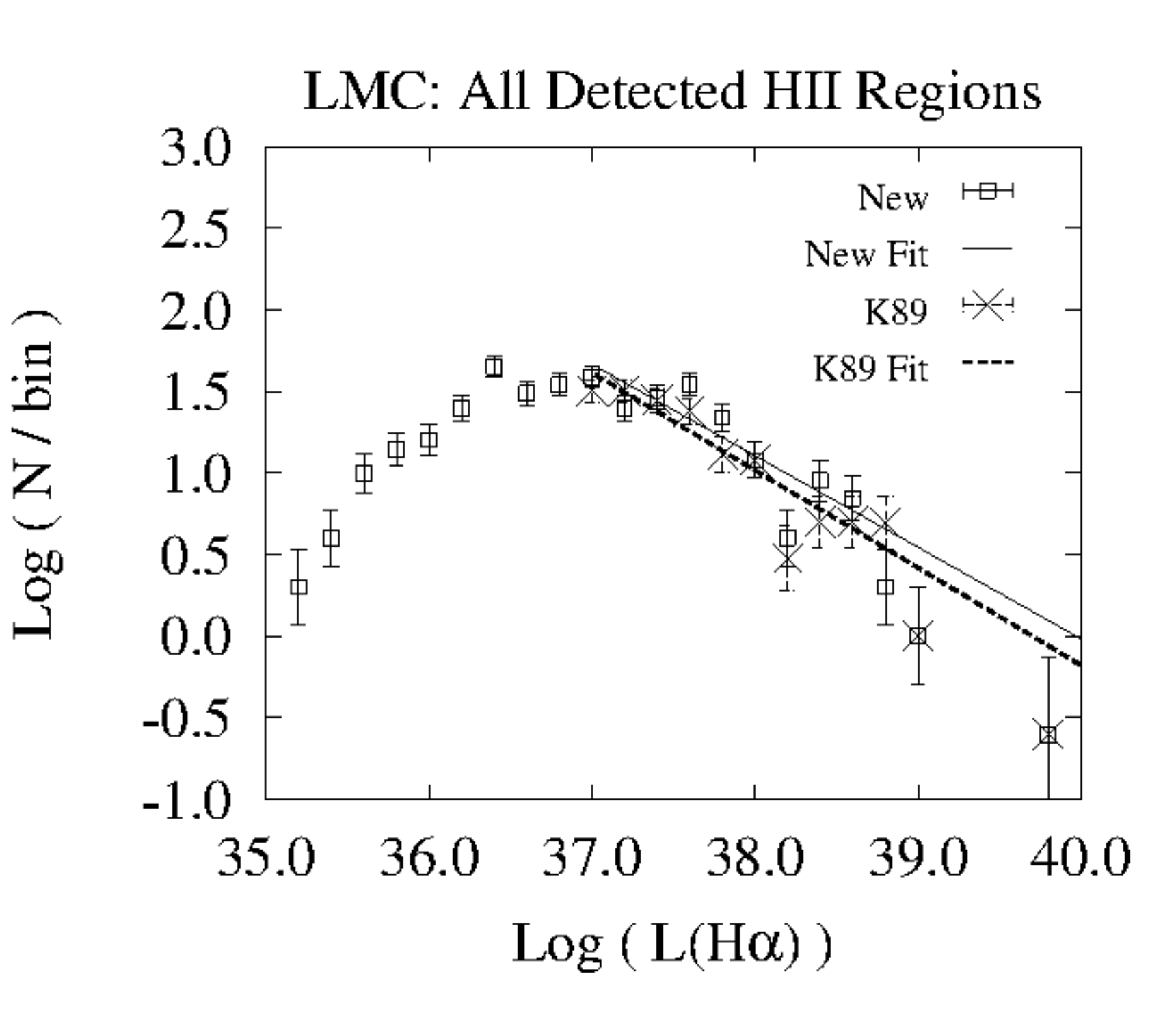}
\label{fig:LMC_LF}
}
\subfigure[SMC]{
  \includegraphics[scale=0.4]{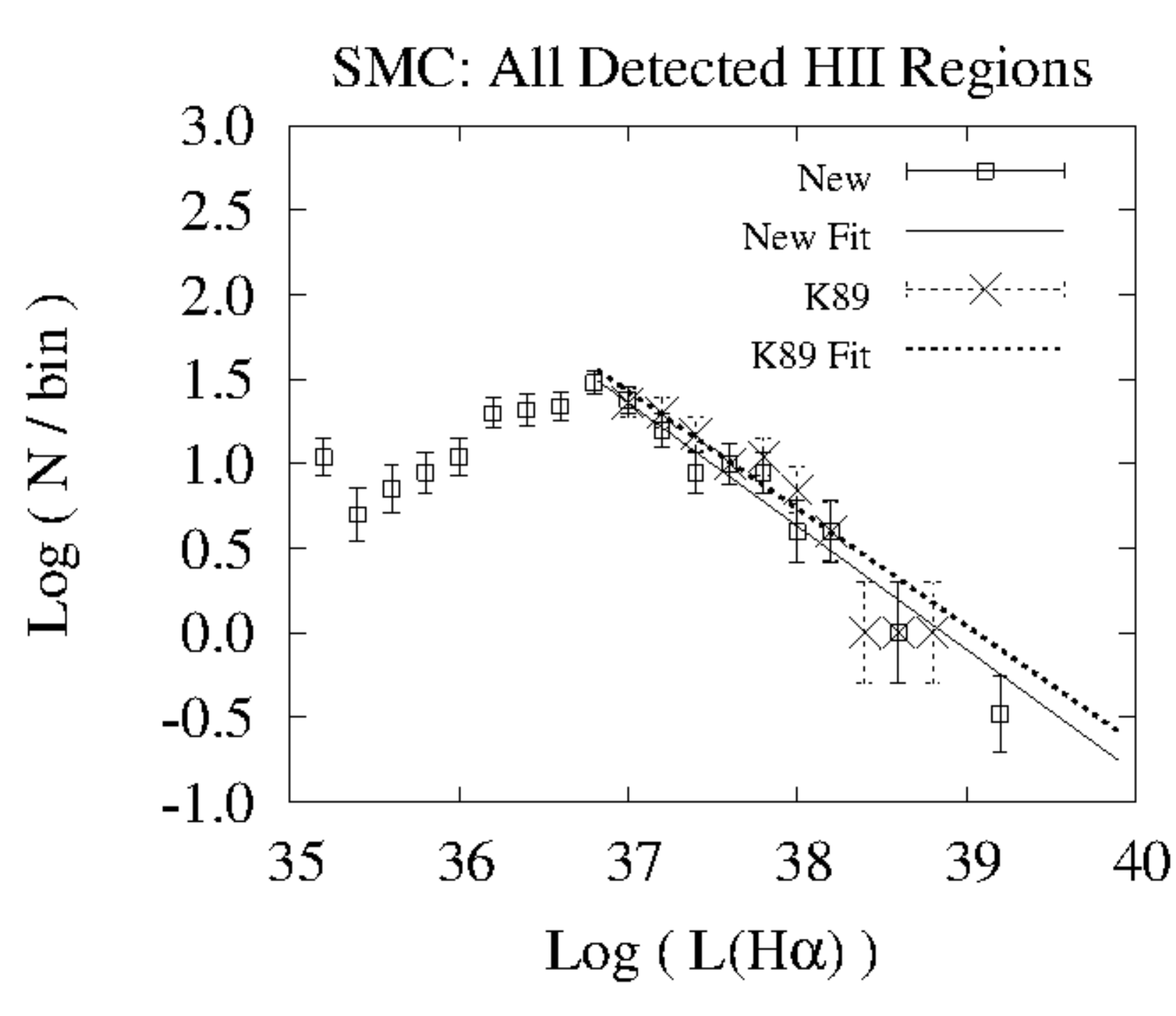}
  \label{fig:SMC_LF}
}
\caption{\footnotesize 
The LMC (panel $a$) and SMC (panel $b$) \HII\ LFs 
for our catalogs of photoionized \HII\ regions (squares), and from
  \citet{Kennicutt1989}(crosses).  Power law fits to the data,
  weighted by inverse error, are shown with the solid and dotted lines,
  respectively, for our data and those of Kennicutt et al.  The error
  bars show the root-$N$ uncertainties.}
\label{fig:LF}
\end{figure}

We find that our \HII\ region boundaries and \Ha\ luminosities
generally agree with those determined in previous, \Ha-only studies by,
e.g., \citet{Kennicutt1989}, including the substructure in most of the DEM
\citep{DEM} and \citet{Henize1956} surveys.  This is especially true
for simple objects with a low local background.  Figures
\ref{fig:LMC_LF} and \ref{fig:SMC_LF} show the differential \HII\
region luminosity functions (\HII\ LF) for our new LMC and SMC
catalogs, respectively (squares), together with those generated by
\citet{Kennicutt1989} (crosses), fitted above $\log\LHa = 37.0$
(where not explicitly stated, the units of \LHa\ are \ergs ).
  The power-law slope of the LMC \HII\ LF reported
by Kennicutt et al. is $B = 1.75 \pm 0.15$, where
\begin{equation}
dN(\LHa)\ \propto \LHa^{-B} d(\LHa) \quad .
\end{equation}
This is statistically consistent with the fitted \HII\ LF slope for our
data, $B = 1.79 \pm 0.08$.  An identical analysis for the SMC, as
shown in Figure~\ref{fig:SMC_LF}, yields an \HII\ LF slope of $B = 1.88\pm
0.09$, for our new catalog, compared to a reported slope of 1.9 from
\citet{Kennicutt1989}.  Thus, although previous measurements of the
\HII\ LF do not use our ionization-based criteria, they result in
essentially identical LF slopes.

Both Magellanic Cloud \HII\ LF slopes flatten around $\log~\LHa =
37.0$, which is equivalent to $Q{\rm (H^0)} = 48.9~\rm s^{-1}$.  This
flattening is observed in other galaxies whose \HII\ LFs probe
$\log~\LHa < 36.0$, including the Milky Way \citep{Paladini2009}, M51
\citep{Lee2011} and M31 \citep{Azimlu2011}.  The observed flattening
of the \HII\ LF in this regime was predicted in Monte Carlo
simulations by \citet{OeyClarke1998} and \citet{Thilker2002}, and it
is caused by stochastic ionizing populations at these low
luminosities.  For comparison, $\log~\LHa = 37.0$ is the luminosity of
the Orion Nebula, whose parent ionizing cluster has a mass of
4500~\Msun \citep{Hillenbrand1998}, dominated by a single O6.5 V star.

A final caveat:  at the lowest luminosities, there is a decrease in the
\HII\ LFs. This is clearly established for $\log~\LHa < 36.5$ in both
galaxies.  The drop in source counts could be an
indication that below this \LHa\ we are not complete. Alternatively,
since the stellar ionizing fluxes plummet strongly for stars later
than early B spectral types, this turnover in the \HII\ LF signals an
intrinsically different class of ionizing sources and nebular objects.
These must include individual \HII\ regions of later B-type stars, and
perhaps some faint, optically thin nebulae that are intrinsically weak
in recombination lines due to their low optical depth.  There also may
be some shock-heated filaments, although we tried to avoid most of
these.  Planetary nebulae should not be important above $\LHa \sim
\rm 5e35~ \ergs$ \citep{Azimlu2011}.

\subsection{An enigmatic, highly ionized region}
\begin{figure}
\centering
\includegraphics[scale=0.24]{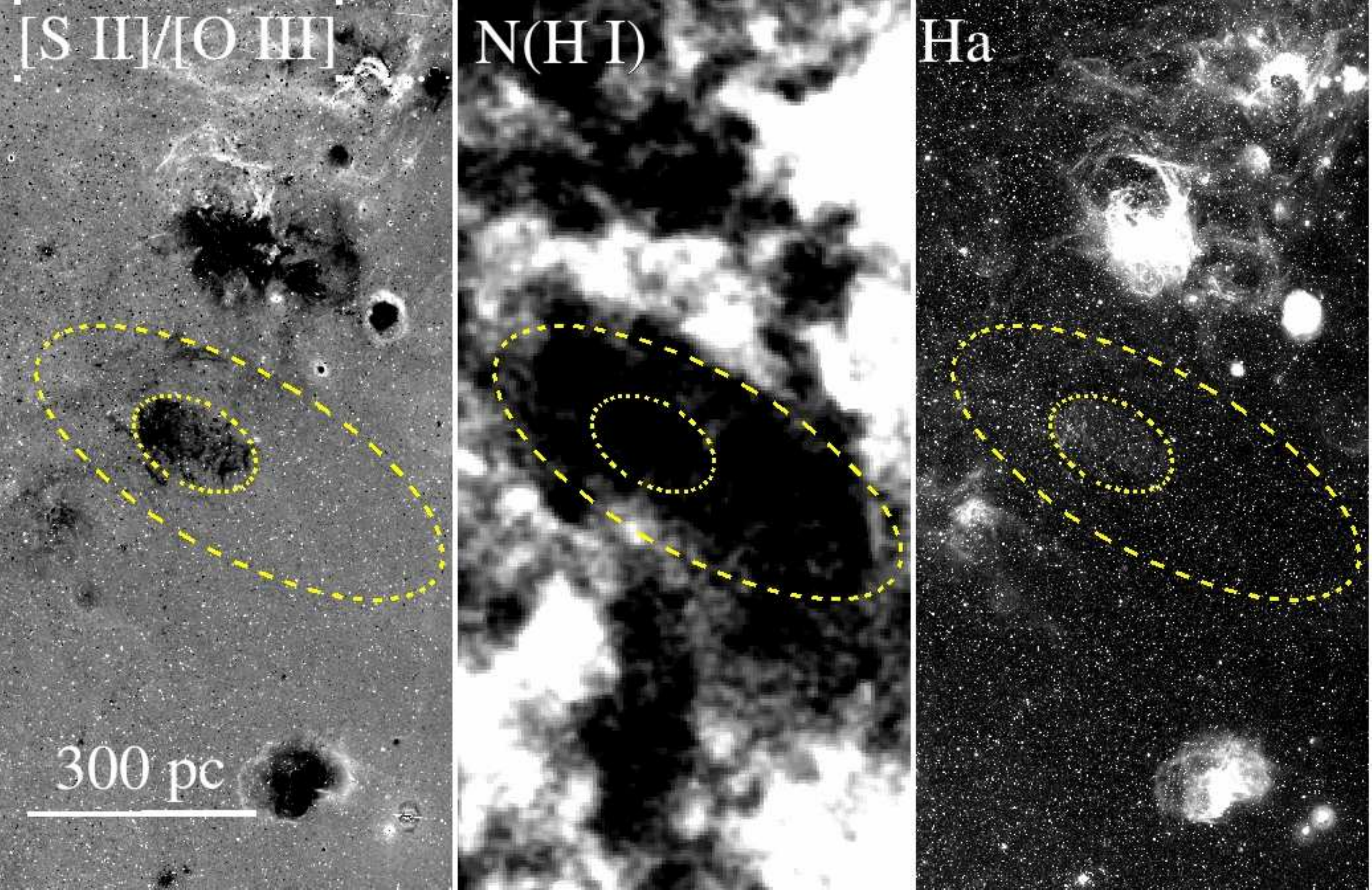}
\caption{A large, highly ionized bubble with no known ionization source.
  Left: \SII/\OIII\ ratio map;  Center:  \HI\ column density map of the
  region, with white showing \HI\ emission;  Right: The region in \Ha.
  All panels are shown with the
  same scale, where north is up and east left.  The central
  contour marks the boundary where the object is indistinguishable
  from the background, while the larger contour marks the rim of the
  bubble seen in \NHI.}
\label{fig:Mystery}
\end{figure}

The technique of ionization-parameter mapping is effective at
highlighting large, extremely faint structures.  The \SII/\OIII\ ratio
map of an LMC object at $\alpha=$ 04:55:50 $\delta=$ 67:30:50 (J2000) is shown
in the left panel of Figure \ref{fig:Mystery}, with an inner ellipse
to mark the extent of highly ionized and filamentary gas.  The middle
panel shows a larger contour that highlights an HI cavity seen in the
\HI\ data of \citet{Kim2003} with a major axis of 550 pc and \HI\
column density \NHI~$=0.6\times 10^{21} ~{\rm cm^{-2}}$.  Both
ellipses have the same orientation suggesting they are related.  The
right panel of Figure~\ref{fig:Mystery} shows faint \Ha\ emission is
co-spatial with the \OIII, while no \SII\ was detected.  This
indicates that the optically emitting gas is fully ionized.  This
structure is intriguing because the \SII/\OIII\ morphology is similar
to optically thin nebulae ionized by OB stars, yet no ionization
sources are known in the region, nor is there evidence of a prior
supernova or shocked gas.  The size and faintness of this highly
excited region, together with the lack of an ionization source, make
this object unique.  Further observations to identify its nature and
origin are required.

\section{Optical Depth of the H~II Regions}
\label{sec:HIItau}

From the diagnostics based on ionization structure as described in \S
\ref{sec:method}, we classify the optical depth of our individual,
catalogued \HII\ regions into the following categories: (0)
indeterminate, (1) optically thick, (2) blister, (3) optically thin,
and (4) shocked nebulae.  These are given in Column 4 of
Tables~\ref{tab:LMCObjs} and \ref{tab:SMCObjs}.  Class 0 objects, with
indeterminate optical depth, fall into two categories: those which
lack \OIII\ emission, causing a high \SII/\OIII\ ratio, with little
ionization structure; and large scale, diffuse structures.  This
latter category is difficult to define morphologically, but since many
objects in the DEM LMC catalog include these features, we have
attempted to catalog them as well.  

We define optically thick objects (class 1) to be those showing classic, low-ionization
envelopes enclosing at least 2/3 of the central, high-ionization
regions in projection, as described in \S \ref{sec:method}.  Blister nebulae (class
2) are defined by a low-ionization envelope that surrounds between 1/3
and 2/3 of the observed object; additionally, objects having complex
internal ionization fronts with extended [O III] emission are treated
as blisters.  Optically thin (class 3) objects show low \SII/\OIII\
  throughout, with low-ionization envelopes covering $<$
  1/3 of the highly ionized gas (see Figures~\ref{fig:DEMS38_S2O3} and
  \ref{fig:DEMS159_S2O3}).  Shocked
objects (class 4) are characterized by an ionization structure which
is inverted relative to photoionization, i.e. these objects are have
enhanced \OIII\ emission surrounding strong \SII.  Our survey 
is not intended to be complete with respect to shocked objects, and we typically
avoided cataloguing them.  Since these are not photoionized, they are
excluded from further consideration.  Additional general classification
criteria include gradients in 
  ionization parameter, the detection of ionization fronts distinct
  from the background, and the ionized extent of the object.  Radial
  projections of the \SII/\OIII\ ratio, were made to assess the
  significance of specific individual features that were identified.
  Three of the authors (EWP, JZ, and AEJ) used these criteria to carry
  out independent classifications of all the objects.  To arrive at a
  final catalog, we resolved the differences by discussing specific
  key features and quantitatively measuring the optically thin
  covering fraction.

We roughly estimate the LyC escape fraction $\fesc$ for each object in
classes (1) -- (3) as follows: optically thick objects are assigned
$\fesc = 0$; optically thin objects are assigned $\taulyc = 0.5$,
which corresponds to $\fesc = 0.6$; and blister objects are assigned a
value of $\fesc$ that is half that for the optically thin objects,
namely, $\fesc = 0.3$.
Since we have observations in only two diagnostic ions, some of the
class 3 objects in reality may be quite optically thin (\S
\ref{sec:limitations}), and so we compare these estimates 
to direct measurements of optical depths using data from
\citet{Voges2008}, who compared the observed \LHa\ to predicted values
based on the spectral types of individual ionizing stars for a sample
of LMC \HII\ regions.  Although, as
  mentioned earlier, there is considerable uncertainty in optical depth
  estimates based on this method, it remains the most direct, quantitative
  way to check our results based on ionization-parameter mapping.

\citet{Voges2008} adopted ionizing fluxes
from the WM-basic models of \citet{Smith2002}. These SEDs are
intermediate in hardness among the different available, modern codes,
and they best fit the observed nebular emission-line spectra
\citep{Zastrow2011a}.  Due to the large uncertainty in determining
spectral types of O stars from photometry alone, we restrict our
comparison with the \citet{Voges2008} sample (their Tables 1 and 3) to
objects having stellar spectral types determined at least in part by
spectroscopic classifications, with a further requirement that at
least half of the derived ionizing luminosity is attributed to stars
with spectroscopic spectral types.  These include a reanalysis of
\HII\ regions from \citet{Oey1997}, which is based entirely on
spectroscopic classifications with individually measured reddenings,
listed in Table~1 of Voges et al.  We exclude DEM~L~7, L~9 and L~55
because they are class 0 objects, and DEM~L~229, which
displays evidence of shock excitation.

Using the predicted \Ha\ luminosities from \citet{Voges2008}, and our
new observed \Ha\ luminosities, we calculate individual \fesc\ values
for each nebula according to
\begin{equation} 
\label{eq:Vogesfesc}
f_{\rm esc} = 1 - L_{\rm obs}/L_{\rm predicted} \quad .
\end{equation}
The predicted luminosity is derived from the expected rate of ionizing
photons $Q{\rm (H^0)}$ assuming each absorbed LyC photon will result
in 2.2 \Ha\ photons.  In Table~\ref{tab:VogesComp}, we present the
\fesc\ values for the 13 \HII\ regions, comparing the rough estimates
obtained from ionization-parameter mapping as described above with
measurements based on the data of \citet{Voges2008}. Column~1 gives
the DEM identifier for each object, and column 2 gives our crude
\fesc\ estimated from ionization-parameter mapping, as described
above.  Column 3 gives \fesc\ estimates based on data from
\citet{Voges2008} with values derived from known stellar spectral
types.

\begin{figure}[b]
\centering
\includegraphics[scale=0.4]{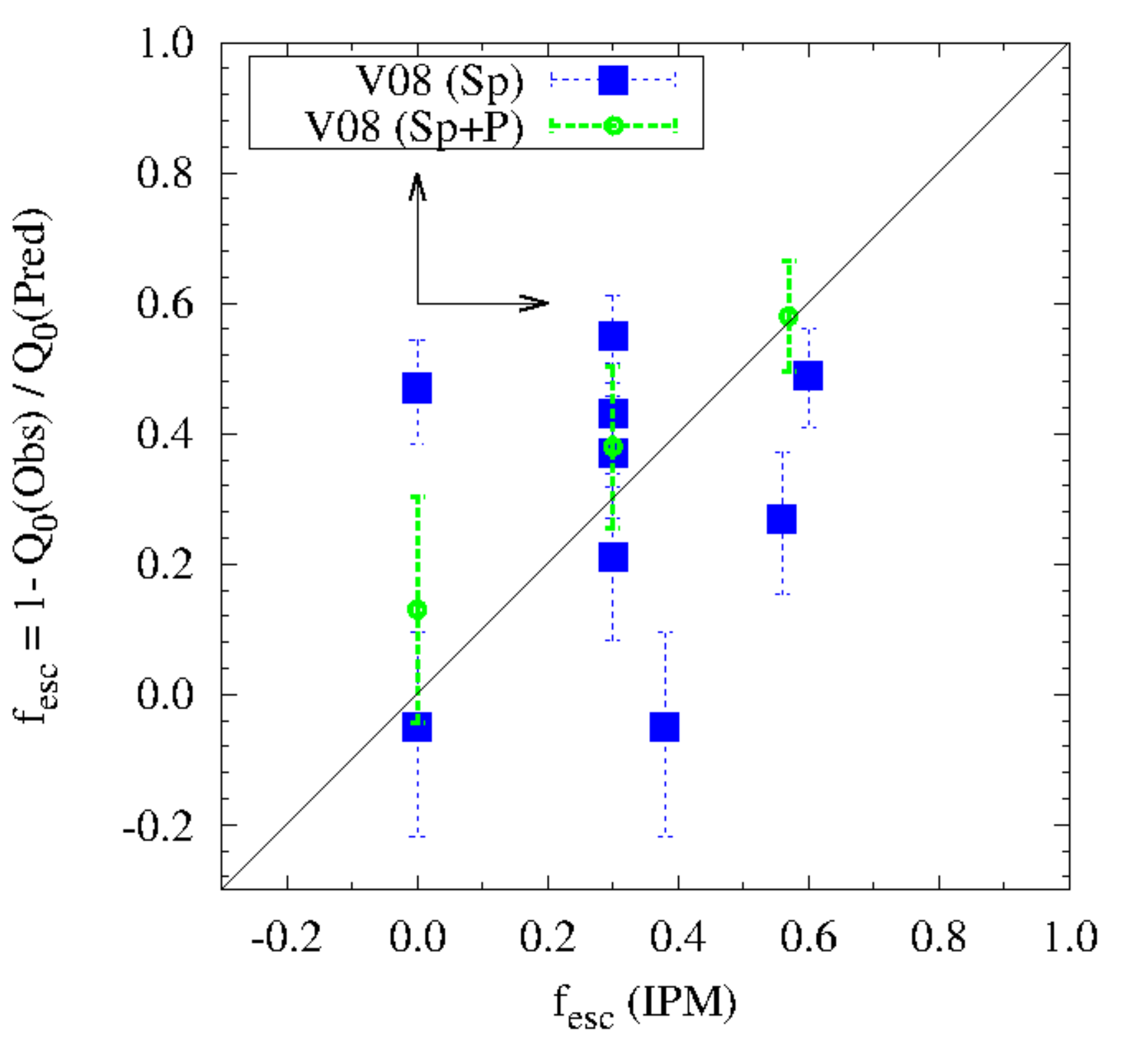}
\caption{\footnotesize Our \fesc\ values estimated from
  ionization-parameter mapping plotted against \fesc\ calculated from
  observed and predicted ionization rates from \citet{Voges2008}.
  The values of \fesc\ derived from both methods are lower limits,
  illustrated by the orthogonal set of arrows.}
\label{fig:VogesComp}
\end{figure}

Figure~\ref{fig:VogesComp} plots the comparison between $\fesc$
estimated from our classification of optical depth based on
ionization-parameter mapping and the measured values based on the data
of \citet{Voges2008}.  There is a general agreement between our crude
estimates for \fesc\ based on ionization-parameter mapping and the
measured values based on the observed ionizing stars, for all but 1
object; the standard deviation from the identity relation is $\sigma
=0.23$, excluding DEM L 293 (see below).  Although, as discussed in \S
\ref{sec:limitations}, our values for \fesc\ are all lower limits,
especially for objects categorized as optically thick, and the
\fesc\ values derived using \citet{Voges2008} are also lower limits,
the surprisingly good correspondence suggests that both methods
actually yield reasonable estimates of the optical depth.

The Voges et al. escape fraction of DEM L 293 is
--1.52, which is an unphysical value, placing it far beyond the bounds of
the plot in Figure~\ref{fig:VogesComp}.  The predicted ionizing
luminosity in DEM L 293 is observationally attributed to only a single
O3~III star.  However, given the 
typical cluster mass in which O3~III stars form, additional, obscured
or overlooked ionization sources in this cluster are likely to be
present.  In particular, \citet{Walborn2002} identified an odd
semi-stellar source within DEM L 293 which is brighter than the single
O3~III star.  If the ionizing luminosity of this source is equal to an
O3~III star then the V08 \fesc\ would then be -0.52, much closer to
the \fesc\ value derived from ionization-parameter mapping.  


Overall, however, our extremely crude estimates of \fesc\ based on
ionization-parameter mapping show surprisingly good agreement with the
empirically measured \fesc.  In spite of the fact that our estimates
tend to yield lower limits, the general agreement confirms that
objects appearing to be optically thick indeed tend to be
radiation-bounded.  As mentioned in \S \ref{sec:evalnebtau}, this is
also supported by their morphologies, which generally resemble smooth,
Str\"omgren spheres.  Furthermore, the occurrence of optically thick objects that
appear to be density-bounded is extremely rare, since this only happens
occasionally for the very hottest spectral types (e.g.,
Figure~\ref{fig:LMC_Models_nH_10}).  Figure~\ref{fig:VogesComp} thus demonstrates the
general viability of ionization-parameter mapping as a diagnostic of
nebular optical depth.

\begin{deluxetable}{lcc}
  \tabletypesize{\footnotesize} 
\tablecaption{Comparison of \fesc\ estimates}
\tablehead{\colhead{Object}&\colhead{\fesc(IPM)$^a$}&
\colhead{\fesc(V08)$^b$}}
\startdata
DEM L 10B$^c$&0.38&-0.05\\
DEM L 13 &0.30&0.38\\
DEM L 31 &0.60&0.49\\
DEM L 34 &0.30&0.55\\
DEM L 68$^c$ &0.59&0.58\\
DEM L 106&0.30&0.37\\
DEM L 152+156&0.56&0.27\\
DEM L 196$^c$&0.03&0.13\\
DEM L 226&0.00&-0.05\\
DEM L 243&0.00&0.47\\
DEM L 293&0.30&-1.52\\
DEM L 301&0.30&0.21\\
DEM L 323+326&0.30&0.43

\enddata 
\label{tab:VogesComp}
\tablecomments{\\
  $^a$ \fesc\ derived via ionization-parameter mapping.\\
  $^b$ \fesc\ derived from predicted \Ha\ fluxes from \citet{Voges2008} and equation \ref{eq:Vogesfesc}.\\
  $^c$ Composite objects whose \fesc\ values reflect the
  luminosity-weighted contributions of individual optically thin and
  thick subregions.  }
\end{deluxetable}

\subsection{Optical depth and \Ha\ luminosity}
\label{sec:LHa}

\begin{figure}[b]
\centering
\includegraphics[scale=0.4]{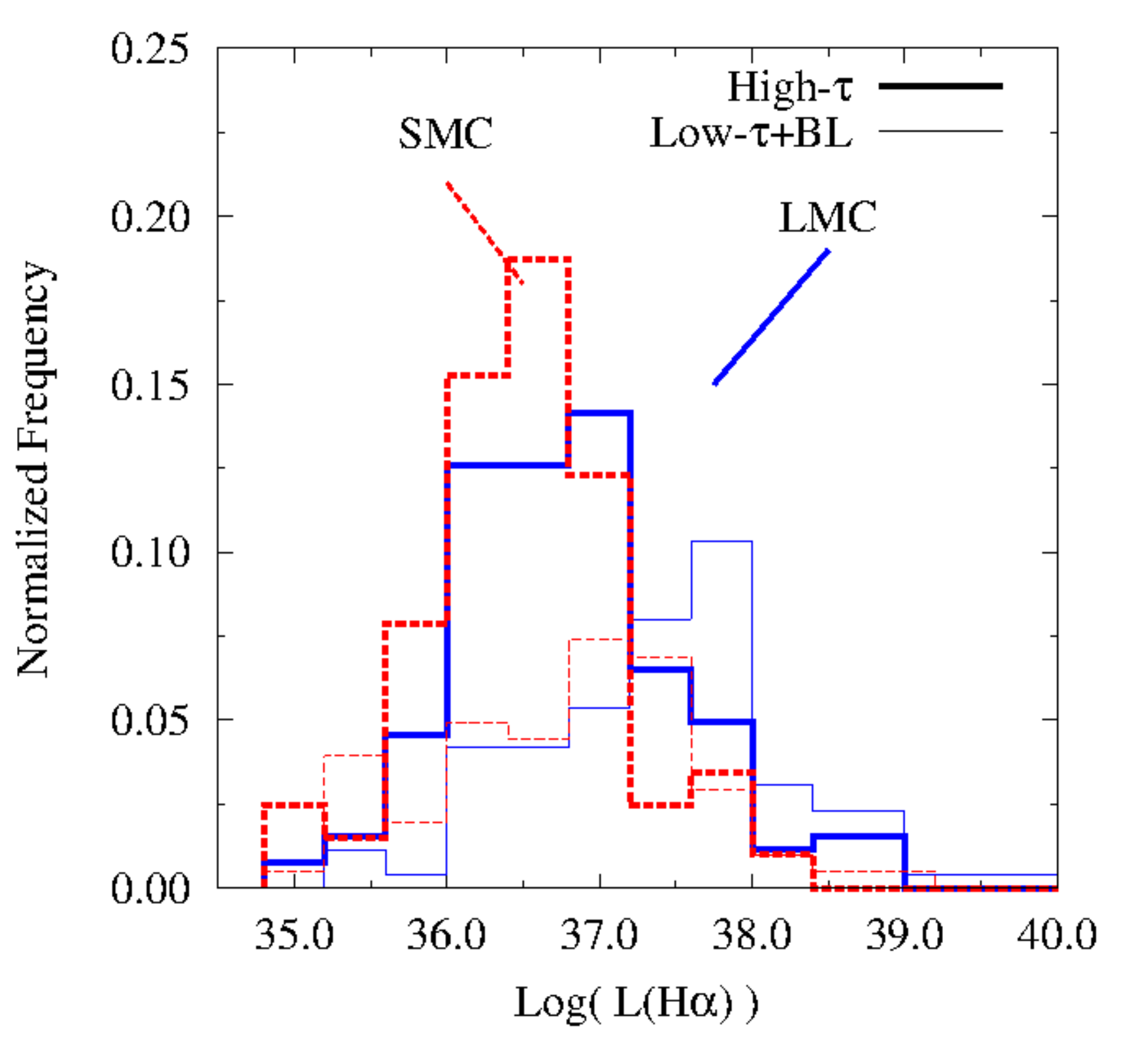} 
\caption{\footnotesize The LFs for optically thin 
  nebulae, including blister objects; and optically thick nebulae,
  shown with thin and thick lines, respectively.  The LMC
  distributions are shown with solid lines (blue) and the SMC are
  shown with dashed lines (red).}
\label{fig:HaDist}
\end{figure}

\begin{deluxetable*}{lcccc|ccccc}
\tabletypesize{\footnotesize}
\tablecaption{Median H~II Region Properties}
\tablehead{
  \colhead{}      &
  \multicolumn{4}{c}{LMC}   &
  \multicolumn{4}{c}{SMC}\\
  \hline
  \colhead{Class}               &
  \colhead{No.} &
  \colhead{\%$^a$} &
  \colhead{$L$}        &
  \colhead{\NHI}     &
  \colhead{No.} &
  \colhead{\%$^a$} &
  \colhead{$L$}        &
  \colhead{\NHI}     & \\
 & & & $10^{36}\ \rm erg\ s^{-1}$ & $10^{21}\ \rm cm^{-2}$ &
& &  $10^{36}\ \rm erg\ s^{-1}$ & $10^{21}\ \rm cm^{-2}$ 
}
\startdata
(0) Indeterminate         & 130 & \nodata   & 3.7  & 1.8 & 7   & \nodata & 0.05 &6.5 \\
(1) Opt Thick             & 158 & 60        & 3.6  & 2.8 & 132 & 62      & 2.1 & 6.4 \\
(2) Blister               & 58  & 18        & 18.8 & 1.9 & 41  & 19      & 4.7 & 5.1 \\
(3) Opt Thin              & 46  & 22        & 19.7 & 2.0 & 30  & 14      & 4.0 & 6.2 \\
(4) Shocked               & 9   &\nodata    & 15.6 & 1.8 &  4  & 2       & 1.5 & 5.7 \\
(2) $+$ (3)               & 104 & 40        & 19.2 & 1.9 & 71  & 33      & 4.5 & 5.3 \\
(1)$+$(2)$+$(3)           & 262 &100        & 5.5  & 2.5 & 203 & 100     & 2.6 & 5.9
\enddata 
\label{tab:medHII}
\tablecomments{$^a$Percentages are calculated for photoionized
  objects, based on total values in the bottom row.} 
\end{deluxetable*}

Table~\ref{tab:medHII} summarizes the median nebular properties for
each optical depth class as catalogued in Tables~\ref{tab:LMCObjs} and
\ref{tab:SMCObjs}.  Column 2 gives the number of objects in each
class in the LMC.  Column 3 gives the corresponding percentage of the total
number of objects that are clearly photoionized, thus excluding class
0 and class 4 objects from the total numbers of photoionized nebulae.
Columns 4 and 5 give the median \LHa\ and median \NHI\ (see \S
\ref{sec:NHI} below) associated with the objects, respectively.  
Columns 6 -- 9 list the same quantities for the SMC.

Figure~\ref{fig:HaDist} shows the \LHa\ distributions for the LMC
(solid, blue line) and SMC (dotted, red line), of optically thin
nebulae (thin lines) and optically thick nebulae (thick lines).  The
optically thin data include blister \HII\ regions as defined above.
The distributions are normalized by the number of \HII\ regions in the
last row of Table \ref{tab:medHII}.  Figure~\ref{fig:HaDist} and
Table~\ref{tab:medHII} show that for both galaxies, the \LHa\
distributions for optically thick objects peak at lower luminosities
than those for the optically thin ones.  This difference is larger in
the LMC, producing a bimodal distribution, with the median
\LHa\ for optically thin nebulae 5 times brighter than for the
optically thick ones (Table~\ref{tab:medHII}).  The median \LHa\ of
optically thin SMC nebulae is only twice that of optically thick ones, as
expected given the lower star-formation rate in that galaxy, and fewer
luminous \HII\ regions.  However, the \LHa\ distributions for the
different classes are similar between the two galaxies, showing peaks
near similar values and similar ranges in luminosity.  The role
  of dust in these trends is unclear.  It is possible that two \HII\
  regions with similar ionizing luminosities will have different
  \fesc\ if one has more dust than the other.  This could explain the
  coexistence of optically thin and thick regions in the same
  luminosity bin.

\begin{figure}[t]
\centering
\subfigure[LMC]{
\includegraphics[scale=.3]{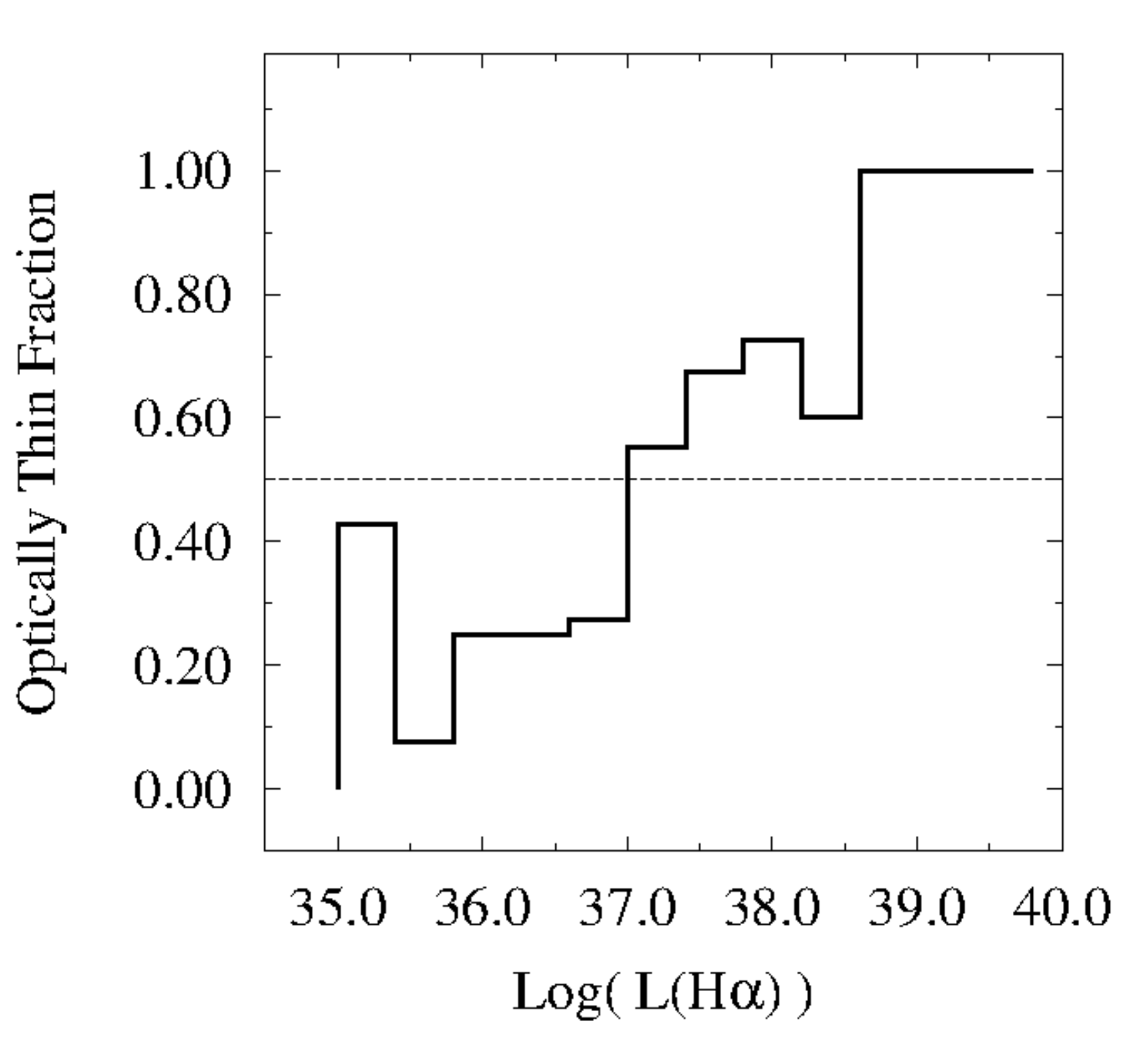}
\label{fig:LMC_frac_Ha}
}
\subfigure[SMC]{
\includegraphics[scale=0.3]{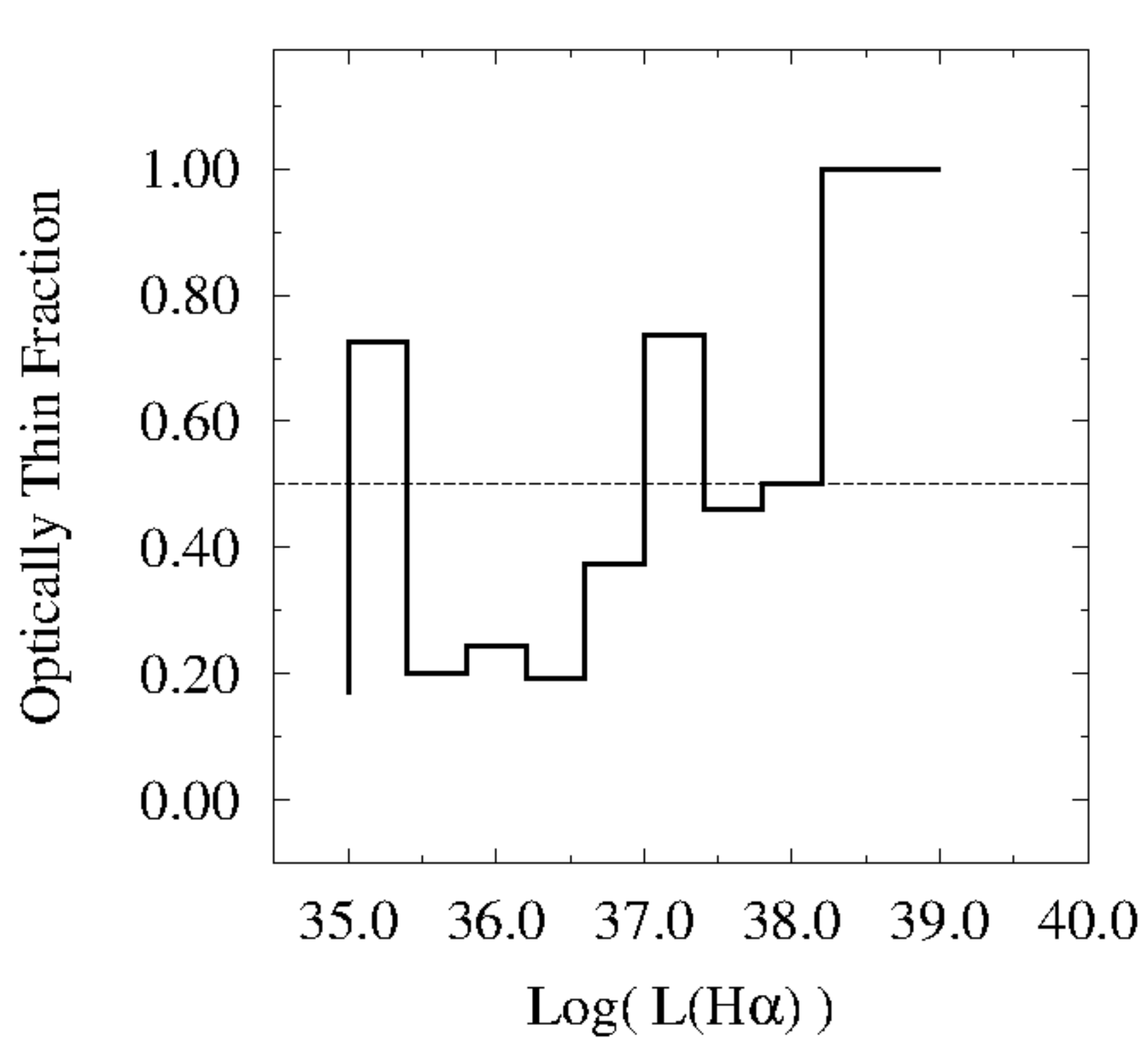}
\label{fig:SMC_frac_Ha}
}
\caption{The fraction of optically thin \HII\ regions as a function of
  \LHa\ is shown for the LMC and SMC in panels \subref{fig:LMC_frac_Ha} and
  \subref{fig:SMC_frac_Ha}, respectively.  Optically thin objects
  dominate above a fraction of 0.5, indicated by
  the horizontal dotted line.  The lowest-$L$ bins likely include
  contamination by planetary nebulae.
}
\label{fig:frac_Ha}
\end{figure}

In Figure~\ref{fig:frac_Ha}, we plot the frequencies of optically thin
nebulae as a function of \LHa.  Both galaxies exhibit a clear increase
in the frequency of optically thin nebulae with increasing \Ha\
luminosity.  However, we stress that both optically thick and thin
objects are found at almost all luminosities having $\log~\LHa <
39.0$.  Table~\ref{tab:medHII} also shows that the frequencies of
optically thin objects,
including blisters, are similar between the two galaxies, 40\% and
33\% in the LMC and SMC, respectively.  Furthermore,
there is a transition luminosity above which optically-thin nebulae
dominate, $\log~\LHa = 37.0$, occurring at the same luminosity in both
galaxies.  \citet{Beckman2000} speculated that such a transition is
responsible for possible discontinuities observed near $\log~\LHa =
38.6$ in extragalactic \HII\ LFs, but our data clearly show that
optically thin objects dominate at luminosities a full 1.6~dex lower
in \LHa. 

It will be interesting to see how strongly our transition value of
$\log~\LHa = 37.0$ depends on galaxy properties.  This relatively low
luminosity corresponds to nebulae ionized stochastically by single O
stars or substantially evolved associations and clusters
\citep{OeyClarke1998}.  Thus, most of the objects typically apparent
in Figures \ref{fig:LMCS2O3} and \ref{fig:SMCS2O3}, as well as those
typically detected in local surveys (e.g. \citealt{Thilker2002}) are
the more luminous \HII\ regions, which are mostly, but not all,
optically thin.  The most luminous objects have the highest likelihood
of being optically thin, including 30~Doradus in the LMC, and the N66
in the SMC.  These are indeed found to be optically thin in our study,
a result consistent with the findings of \citet{Pellegrini2011} in 30
Dor, and the low optical depths found for other giant extragalactic
\HII\ regions \citep[e.g.,][]{Castellanos2002}.

\subsection{Relation with the Neutral ISM}
\label{sec:NHI}

The neutral ISM represents the default environment into which ionizing
photons from optically thin regions are deposited, and its properties
are fundamental to the radiative transfer of the Lyman continuum.  The
Magellanic Clouds were mapped in \HI\ with the Australia Telescope
Compact Array by \citet{Kim2003} (LMC) and \citet{Stanimirovic1999}
(SMC).  The LMC \HI\ data
have 60 arcsec resolution over the ${\rm
  11.1^{\circ}\times12.4^{\circ}}$ survey area; the SMC \HI\ data have
a resolution of 98 arcsec over the ${\rm 5^{\circ}\times5^{\circ}}$
field.  The LMC is a face-on disk galaxy, while the SMC has a more
amorphous, three-dimensional irregular morphology.
We now explore the relationship between nebular optical depth
and the neutral ISM.  

\begin{figure*}
\centering
\subfigure[LMC]{
\includegraphics[scale=0.35]{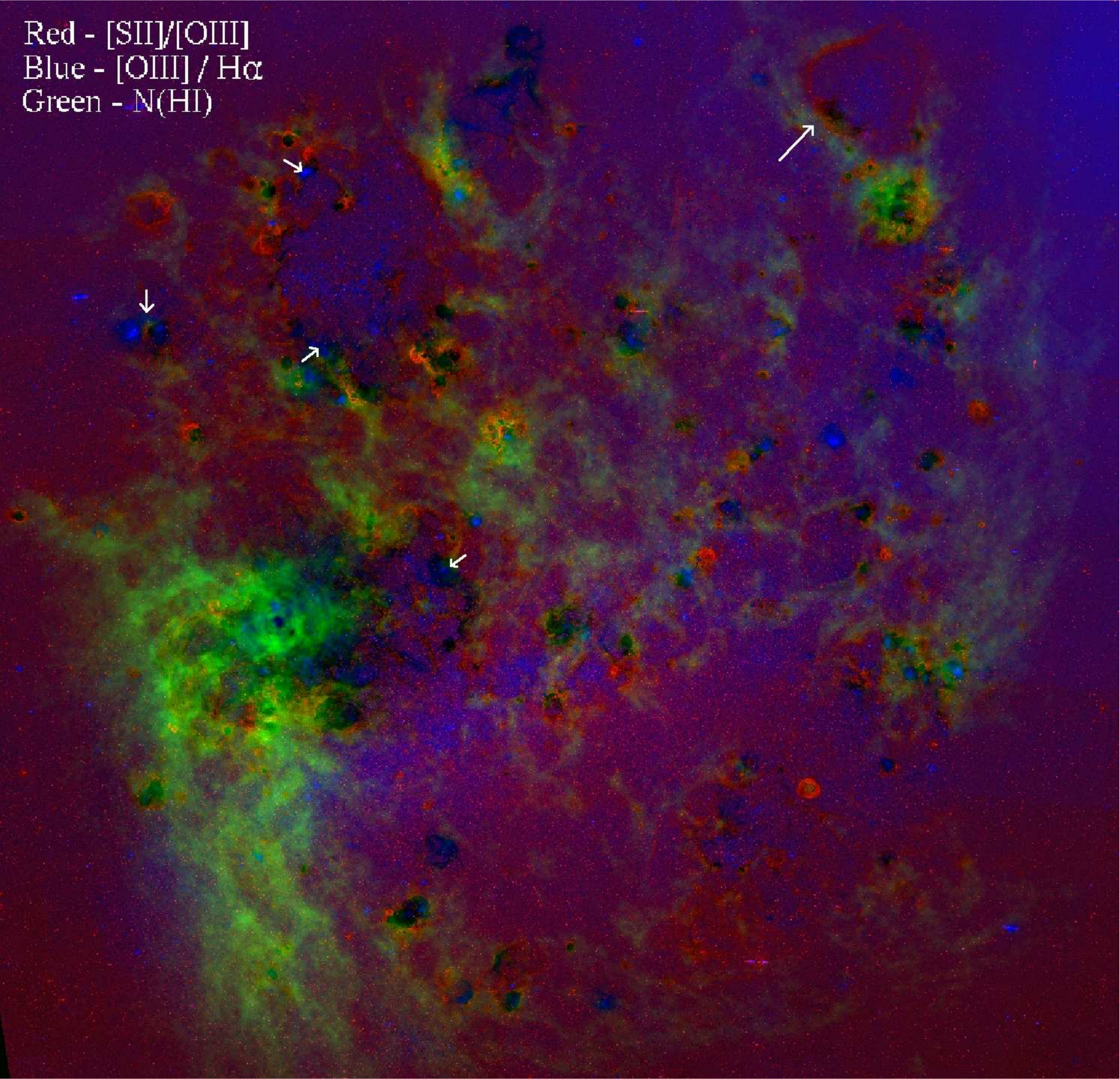}
\label{fig:LMCrgb}
}
\subfigure[SMC]{
\includegraphics[scale=0.35]{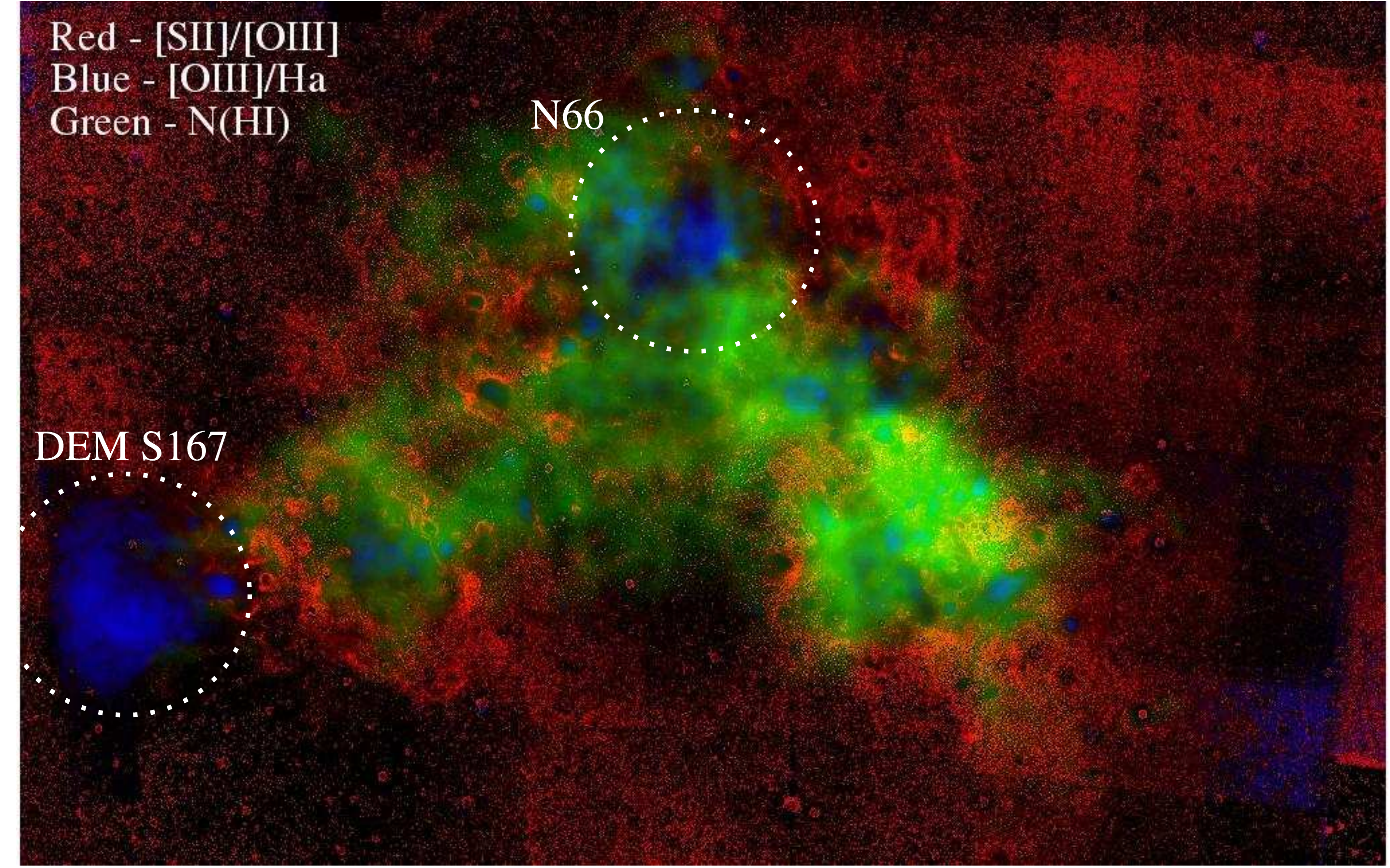}
\label{fig:SMCrgb}
}
\caption{\footnotesize Composite images of the LMC (top) and SMC (bottom), which
  contrast the different ISM structure seen in the two galaxies.
  \OIII/\Ha\ (blue), \SII/\OIII\ (red), and \NHI\ (green) respectively trace
  high-excitation nebulae, low-excitation gas, and neutral gas
  along the line of sight.
}
\label{fig:3phaseColor}
\end{figure*}

Figure~\ref{fig:3phaseColor} 
traces the propagation of radiation in the LMC (top) and SMC (bottom)
using \SII/\OIII\ (red), \OIII/\Ha\ (blue) and \NHI\ (green).  
The contrast in ionization morphology between the two galaxies is
striking.  As ionizing radiation enters the diffuse ISM, it encounters
a combination of ionized and neutral gas.  The LMC neutral disk has
been disrupted, forming shells and filaments surrounding the ionized
gas \citep{Kim1998}.  These structures are believed to be the result
of stellar feedback acting on the ISM (e.g. \citealt{OeyClarke1997}).
Often, optically thin \HII\ regions line the edges of large \HI\
shells, radiating LyC photons into their interiors.  A few examples
are highlighted with arrows in the LMC (Figure~\ref{fig:3phaseColor}).
These large-scale \HI\ structures appear to allow ionizing
radiation to travel hundreds of pc without being absorbed.

\begin{figure*}
\centering
\includegraphics[scale=0.5]{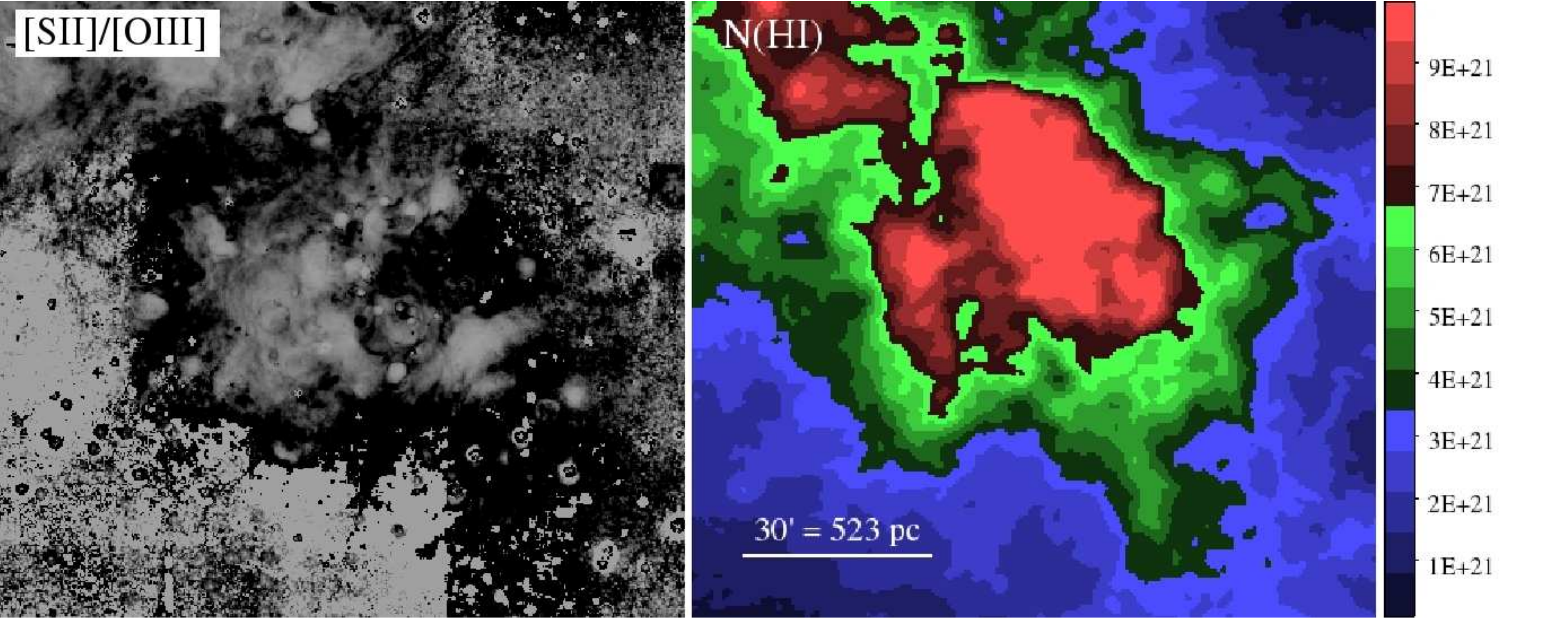}
\caption{\footnotesize The southwest region of the SMC in \SII/\OIII\
  (left), smoothed with an $11\times 11\ \rm arcsec^2$ median filter.
  We have inverted the gray scale to better highlight low-ionization
  regions, thus black corresponds to areas of low ionization.  Surrounding the
  entire complex is an ionization transition zone (black).  This
  transition zone is coincident with the boundaries of a massive \HI\
  cloud seen in the \NHI\ map of \citet{Stanimirovic1999} (right).}
\label{fig:SMC_SW}
\end{figure*}

The SMC, on the other hand, has a less fragmented \HI\ structure
\citep{Stanimirovic1999}.  The neutral ISM in this galaxy is much more
diffuse and less filamentary than that in the LMC \citep{Oey2007}.
Intense and vigorous star formation is at the center of the two most
prominent \HI\ masses.  The first is coincident with N66 near the
northern boundary of the galaxy.  Lines of sight toward this optically
thin region show that \NHI\ anti-correlates with highly ionized gas
(Figure~\ref{fig:3phaseColor}).  Thus, in this region, \HI\ is being
disrupted by the ionizing radiation entering the diffuse ISM.  The
second region is located in the SW portion of the galaxy.  Despite a
high \NHI, this region contains many optically thin nebulae, which
form a large complex filled with a highly ionized DIG.  To improve our
sensitivity to ionization transitions in the DIG, we applied an
$11\times11~\rm arcsec^{2}$ median filter to the \SII\ and \OIII\
data, creating a smoothed \SII/\OIII\ map.  The region is seen in
Figure~\ref{fig:SMC_SW} in the inverted map (left) and \NHI ~(right).
The enhanced sensitivity reveals an ionization transition zone
coincident with the edge of the \HI\ distribution.  Thus, the \HI\ gas
appears to be trapping the ionizing radiation, while individual
nebular \fesc\ depends on the detailed morphology.

Given the strong morphological contrast between the two galaxies, in
the \HI\ properties and 
star-formation intensity, the quantitative similarities in the
nebular optical depths found above in \S \ref{sec:HIItau} are
surprising.  In particular, despite expectations that the reduced
\NHI\ and higher star-formation intensity of the LMC would lead to
more optically thin nebulae, we saw above that the relative frequency of
optically thin and thick objects are similar between the two
galaxies (Table~\ref{tab:medHII}).  It will be important to
see whether other galaxies also yield similar relative frequencies.

\subsection{Optical Depth and \NHI}
\label{sec:tauNHI}

\begin{figure}[b]
\centering
\includegraphics[scale=0.3]{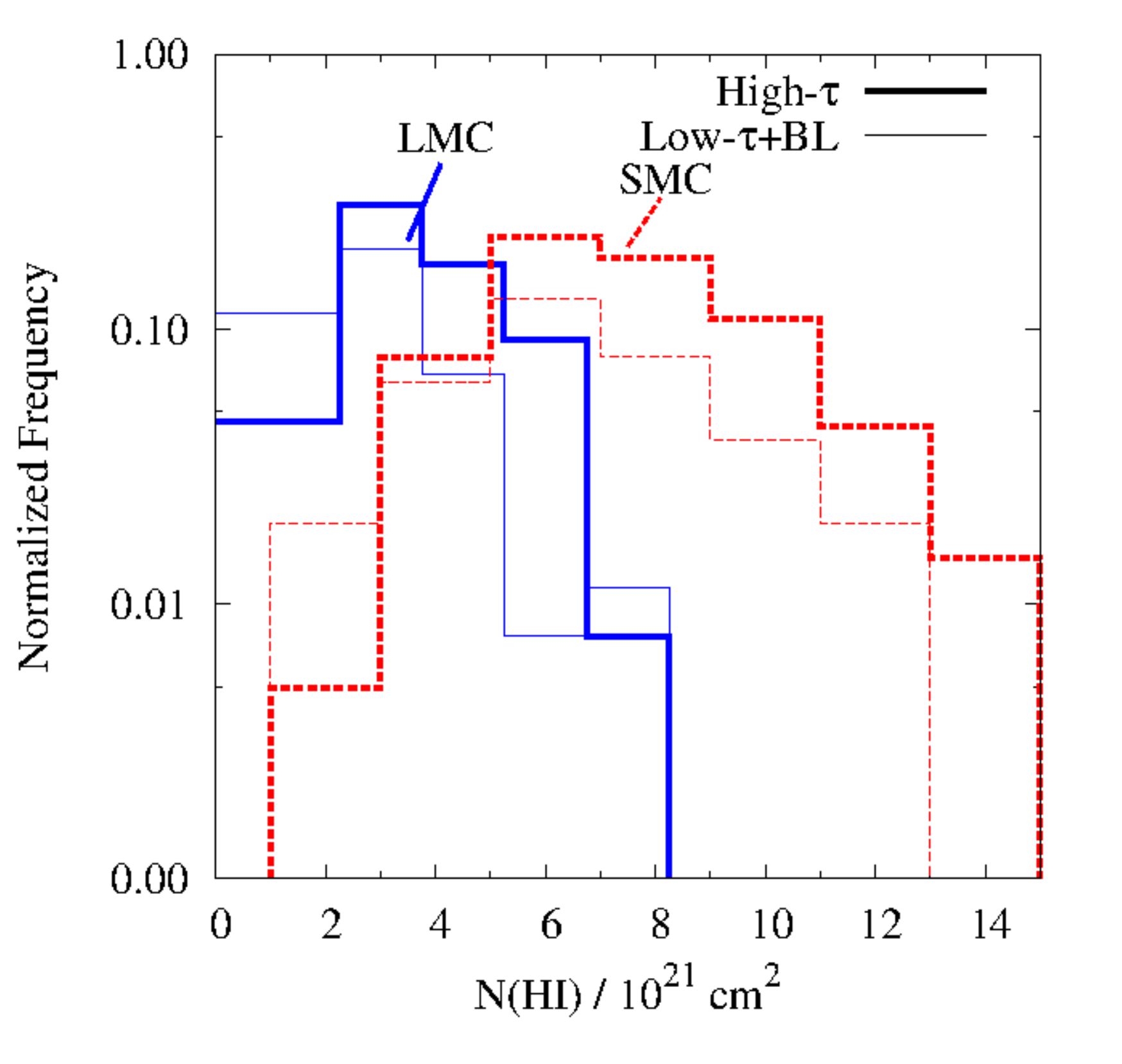} 
\caption{\footnotesize The normalized \NHI\ distributions for 
 optically thin and optically thick nebulae, shown with thin and thick
 lines, respectively.  The LMC and SMC distributions are shown with solid
 lines (blue) and dashed lines (red), respectively.  These have been
 scaled by the total number of objects in each group.}
\label{fig:HIDist}
\end{figure}

Figure~\ref{fig:HIDist} shows the \NHI\ distribution for optically
thin (thin lines) and thick nebulae (thick lines) for the LMC (solid
blue lines) and SMC (dotted red lines).  We used the \NHI\ maps of
\citet{Kim2003} for the LMC and \citet{Stanimirovic1999} for the SMC,
to average the \NHI\ within the individual \HII\ region apertures for
each object, as defined in \S \ref{sec:IBHIIcat}.  We caution that
these measurements correspond to the \NHI\ along the line of sight
toward the objects.  Because the SMC has a more three-dimensional
geometry than the LMC, which is an almost face-on disk, the \NHI\
measurements for the SMC include a larger contribution from foreground
and background ISM than in the LMC.  However, we note that the SMC
metallicity and dust content are only one-fifth that of the LMC; thus,
for the same \NHI, \taulyc\ will be lower in the SMC.  Hence the less
abundant dust may somewhat offset the effect of increased \NHI\ in
this galaxy.  Still, because of the contrasting galaxy morphologies,
the difference in median \NHI\ between optically thin and thick
populations within each galaxy is smaller than the difference in
global median \NHI\ between the two galaxies.  Specifically, the
median value of all LMC \HII\ regions is 2.4 times lower than in the
SMC (Table~\ref{tab:medHII}), while the ratio of the median \NHI\ for
optically thick (class 1) to thin (class 2 $+$ 3) objects is
1.5 and 1.2 in the LMC and SMC, respectively.
Figure~\ref{fig:HIDist} shows that, for both galaxies, the \NHI\
distributions for optically thin and thick nebulae are similar to each
other, but that the former are weighted more toward lower columns, as
expected.  However, we note that even
in the LMC, which has minimal line-of-sight projection effects, the
\NHI\ distribution for optically thin objects extends up to \NHI\
$=8\times 10^{21} {\rm cm^{-2}}$, a value as high as that for the
optically thick ones.

\begin{figure}[t]
\centering
\subfigure[LMC]{
\includegraphics[scale=0.30]{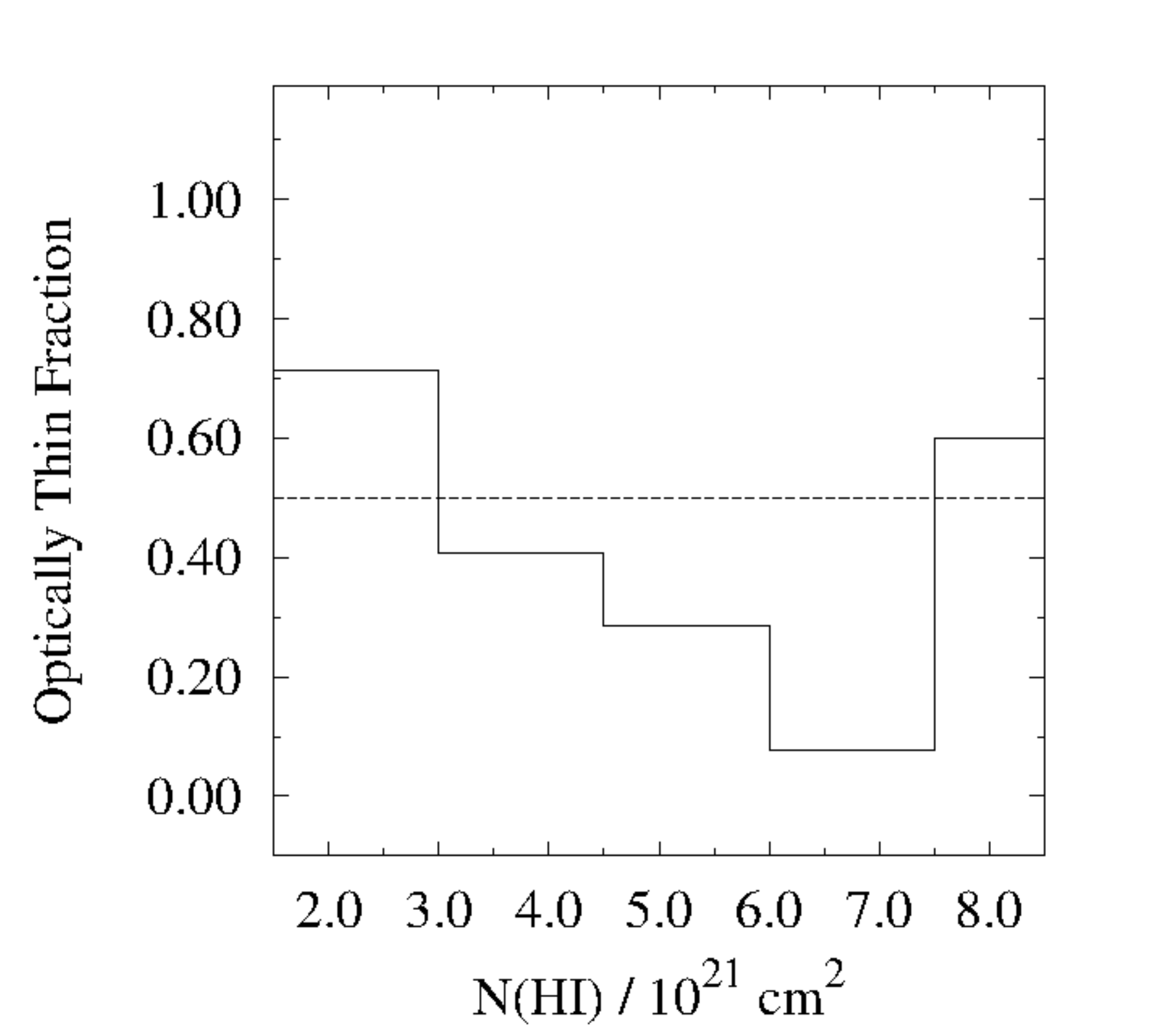}
\label{fig:LMC_frac_HI}
}
\subfigure[SMC]{
\includegraphics[scale=0.30]{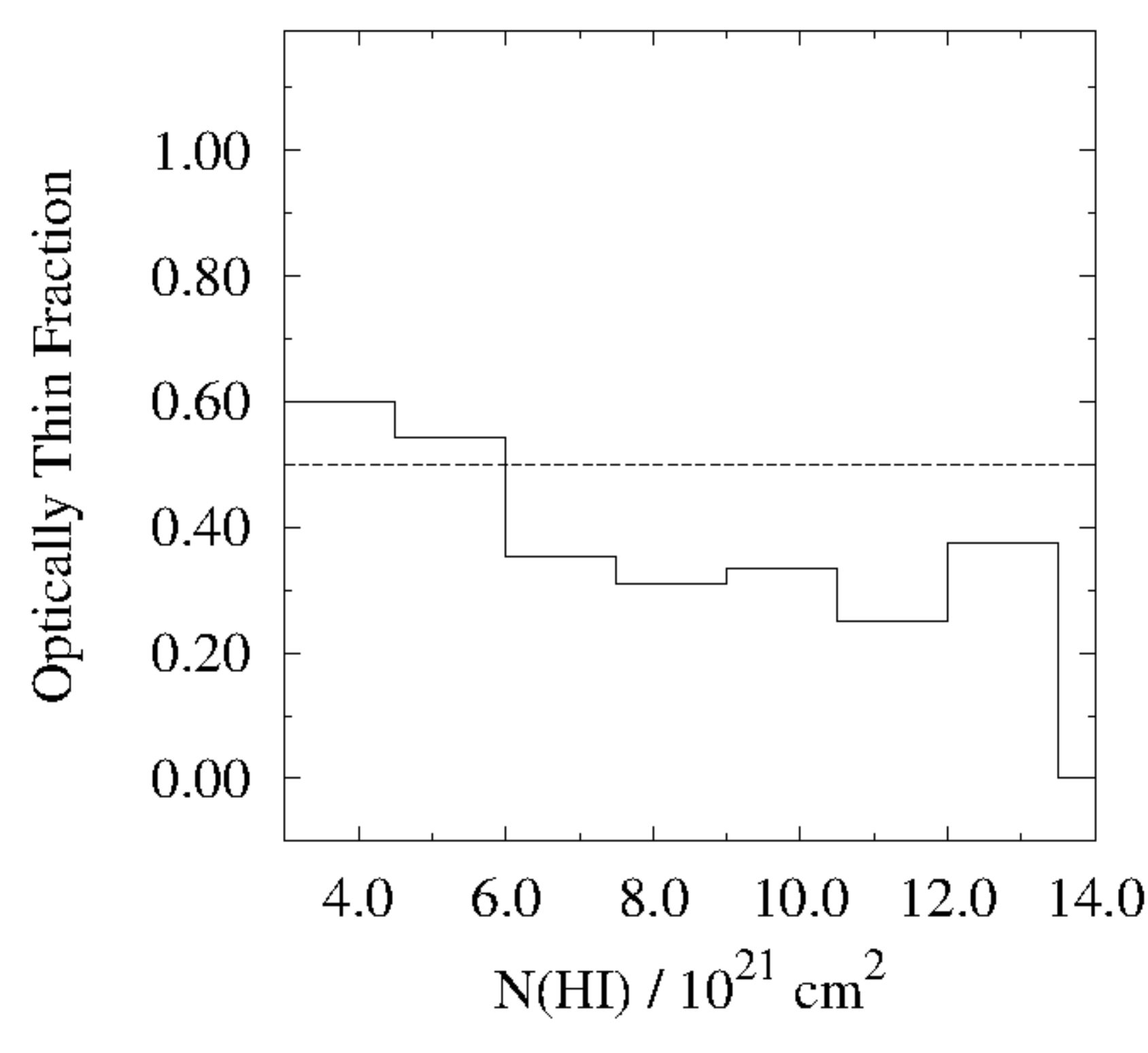}
\label{fig:SMC_frac_HI}
}
\caption[]{The frequency of optically thin \HII\ regions vs
  line-of-sight \NHI\ within the object apertures, for the LMC and SMC
  in panels \subref{fig:LMC_frac_HI} and \subref{fig:SMC_frac_HI},
  respectively.  Optically thin objects dominate above a fraction of
  0.5, indicated by the horizontal dotted line.  
}
\label{fig:frac_HI}
\end{figure}

In Figure \ref{fig:frac_HI} we show the relative frequency of
optically thin nebulae as a function of \NHI\ in the LMC and SMC.
As expected, we see a strong decrease in the frequency of optically thin
objects with increasing \HI\ column, although as emphasized above,
there are still optically thin objects found near the highest \NHI.
In both galaxies, there is a transition \NHI\ below which
optically-thin nebulae constitute the majority, at $3 \times 10^{21}~{\rm
  cm^{-2}}$ in the LMC, and $6 \times 10^{21}~{\rm cm^{-2}}$ in the
SMC.  The SMC transition \NHI\ is two times higher than in LMC, which
is consistent with the 2.4 times higher median \NHI\ value
for all \HII\ regions in the LMC relative to the SMC, due to
line-of-sight ISM projection.  Thus, in spite of the very different
\HI\ morphology between the two galaxies, the quantitative
relationship between nebular optical depth and \HI\ column is
remarkably similar.

\section{Global Escape Fractions}
\label{sec:GlobFesc}

While the frequency of optically thin vs thick \HII\ regions is
similar in both galaxies, it is the structure of the diffuse ISM that
ultimately determines how many ionizing photons heat the galaxy, and
how many escape into the IGM, a quantity crucial to our understanding
of cosmic evolution.  Figure~\ref{fig:3phaseColor} highlights how, in
comparison to the SMC, the evacuated ISM of the LMC allows the
radiation produced in these regions to travel farther, and perhaps
leave the galaxy. In the SMC much of the ionizing radiation escaping
\HII\ regions is unable to penetrate the higher apparent \HI\ column.
We now explore the global escape fractions for the Magellanic
Clouds.

\subsection{HII Region Location}
\label{sec:HIILocation}

\begin{figure}
\centering
\includegraphics[scale=0.28]{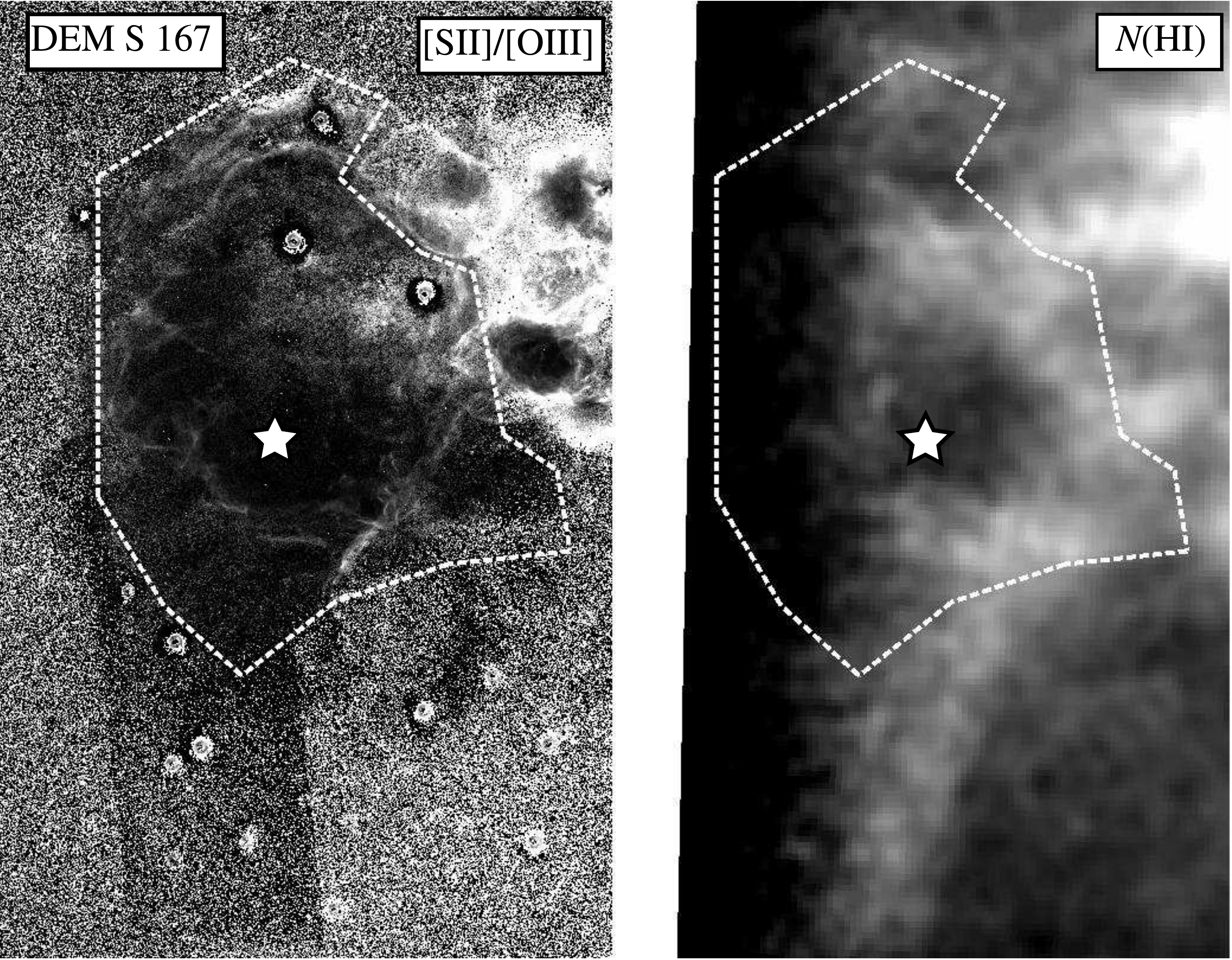}
\caption{\footnotesize The optically thin \HII\ region DEM~S~167
  located at the southeast boundary of the SMC.  The left panel,
  showing \SII/\OIII, shows a strip running N-S with spuriously
  enhanced levels of \OIII.  However, extended \OIII\ emission (black)
  is clearly detected beyond the southern nebular boundary, in excess
  of the artifact's signal.  The
  right panel shows the \HI\ column density map from
  \citep{Stanimirovic1999}, demonstrating that the \OIII\ emission
  extends beyond the \HI\ shell.  The star indicates the location of
  the WO4+O4 binary, and the ionization-based boundary of the nebula
  is shown as a dashed line.}
\label{fig:DEMS167}
\end{figure}

\citet{Gnedin2008} highlighted the importance of \HII\ region location
on the escape of ionizing radiation from galaxies.  In the LMC and
SMC, the largest and most luminous \HII\ regions are found toward the
galaxy edges, and these objects are apparently optically thin.  Take for
example DEM~S167, seen in the southeast extreme of the SMC
(Figures~\ref{fig:SMCS2O3} and \ref{fig:SMCrgb}). In
Figure~\ref{fig:DEMS167} the transition in ionization is shown in
\SII/\OIII\ (left) coincident with the \HI\ shell SSH97~499
\citep{Stanimirovic1999} (right).  These diagnostics
indicate that part of the region is optically thick.  However,
there is \OIII\ emission extending to the south, well beyond the
ionization transition zone.  The 
existence of extended \OIII\ implies a large nebular escape fraction.
The significance of escaping radiation from DEM~S167 is amplified by
its location near the edge of the SMC (Figure~\ref{fig:SMCrgb}).

Following \citet{Voges2008}, we compare the expected \Ha\ luminosity
from the stellar population to the observed value.  The predicted
\Ha\ luminosity from 7 known O stars, ranging from O4V to O9.5V, was
derived using the observed relation between spectral type and $Q{\rm
  (H^0)}$ from \citet{Martins2005}.  This includes a rare,
well-studied WO4+O4 binary, for which we adopt the luminosities
reported by 
\citet{StLouis2005} equal to 75\% of the total ionizing budget.  The
predicted \Ha\ luminosity is $\log L = 38.137$, implying \fesc $=
27$\%, consistent with our estimate of 30\% from ionization-parameter
mapping.  As the eastern-most known SMC \HII\ region, its blister
opens away from the galaxy, making it a prime candidate to contribute
to the galactic escape fraction.  The escaping UV radiation would be
detectable only from certain directions as predicted by
\citet{Gnedin2008}.

\subsection{\HII\ region luminosity}
\label{sec:LHII}

The most luminous \HII\ \ region in the LMC is 30 Doradus, ionized by
the cluster R136a.  It has a reddened luminosity of $\log~\LHa =
{39.66}$, and it is ionized by hundreds of O stars.
Similarly, at $\log~\LHa = 38.82$, the brightest nebula in the
SMC is N66, ionized by at least 30 O stars in the cluster NGC 346.
We can see in Figures~\ref{fig:LMCS2O3} and \ref{fig:SMCS2O3} that
these luminous objects are strongly optically thin,
based on their very extended \OIII\ emission.
Furthermore, they are not deeply embedded in their
respective galaxies, implying that these massive
regions produce ionizing radiation that may escape into the IGM.

\begin{figure}
\centering
\includegraphics[scale=0.4]{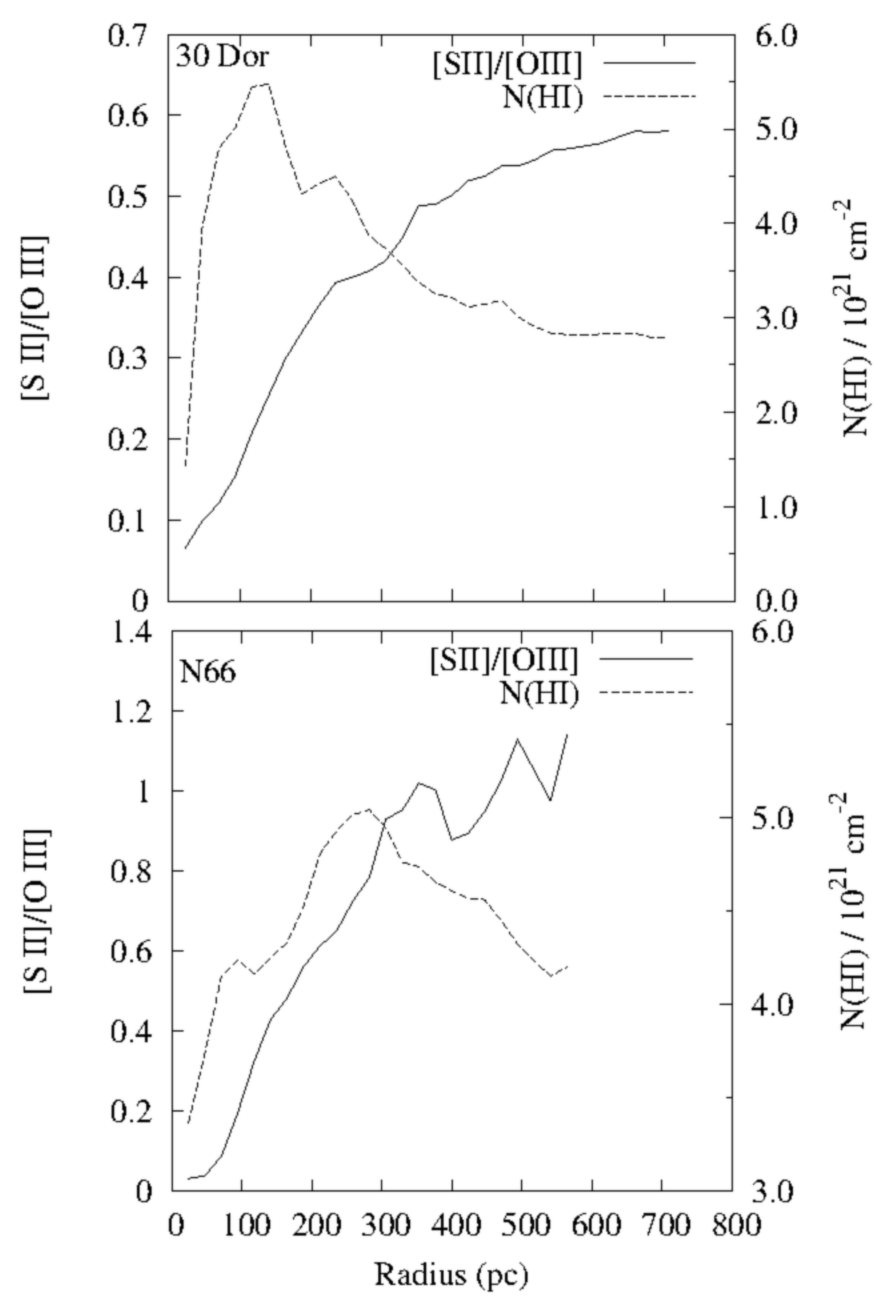}
\caption{\footnotesize The large-scale radial profiles of [S~II]/[O~III] and
  \NHI, averaged in annuli centered on R136a in 30 Dor (top)
  and NGC 346 in N66 (bottom).  Nearby H~II regions falling within the
  annuli were masked.  }
\label{fig:30dor_N66_radial}
\end{figure}

To examine the LyC photon path lengths from these two, luminous
objects, we show the azimuthally averaged \SII/\OIII\ ratio for these
regions in Figure~\ref{fig:30dor_N66_radial}, centered on the main
ionizing clusters R136a (top), and NGC 346 (bottom).  Small, discrete
\HII\ regions that are projected in the line of sight within these
regions are excluded from the azimuthal averages in
Figure~\ref{fig:30dor_N66_radial}.  Figure \ref{fig:30dor_N66_radial}
also shows the average radial \NHI.  The profile of 30 Doradus is
plotted to maximum radii where the gradient of both quantities is
equal to zero, and marks the distance at which the ionizing radiation
from these sources is no longer dominant.  N66 is fainter, so it is
less clear exactly where the influence of its ionizing source ends.
Clearly, the extended \OIII\ emission from the gas surrounding both
objects requires a photoionizing source of high-energy photons from
R136a and NGC 346, which dominate the ionization of the DIG out to at
least 600 pc, whether or not this gas was ever associated with the
\HII\ region, or is just part of the DIG. This is in agreement with
the enhanced \Ha\ surface brightness of 30 Dor out to 850 pc noted by
\citet{Kennicutt1995}.

\begin{figure}
\centering
\includegraphics[scale=0.3]{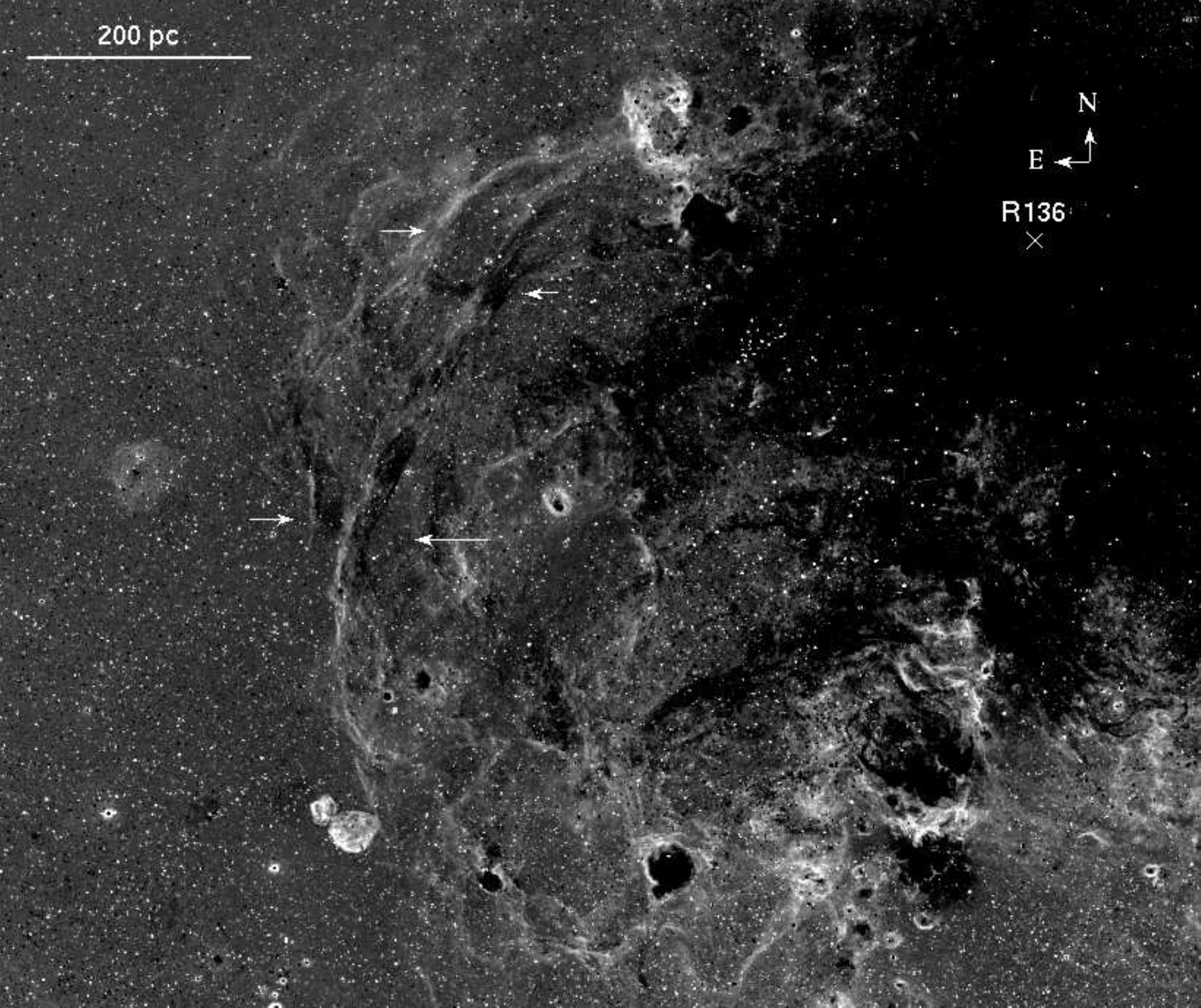}
\caption{\footnotesize The complex region east of R136a seen in
  \SII/\OIII.  R136a is marked in the upper right corner of the image.
  Marked by arrows are ionized filaments which exhibit ionization
  stratification over nearly 500~pc in length.  The
  orientation of the stratification is evidence of a centrally located
  illuminating source at a location consistent with 30 Doradus.}
\label{fig:30dor_filaments}
\end{figure}

Figure~\ref{fig:30dor_N66_radial} shows that the peak \NHI\ associated
with optically thin objects can occur at radial distances that are
well within the radial limits of the photoionized region.  This
suggests that the neutral and molecular ISM is highly inhomogeneous
and clumpy, with large holes or clear areas that allow the escape of
ionizing radiation.  This situation is similar to the ionization cone
detected in NGC 5253 \citep{Zastrow2011b}. In particular, the radial
bins used to produce Figure~\ref{fig:30dor_N66_radial} mask
  important features in the \SII/\OIII\ and \NHI\ distributions
around 30 Doradus.  These include narrow, radial projections
containing continuous regions of highly ionized gas extending 1 kpc in
various directions from 30 Doradus.  There is no evidence for
additional ionizing sources that can explain the extended ionization,
so we argue that the variation in path length is due to variations in
ISM density.  East of the ionizing source, R136a, we see a complex of
edge-on filaments (Figure \ref{fig:30dor_filaments}), associated with
the giant \HI\ shell LMC~2 \citep{Meaburn1980}.  These filaments form
a continuous arc over 500~pc in length, at a distance ranging from 0.6
and 0.8~kpc from R136a, with strong \OIII\ facing the ionizing
cluster, and strong \SII\ facing away from it, as shown by the arrows
in the Figure.  This ionic stratification confirms that the ionizing
photons striking these filaments or sheets originate from R136a.
Similar filaments are detected west of N66 (Figure~\ref{fig:SMCrgb}),
opposite the bulk of the SMC.  Unfortunately, our data do not extend
far enough to look for filaments beyond DEM S167 in the direction away
from the SMC.  However, we see that both the location and luminosity
of these giant \HII\ regions strongly influence the likelihood of LyC
radiation escaping from the host galaxies.

\subsection{Integrated \HII\ region escape fractions}
\label{sec:Aggregate}

To understand the Lyman-continuum radiation transfer within galaxies,
it is of central interest to evaluate the luminosity-weighted,
mean LyC escape fraction \fescave\ of all the nebulae within each
galaxy.  We first calculate the total \HII\ region ``escape
luminosity'' \Lesc\ in terms of individual, observed \HII\
region luminosities using
\begin{equation}
\label{eq:L_esc}
\Lesc = \sum_{i} \left( { \LHa_{i} \times \frac{f_{{\rm esc},i}}{1-f_{{\rm esc},i}} } \right)
\end{equation}
where $i$ represents the $i^{th}$ object in the given galaxy.  Note
that \LHa\ is the observed luminosity, as before, which is related to
the total ionizing
luminosity $L_{\rm tot}$ by $\LHa=L_{\rm tot}\bigl(1-\fesc\bigr)$.  
We again adopt $\fesc= 0.6$ for optically thin nebulae and 0.3 for
blister regions.  Optically thick nebulae contribute no escaping
radiation, but add to the total observed \Ha\ luminosity.  The total
escape luminosities for the individual object classes are listed in
Table~\ref{tab:fesc_Results}.  The total \Lesc\ from all \HII\ regions
in the galaxies are $\log$ \Lesc $= 40.1$ in the LMC, and $\log$ \Lesc
$= 39.2$ in the SMC.

Next, we calculate the luminosity-weighted \HII\ region escape
fraction in each galaxy according to,
\begin{equation}
\label{eq:f_esc_galaxy}
\fescave\ =
\frac {\sum_{i} \LHa_{{\rm esc},i}}{ \sum_{i} ( \LHa_{{\rm esc},i}+
  \LHa_{i} )} \quad .
\end{equation}
We find the lower limit on \fescave\ in the LMC and SMC to be
0.42$-$0.51 and 0.40, respectively.  The lower LMC value
corresponds to a scenario where indeterminate, class 0 objects are
optically thick, while the upper limit assumes they are optically
thin.  In the LMC these objects account for $< 14$\% of the total
\HII\ region luminosity, while they do not make any significant
contribution in the SMC.  We have not included the uncertainty due to
photometry, which is 20\% for individual objects, and introduces an
error of 22\% to our \fesc\ calculations.  Therefore, our final lower
limits on \fescave\ in the LMC and SMC are 0.42$\pm$0.09 and
0.40$\pm$0.09 respectively.

Because we are using only two line ratios to constrain \fesc, we again
note that these estimates for \fescave\ are lower limits, although as
discussed above, they are not strong lower limits.  
We can compare our results to the estimated \fescave\ for the
Magellanic Clouds by \citet{Kennicutt1995}, who adopted the DIG 
luminosity for \Lesc.  They found \fescave\ = 0.35 and 0.41 in the
LMC and SMC, respectively, which agree well with
our estimates.  Table~\ref{tab:fesc_Results} gives the total \LHa,
\Lesc, and \fescave\ for the \HII\ region populations listed in column
1, with LMC and SMC values shown on the left and right side of the
Table, respectively.

\begin{figure}[t]
\centering
\includegraphics[scale=0.35]{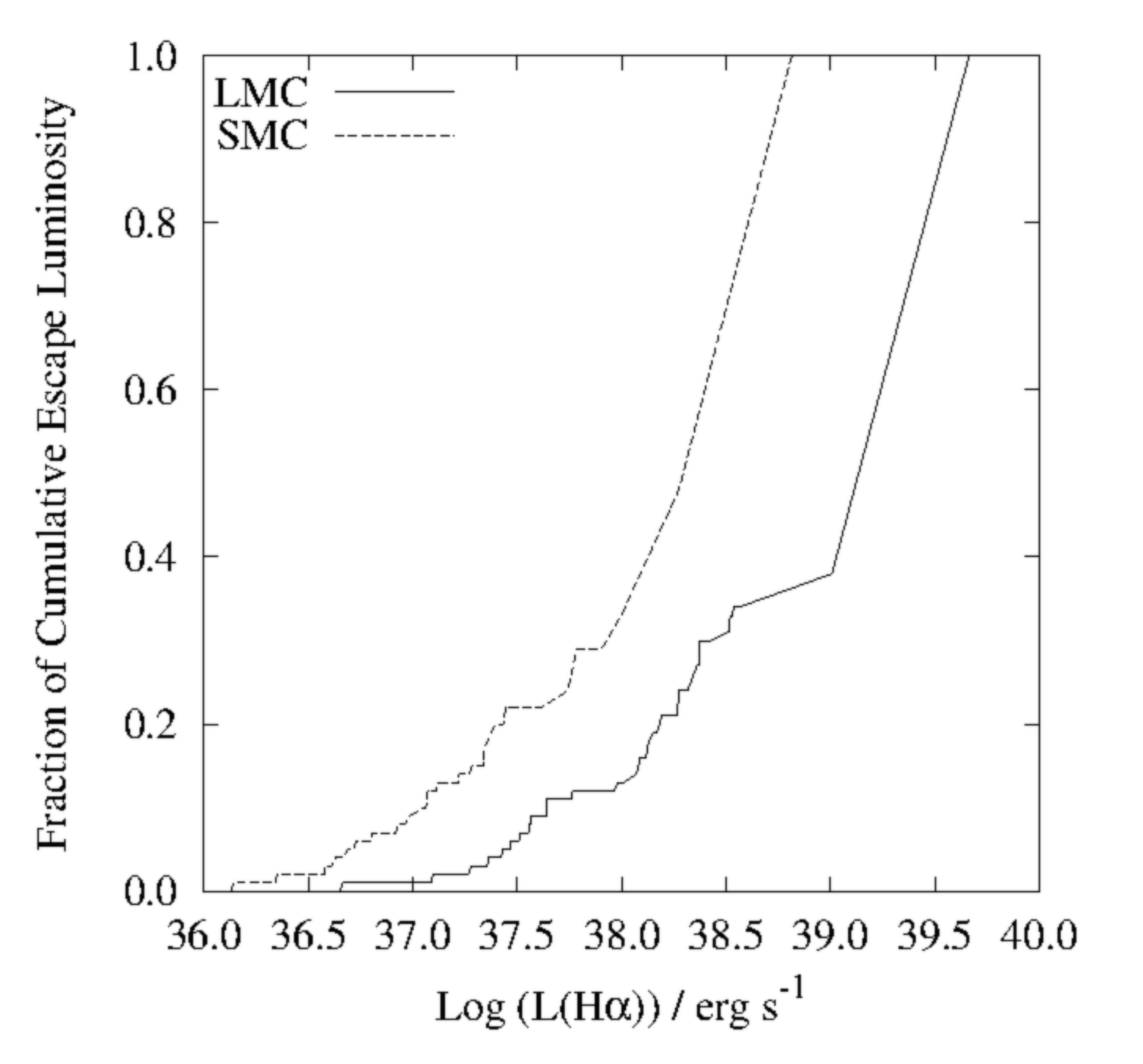} 
\caption{\footnotesize The cumulative fractional \Lesc\ as a function
  of $\log~\LHa$ for the LMC (solid line) and SMC (dashed line).} 
\label{fig:cumfrac}
\end{figure}

The similarities in \fescave\ between the two galaxies are not due to
\HI\ distributions, which, as we saw above, differ strongly.  Instead,
they apparently result from the brightest optically thin nebulae.  In
the LMC, 30 Dor contributes nearly 60\% of the total LMC escape
luminosity.  There is a similar situation in the SMC, where N66,
ionized by the cluster NGC~346, contributes an estimated 50\% of the
total escape luminosity in that galaxy.  Looking at
Figure~\ref{fig:cumfrac}, the cumulative fractional \Lesc\ as a
function of observed \LHa, we find that in the SMC, only 30\% of the
escaping ionizing radiation comes from objects with $\log~\LHa <
38.0$.  In the LMC, the contribution is near 10\% for the same range
of \LHa.  Thus the dominant contribution to the
escape luminosity is from objects more luminous than $\log~\LHa >
38.0$.  Therefore, it is not surprising that the derived
luminosity-weighted \fescave\ is similar between the two galaxies,
when both 1) the total escaping luminosity is dominated by the bright
objects and 2) a single \fesc\ is assumed to describe all optically
thin nebulae.

\begin{deluxetable*}{lccccccc}
\tabletypesize{\footnotesize}
\tablecaption{Global radiative transfer properties of nebular populations}
\tablehead{
  \colhead{}&
  \multicolumn{3}{c}{LMC}&
  \multicolumn{3}{c}{SMC}&\\
  \colhead{~~~~~~~~~}&
  \colhead{$\log \LHa ^a$}&
  \colhead{$\log \Lesc\ ^b$}&
  \colhead{\fescave\ $^b$}&
  \colhead{$\log \LHa ^a$}&
  \colhead{$\log \Lesc\ ^b$}&
  \colhead{\fescave\ $^b$}
}

\startdata
Indeterminate (class 0) &39.3 &\nodata& \nodata  & 36.6 & \nodata& \nodata\\
Opt Thick (class 1) &39.4 &\nodata& 0.0  & 38.9& \nodata&0.0\\
Blister (class 2)   &39.5 & 39.1  & 0.3  & 38.7& 38.4   &0.3\\
Opt Thin (class 3)  &39.9 & 40.0  & 0.6  & 39.1& 39.2    &0.6\\
Classes $1+2+3$     &40.2 & 40.1  & 0.42 & 39.4& 39.2   &0.40\\
\citet{Kennicutt1995}$^c$ & 40.2  & 40.0 & 0.35& 39.5   &39.3&0.41
\enddata 
\label{tab:fesc_Results}
\tablecomments{
\\
$^a$Columns 2 and 5 give the sum of the observed $L$.\\
$^b$\Lesc\ and \fescave\ are lower limits. \\
$^c$\Lesc\ from \citet{Kennicutt1995} corresponds to their
  measured DIG luminosity, $L_{\rm DIG}$.\\
}
\end{deluxetable*}

\subsection{Ionizing the WIM}
\label{sec:ItDIG}

Now, we confront an important question: {Can ionizing radiation
  escaping from optically thin star-forming regions explain the
 luminosity of the WIM?}  Previous studies estimated that $\sim$40\%
  of the WIM (or DIG)
ionization is due to isolated field stars, and optically thin \HII\
regions powered by clusters contribute the remaining 60\%
(e.g. \citealt{Oey2004, Hoopes2000, Oey1997, Miller1993}).  
From Table~\ref{tab:fesc_Results}, we find that \fescave\ and
\Lesc\ from the \HII\ regions alone is enough to balance the observed
LMC DIG recombination rate observed by \citet{Kennicutt1995}, and that
\HII\ regions account for 84\% of the DIG luminosity in the SMC.
Including photometric errors, the DIG of both galaxies can be powered
by optically thin radiation from \HII\ regions alone.
This is not at odds with the previous results when we consider
that not all field stars are sitting naked in the DIG.  Many are
ionizing discrete nebulae, so that simply adding the ionizing
luminosities of all field stars to \Lesc\ will overestimate its value.

We have a unique opportunity to examine this quantitatively in the
SMC: the RIOTS4 survey of \citet{OeyLamb2011} is the only complete
survey to target field massive stars in an external galaxy, and it
includes 115 spectroscopically confirmed field O stars from the
\citet{Oey2004} sample of field OB stars.  Of these SMC field O stars,
60 show no associated nebular emission.  Using $Q(\rm H^0)$ from
\citet{Martins2005} to convert the stellar spectral types from
\citet{OeyLamb2011}, the total ionizing flux from these is equivalent
to 12\% of the DIG ionization rate.  There are an additional 27 stars
whose location inside nebulae is ambiguous. If we assume their
ionizing luminosity also streams into the DIG, field O stars produce a
total ionization rate equivalent to 23\% of the DIG.  If we further
assume no ionizing radiation escapes the galaxy, then the DIG emission
measured by \cite{Kennicutt1995} reflects the combined ionizing
radiation from field stars and from optically thin \HII\ regions, with
77--88\% of the ionizing radiation originating from \HII\ regions.
Using equation~\ref{eq:L_esc}, an SMC aggregate $\fescave \sim 0.38 -
0.41$, instead of 0.40 (Table~\ref{tab:fesc_Results}), would result in
\HII\ regions producing 77--88\% of the DIG luminosity in
Table~\ref{tab:fesc_Results}, which is within our \fesc\ uncertainty caused
by photometry.

\subsection{Galactic escape fractions from the Magellanic Clouds}
\label{sec:GalacticFesc}

With ample evidence of the great distances that ionizing radiation
travels from massive star forming regions, we can make an initial
quantitative estimate of \fescgal, the galactic escape fraction, by
comparing the aggregate escape luminosities from
equation~\ref{eq:L_esc} to the DIG luminosity $L_{\rm DIG}$, where
\begin{equation}
\fescgal = (\LHa_{\rm esc}-\LHa_{\rm DIG})/\LHa_{\rm tot} \quad .
\end{equation}
The quantities needed to calculate \fescgal\ are listed in
Table~\ref{tab:fesc_Results}, where $\LHa_{\rm tot}$ is the sum of
observed $L$ and $\Lesc$, as before.  We see that in the
SMC, \fescgal $=0$.  However, as discussed above, this neglects the
contribution from a known population of massive field stars.  As
calculated in \S \ref{sec:ItDIG}, the ionizing radiation from truly
isolated stars in the SMC is 12\% -- 23\% of the DIG luminosity since
about half of these field stars reside in \HII\ regions.  In the SMC,
this yields a lower limit to \fescgal\ of 4\% -- 9\%.  For the LMC,
Table~\ref{tab:fesc_Results} shows that the ionizing luminosity
escaping \HII\ regions is also about the same as the value needed to
explain the LMC DIG, without accounting for an unknown population of
field O-stars.  It is
reasonable to assume that the field star ionizing luminosity relative
to \HII\ region $L_{\rm tot}$ is similar (Oey et al. 2004) in the LMC and
SMC (0.05 -- 0.11).  From Table~\ref{tab:fesc_Results} and
equation~\ref{eq:L_esc} we find a lower limit to \fescgal\ in the LMC
is 11--17\%.

It is important to bear in mind that our constraints on nebular
\fescave\ technically are lower limits, as stressed in \S \ref{sec:method}.  The
ranges in galactic escape fraction quoted above only reflect the
uncertainties in the field star population.  Thus it is possible
that \fescgal\ may be underestimated in one or both of the galaxies;
the crude estimates in Table~\ref{tab:fesc_Results} preclude any
conclusive results.  Future work is needed to quantitatively improve
these constraints.  Further efforts may be directed at obtaining more
definitive ionization-parameter mapping by adding imaging in more
ions, or modeling diagnostic emission lines in filaments ionized by
distant sources like those for 30 Doradus and N66, which can constrain
the ionizing photon flux and SED of these dominant objects.

\section{Conclusions}

We have demonstrated the power of spatially resolved,
ionization-parameter mapping to quantitatively probe the optical depth
of \HII\ regions to the Lyman continuum.  Our {\sc Cloudy}
photoionization simulations show that spatially resolved emission-line
ratio mapping reveals the presence or absence of ionization
stratification that diagnoses the optical depth of photoionized
regions.  The technique
also constrains the optical depth in the line of sight.  We show that
ionization-parameter mapping in only \SII\ and \OIII\ is a powerful
and productive technique when studying global nebular properties.
Although there is a degeneracy between optically thin and
weakly ionized regions when using only two radially varying ions, 
the technique works well in the aggregrate, and the degeneracy
is resolved with observations of three sensitive ions.  It
may be possible to develop similar methods using emission from PAHs,
which are easily destroyed in ionized gas, and enhanced by non-ionizing
UV light in ionization fronts.

Our application of ionization-parameter mapping uses the \SII, \OIII,
and \Ha\ data of the LMC and SMC from the MCELS survey.  First, we
used \SII/\OIII\ ratio maps to define new boundaries for photoionized
\HII\ regions.  The \SII/\OIII\ maps reveal the nebular ionization
structure, thereby allowing us to isolate the emission from individual
photoionized \HII\ regions, even if they are overlapping and/or
embedded in large complexes or bright DIG.  We used these data,
together with the \Ha\ surface brightness, to define the boundaries of
401 \HII\ regions in the LMC and 214 in the SMC.  The resulting
\HII\ region luminosity functions are consistent with those published
for these same galaxies \citep{Kennicutt1989}, indicating that the
simpler, \Ha-only boundary criteria do result in statistical
properties that are similar to those for objects defined by our more
physically motivated criteria.

Based on their observed ionization structures, the optical depths of
the individual \HII\ regions were crudely divided into optically
thin, optically thick, and blister classes.  Based on our models, we assign $\fesc =
0.6$ for the population of optically thin regions, 0.3 for blisters,
and 0.0 for the optically thick objects.  These estimates agree
within 23\% with
more direct measurements of the optical depth for a sample of objects
with known spectral classifications for the ionizing stars.  

These rough optical depth classes already yield fundamental new insights into the
quantitative radiation transfer of the nebular population and DIG
ionization in these galaxies.  We find that the frequency of optically
thin nebulae is 40\% in the LMC and 33\% in the SMC.
The luminosity distributions reveal that the
median luminosity of optically thin nebulae is significantly brighter
than for those which are optically thick, by a factor of 2 -- 5.  More
importantly, the frequency of optically thin nebulae increases with
\LHa, such that above $\log~\LHa/(\ergs) = 37.0$, \HII\ regions in both
galaxies are dominated by optically thin objects.  Due to their high
luminosity and significant \fesc, these objects also dominate the
total ionizing radiation leaking into the DIG.  It will be important
to determine whether all star-forming galaxies show a similar
luminosity threshold for the dominance of optically thin objects. 

We also see a correlation in the frequency of optically thick regions
and \HI\ column density, with the median \NHI\ of optically thick
nebulae 1.5 and 1.2 times higher than those of optically thin ones in
in the LMC and SMC, respectively.  In contrast, the
median \NHI\ of all objects measured in the SMC is 2.4 
times higher than in the LMC, probably owing to projection effects.  It
is surprising that despite major differences in the character of the
ambient neutral ISM outside of \HII\ regions, the quantitative
properties of the LyC radiative transfer within the nebulae are
remarkably similar between the two galaxies.  This brings us to an
important conclusion: the large-scale fate of ionizing radiation emitted by
O-stars in the LMC and SMC may be determined by the external,
neutral \HI\ environment, which in the SMC appears more efficient at
trapping radiation once it escapes optically thin \HII\ regions
(e.g. the southwest region of the SMC).

Optically thin nebulae are sufficiently luminous to maintain
the ionization of the DIG in both galaxies, as measured
by \citet{Kennicutt1995}.   
We also consider the global escape fraction of ionizing radiation
from these galaxies into the IGM.  We find evidence that luminous,
optically thin \HII\ regions near the outer edges of both galaxies may
produce ionizing radiation that escapes into the
interstellar environment.  This is evidenced by the kpc-scale path lengths
traveled by ionizing photons from these massive \HII\ regions, shown by
the existence of extended \OIII\ halos opening toward the IGM. 

We find the combined, luminosity-weighted, LyC escape fractions from
all \HII\ regions to be at least 0.42 and 0.40 in the LMC and SMC,
respectively.  The corresponding escape luminosities are at least
$\log$ \Lesc/$({\rm erg\ s^{-1}}) = 40.1$ and 39.2 in the LMC and SMC, respectively.
Considering the existence of field O stars with no nebulae, the
implied total available LyC luminosity is greater than needed to
explain the DIG emission in both galaxies.  These are still crude
estimates, but an excess implies that a fraction of the ionizing
radiation produced leaves the galaxy and may enter the IGM.  We currently
estimate lower limits to the galactic escape fractions of $\fescgal =$ 4 -- 9\% in the SMC,
and 11 -- 17\% in the LMC.  These values are consistent with
\fescgal\ $\sim$10\% -- 20\%, as required for cosmic reionization to be
driven by star forming galaxies at high redshift \citep{Sokasian2003}.
These estimates for \fescgal\ would increase when accounting for
lower optical depth due to absence of dust and metals at high redshift.

\acknowledgments 

We thank Sne\v{z}ana Stanimirovi\'{c} and Sungeun Kim
for access to the SMC and LMC \NHI\ data, respectively.  We thank the
anonymous referee for helpful contributions.  We also thank Joel Lamb,
and Mark Reynolds for help checking this manuscript.  MSO, EWP, and JZ
acknowledge support from NSF grant AST-0806476, and a Margaret and
Herman Sokol Faculty Award to MSO.  PFW acknowledges support from the
NSF through grant AST-0908566, and AEJ acknowledges an NSF Graduate
Research Fellowship.

\clearpage

\appendix
\section{A.  SurfBright:  Modeling surface brightness using {\sc Cloudy}}

The comparison of 1-D photoionization models to observations is most
commonly done with spatially integrated emission-line fluxes.  This
approach is used because a direct comparison to the observed
surface-brightness profile of a nebula involves the convolution of
nebular geometry with a changing volume emissivity.  While computing
predicted surface brightnesses is more difficult, it does provide
important, additional constraints on a model.

We developed the Perl routine {\sc SurfBright}, which calculates the
projected, 2-D surface brightness of isotropically emitted emission
lines in a nebula.  This code uses the results of 1-D {\sc Cloudy}
simulations and a user specified 3-D geometry. The code works in
Cartesian coordinates relative to the nebula with {\bf z} parallel to
the line of sight from observer to object. This geometry is defined at
each $x,y$ coordinate (projected on the sky) by the minimum and
maximum radial distance of the cloud along the line of sight
$z_{min}(x,y),\ z_{max}(x,y)$. A specified physical geometry allows
for the prediction of absolute nebular surface brightness. With the
object-observer distance and 2-D spatial resolution $\Delta x$ and
$\Delta y$, observations can be simulated for any arcsec, pc or cm
pixel scale.  Line strength is given in units of surface brightness,
which is is independent of the distance between observer and object,
for resolved nebulae.

\begin{figure}[t]
  \centering
  \subfigure[Sphere]{
    \includegraphics[angle=0,scale=0.6]{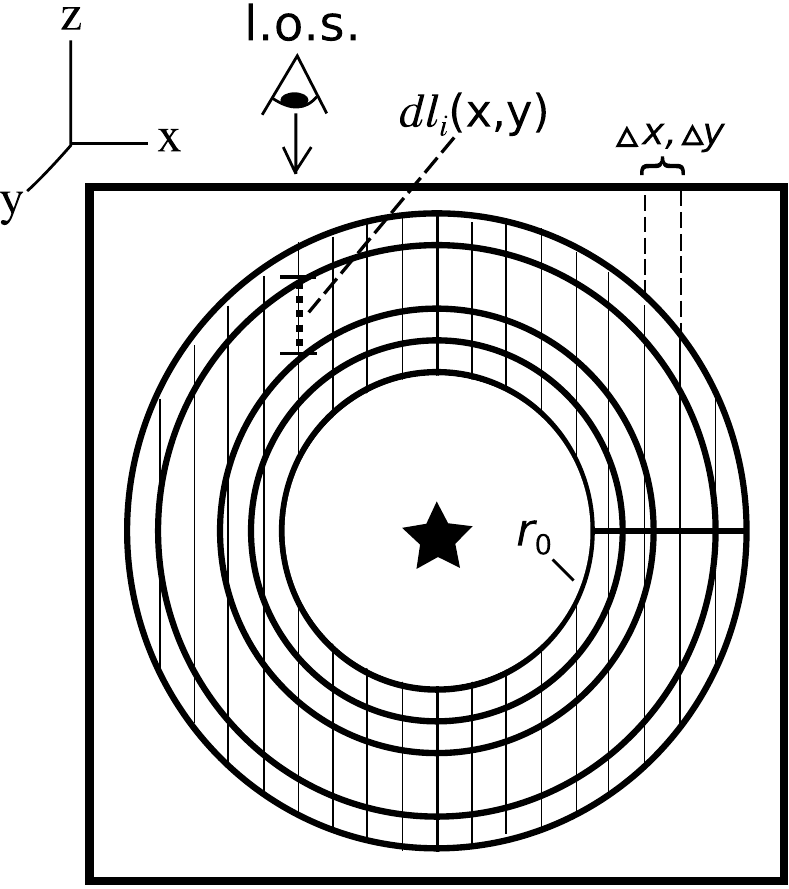} 
    \label{fig:SurfSphere}
  }
  \subfigure[Truncated Sphere]{
    \includegraphics[angle=0,scale=0.6]{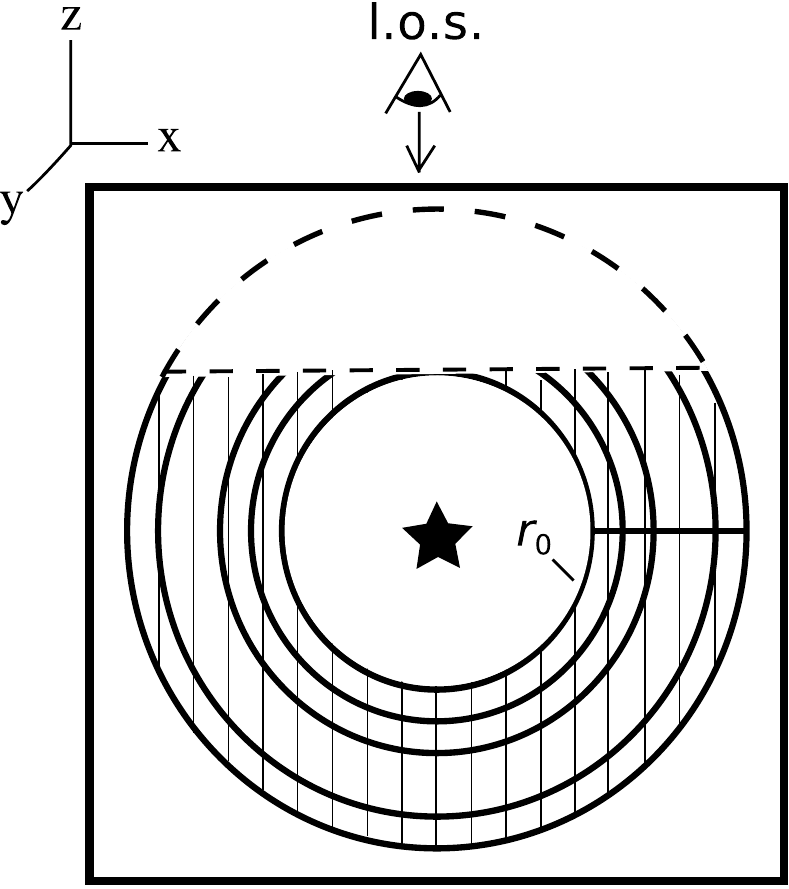} 
    \label{fig:SurfTruncSphere}
  }
  \caption{\footnotesize Schematic representation showing the 
    conversion from a 1-D {\sc Cloudy} simulation into a 2-D {\sc SurfBright}
    surface brightness model.  Vertical lines represent lines of
    sight along the $z$-axis through the nebula at various projected $x,y$
    positions.  The shell 
    structure results from the 1-D {\sc Cloudy} calculation.
    Panel ($a$) shows spherically symmetric geometry, and panel ($b$)
    shows an object with truncated spherical geometry.}
    \label{fig:SurfBright}
\end{figure}

A schematic of a {\sc Cloudy} simulation is shown in Figure
\ref{fig:SurfBright}$a$.  The uneven spacing of shells around the
ionizing star reflects the individual zones calculated within a {\sc
  Cloudy} simulation, which are set by changes in the physical
conditions of the gas as a function of depth.  The {\sc Cloudy}
simulation calculates the isotropic volume emissivity $\epsilon$ of
each emission line in each radial zone.  At each $x,y$ coordinate
projected in the plane perpendicular to the line of sight $z$, {\sc
  SurfBright} calculates the observed surface brightness in the
$i^{th}$ emission line from all {\sc Cloudy} shells according to,
\begin{equation}
S_{i} = \sum_{j} dl_{j}(x,y) \times \epsilon_{i}(x,y,z) / 4\pi \times 2.3504
\times 10^{-11}\ \rm \ergs cm^{-2}~arcsec^{-2}  \quad ,
\end{equation}
where $dl_{j}$ is the path length along the line of sight ($z$-axis)
through shell $j$ at the projected position $x,y$; and $\epsilon_{i}$
are the local volume emissivities taken from the 1-D {\sc Cloudy}
simulation where the 1-D radius $r$ in the model is equal to the
magnitude of the radius of the geometrically defined cloud {\bf
  r}($x,y,z$).

Our method of integrating the emission along the line of sight, from
the far to near side of the \HII\ region, also allows us to include
the effects of internal extinction, which is important for nebulae
with a high observed column density, such as photodissociation regions
and molecular clouds.  \citep{Pellegrini2009}.  Computationally, we
first determine the line flux entering a given shell, which is 
added to the diffuse flux in that shell.  The sum is locally
extinguished by the internal extinction in a shell calculated by {\sc
  Cloudy} scaled by $dl/dr$, where $dr$ is the shell
thickness.  This determines the flux entering the next shell along a
line of sight toward the observer.

{\sc SurfBright} currently includes generic geometrical configurations
for planar slabs and simple spheres.  The orientation, length and
inclination of the slab are free parameters.  Whole and truncated
spheres as in Figure~\ref{fig:SurfBright}$b$ are also possible.
Future support will include completely arbitrary geometric
configurations. Currently the code uses only one {\sc Cloudy} model to
determine the emissivity as a function of {\bf r}.  Later we will also
add the ability to first define a geometry with initial density
parameters and calculate the needed {\sc Cloudy}
simulations from the specified geometry when predicting
observations.  This will make it possible to accurately model the
emission from complex, irregular nebulae with different ionization parameters.

\section{B.   New LMC and SMC \HII\ Region Catalogs}
Tables~\ref{tab:LMCObjs} and \ref{tab:SMCObjs} present our new \HII\
region catalogs, with the nebular boundaries defined by
ionization-parameter mapping as described in \S \ref{sec:IBHIIcat}.
Some objects are defined within the boundaries of larger,
background objects, and their flux is not included in the luminosity
for the larger objects.  The \HII\ regions are
classified by optical depth as described in \S \ref{sec:HIItau}, into types 0 -- 4, corresponding
to (0) indeterminate, (1) optically thick, (2) blister, (3) optically
thin, and (4) shocked.  The object ID's are listed in columns 1 and 2,
with our new designations in the former and identifications
from existing catalogs in the latter.  We designate new, independent substructures
within a previously catalogued object by appending 
numbers, for example, DEM~L173-1 and DEM~L173-2.
The object coordinates are listed in columns 3 and 4; optical depth
classifications are given in column 5.  Columns 6 and 7 respectively give the
average \NHI\ measured in the line of sight within the nebular
aperture, and the \Ha\ luminosity \LHa.  

\begin{figure*}
\centering
\subfigure[MCELS-L001]{
\includegraphics[scale=0.1]{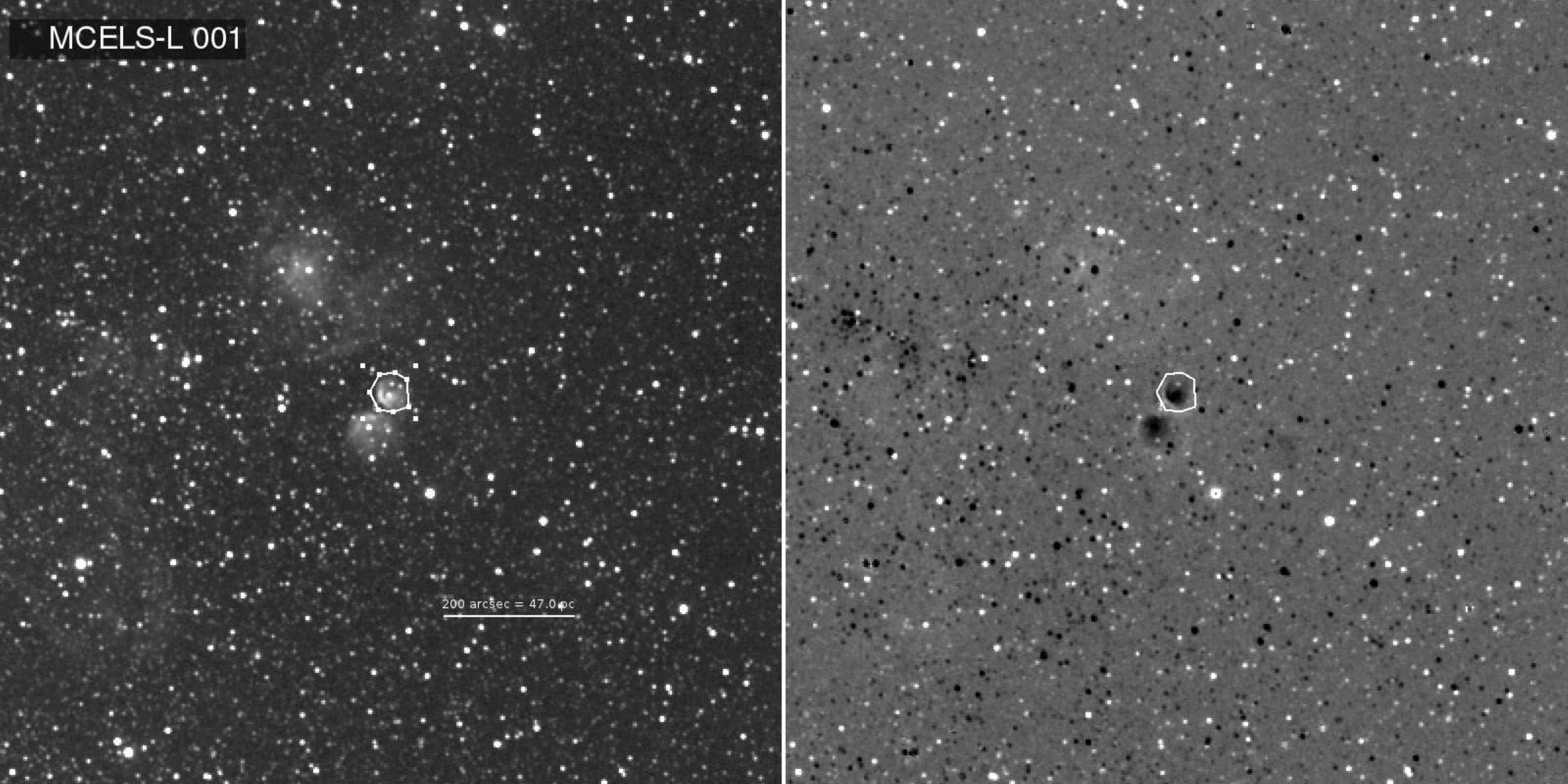}
\label{fig:MCELS-L_1}
}
\subfigure[MCELS-L002]{
\includegraphics[scale=0.1]{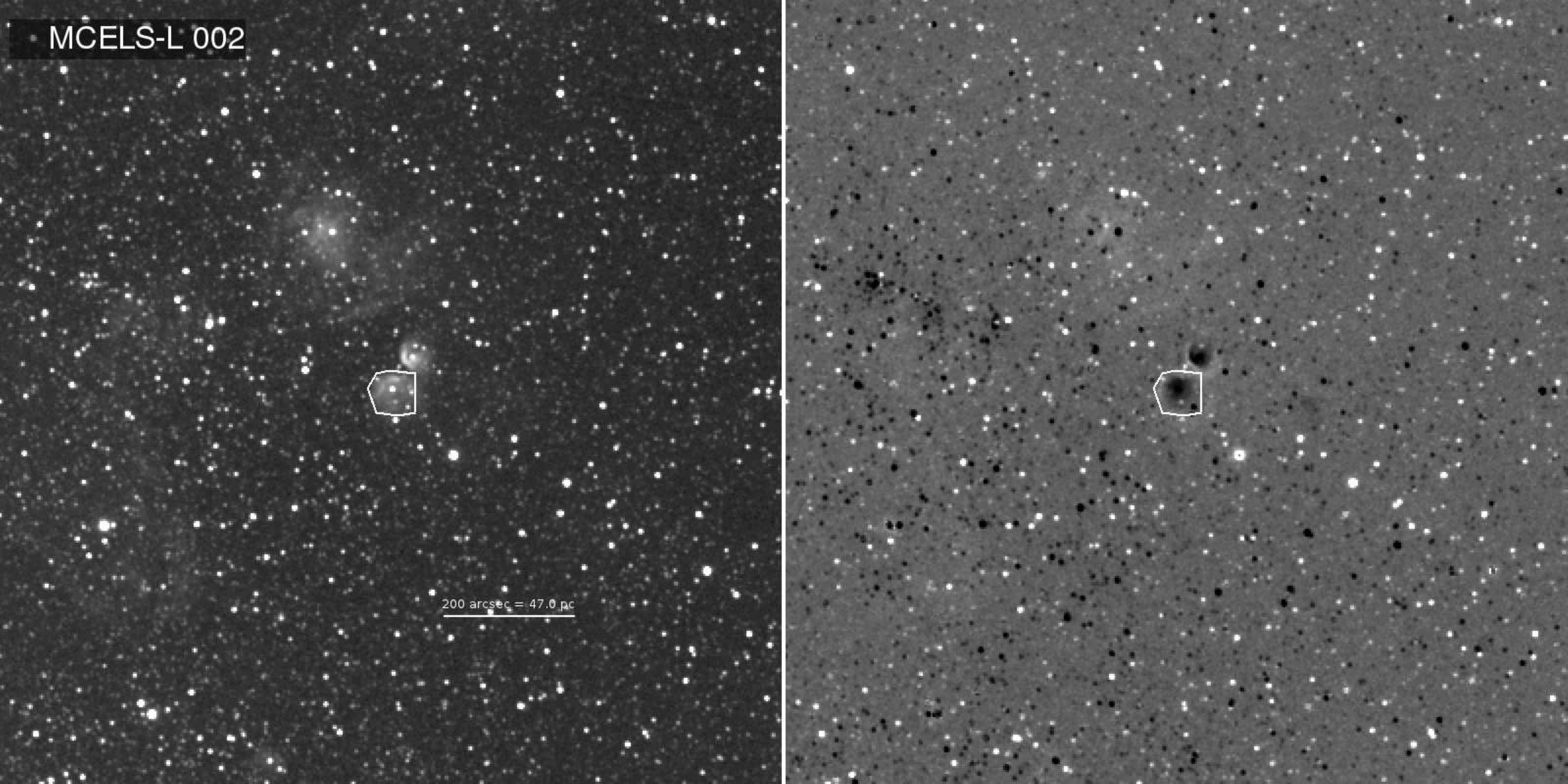}
\label{fig:MCELS-L_2}
}
\subfigure[MCELS-L003]{
\includegraphics[scale=0.1]{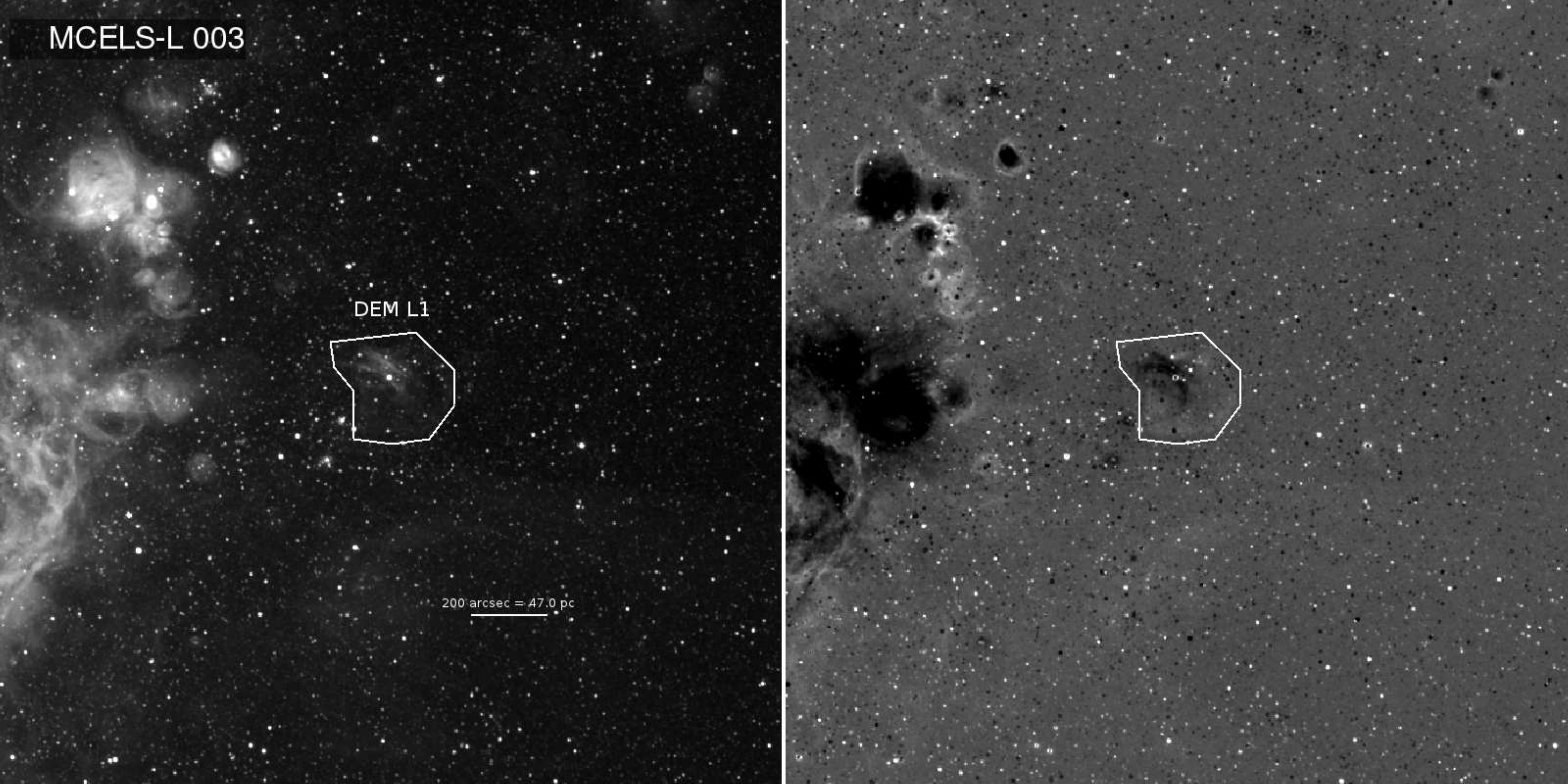}
\label{fig:MCELS-L_3}
}
\subfigure[MCELS-L004]{
\includegraphics[scale=0.1]{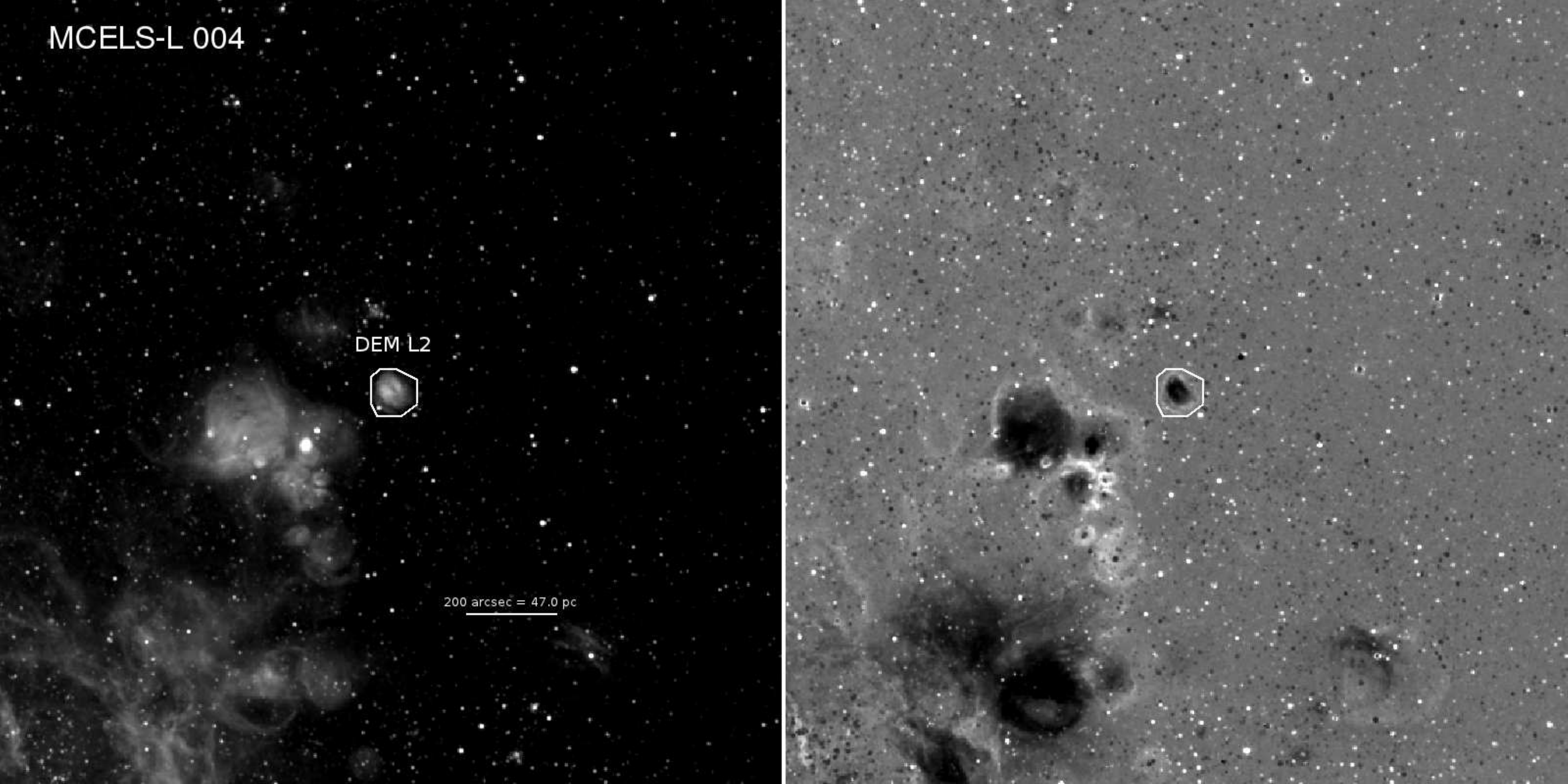}
\label{fig:MCELS-L_4}
}
\subfigure[MCELS-L005]{
\includegraphics[scale=0.1]{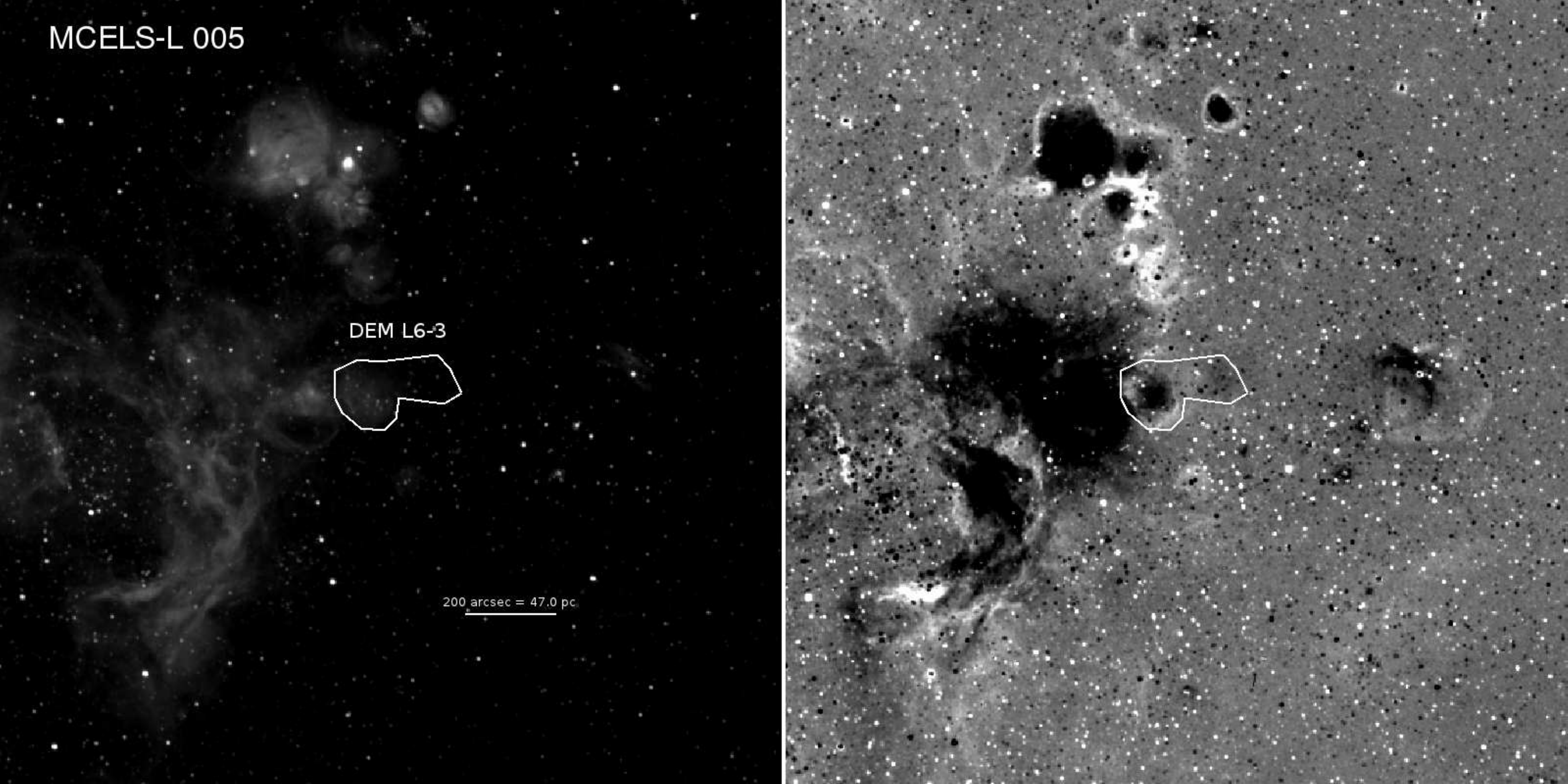}
\label{fig:MCELS-L_5}
}
\caption{\footnotesize 5 LMC \HII\ regions from the MCELS catalog in
  \Ha (left) and \SII/\OIII (right).}
\label{fig:LMC_ImageCatalog}
\end{figure*}

\begin{figure*}
\centering
\subfigure[MCELS-S~001]{
\includegraphics[scale=0.1]{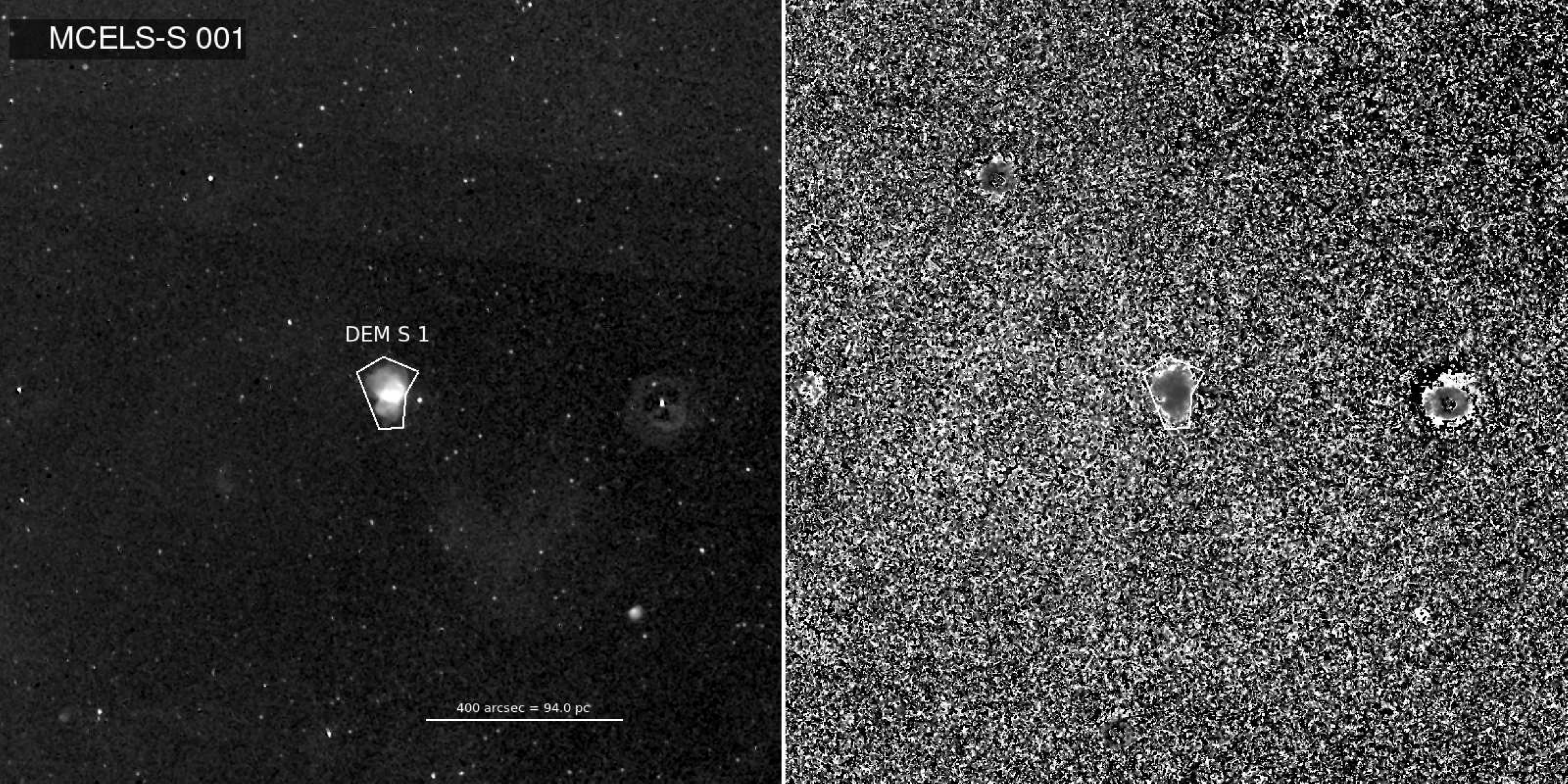}
\label{fig:MCELS-S_001}
}
\subfigure[MCELS-S~002]{
\includegraphics[scale=0.1]{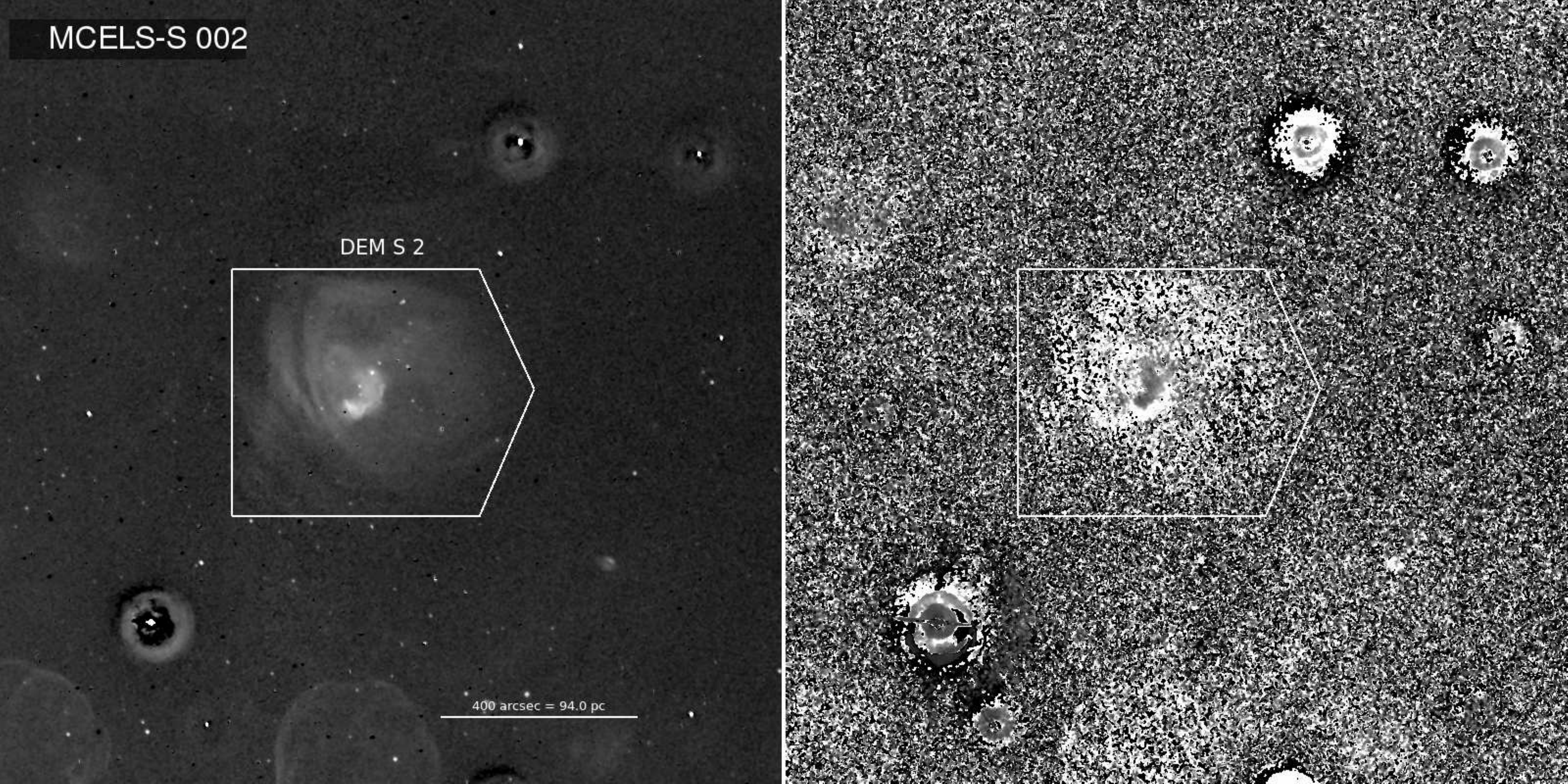}
\label{fig:MCELS-S_002}
}
\subfigure[MCELS-S~003]{
\includegraphics[scale=0.1]{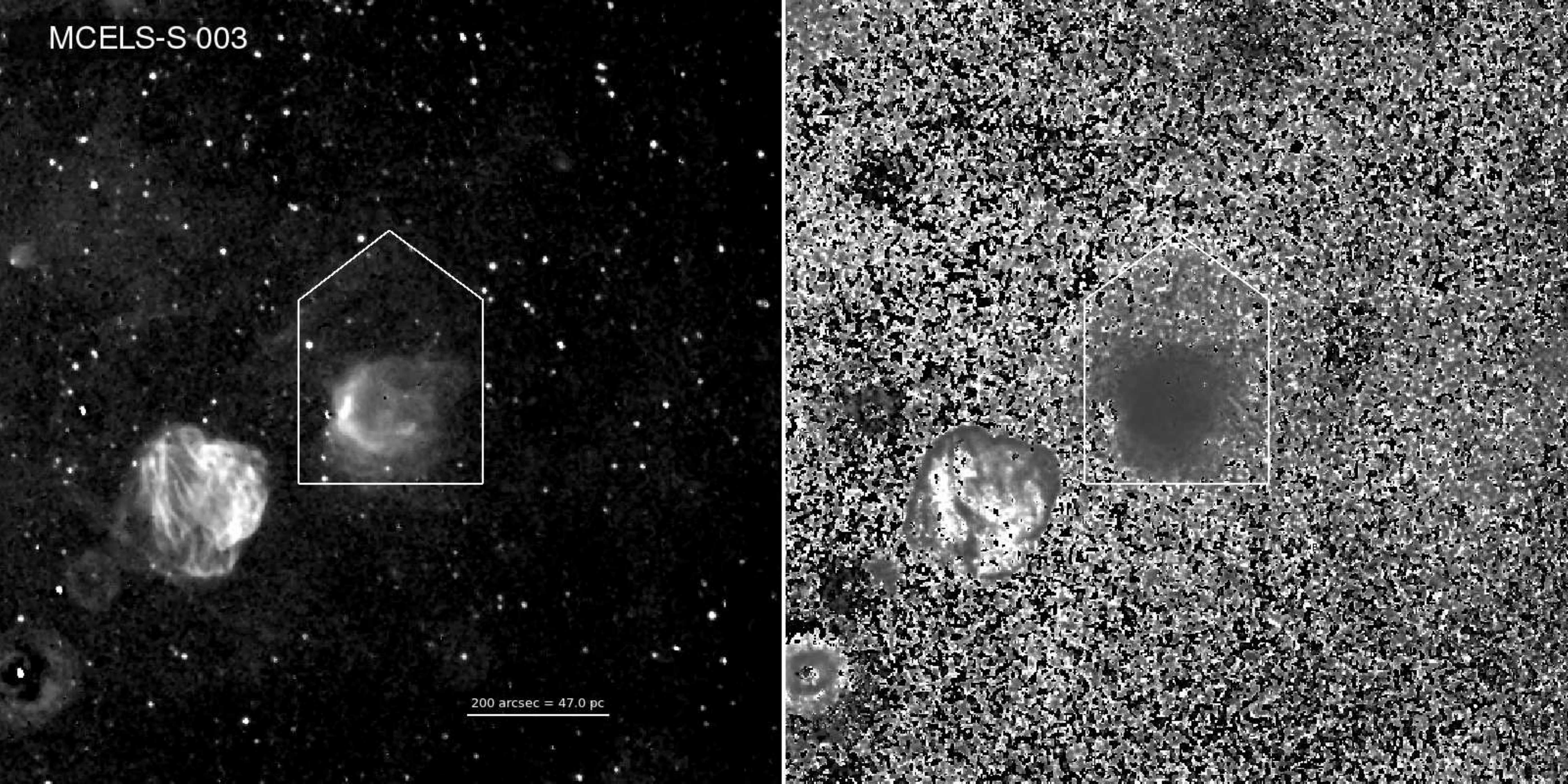}
\label{fig:MCELS-S_003}
}
\subfigure[MCELS-S~004]{
\includegraphics[scale=0.1]{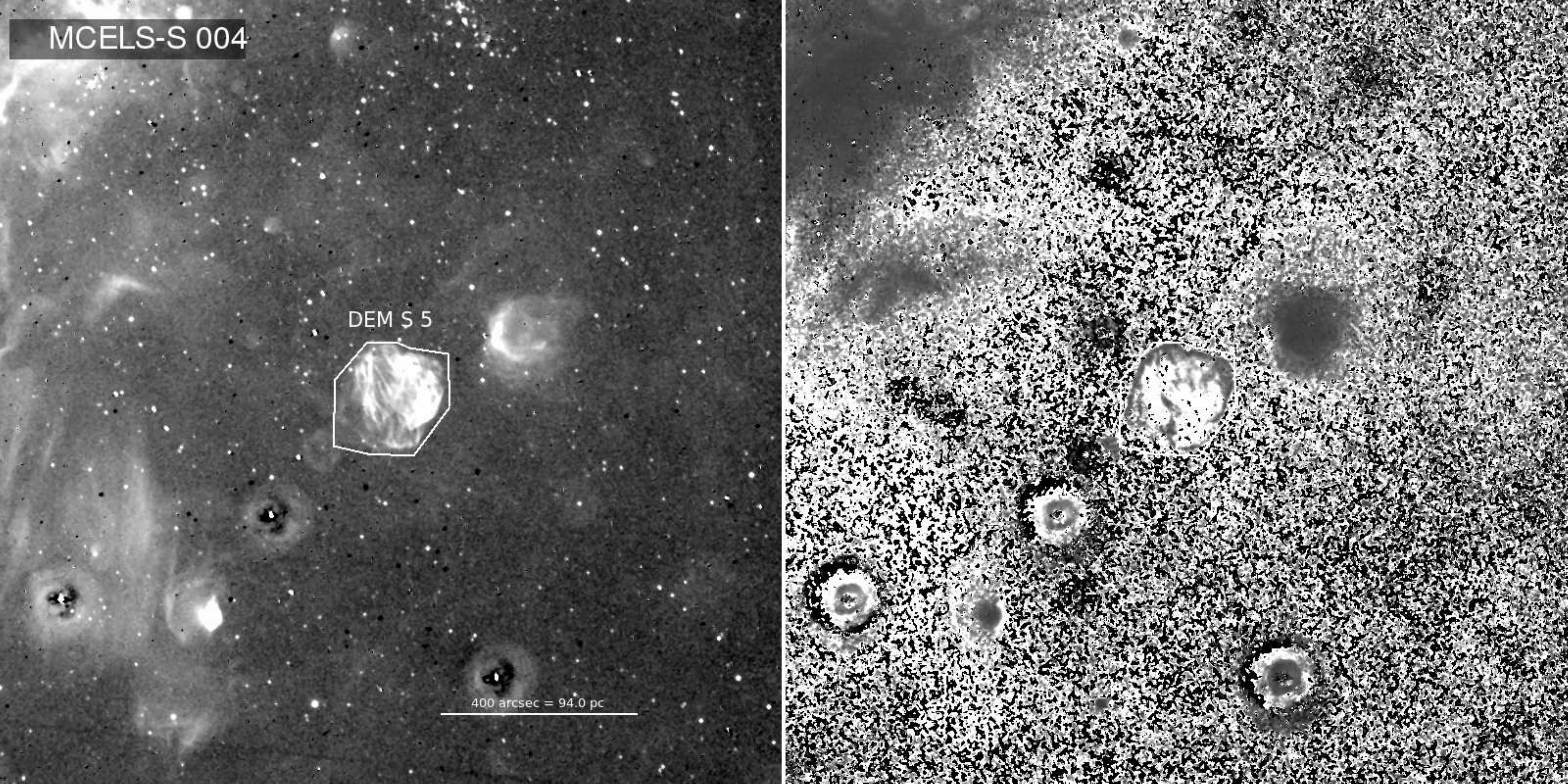}
\label{fig:MCELS-S_004}
}
\subfigure[MCELS-S~005]{
\includegraphics[scale=0.1]{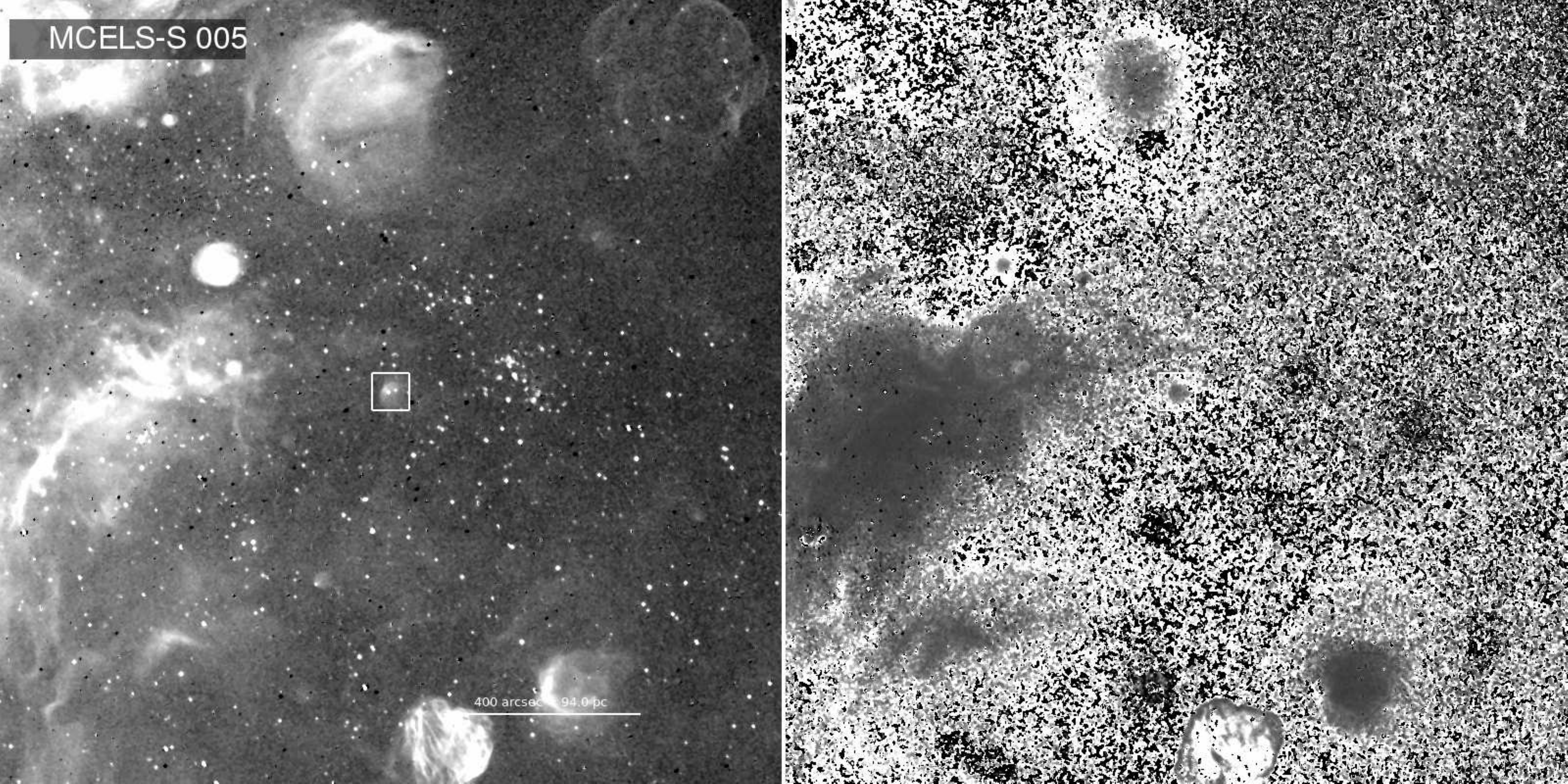}
\label{fig:MCELS-S_005}
}
\caption{\footnotesize 5 SMC \HII\ regions from the MCELS-S catalog in
  \Ha\ and \SII/\OIII.}
\label{fig:SMC_ImageCatalog}
\end{figure*}

In Figures~\ref{fig:LMC_ImageCatalog} and \ref{fig:SMC_ImageCatalog},
we present representative images in \Ha\ and \SII/\OIII, of five
objects in each galaxy.  Each image is centered on the coordinates
listed in Tables~\ref{tab:LMCObjs} and \ref{tab:SMCObjs}.  The images
show the aperture used to measure the \Ha\ flux, as well as a scale
bar in both arcsec and pc.  When shown with dashed lines, apertures
have been enlarged to reveal underlying structure.  There are two
labels in each image: our MCELS catalog ID in the upper left, and,
centered above the object, an alternate catalog ID, unless none
exists.  The complete set of images can be found in the online
edition of this article.

\renewcommand*\thetable{5}
\LongTables
\begin{longtable}{lcccccc}
\caption{REVISED - MCELS LMC HII Region Catalog - REVISED}\\
\hline \hline \\[-2ex]
\multicolumn{1}{l}{Object ID}&\multicolumn{1}{c}{Other ID$^a$}&\multicolumn{1}{c}{RA
    (J2000)}&\multicolumn{1}{c}{Dec}&\multicolumn{1}{c}{Type$^b$}&\multicolumn{1}{c}{$N$(HI)}&\multicolumn{1}{c}{$L$(\Ha)}\\
\multicolumn{1}{c}{}&\multicolumn{1}{c}{}&\multicolumn{1}{c}{h:m:s}&\multicolumn{1}{c}{d:m:s}&\multicolumn{1}{c}{}&\multicolumn{1}{c}{$10^{21}
  cm^{-2}$}&\multicolumn{1}{c}{\ergs}\\[1.5ex] \hline
\endfirsthead

\hline \hline \\[-2ex]
\multicolumn{1}{l}{Object ID}&\multicolumn{1}{c}{Other ID$^a$}&\multicolumn{1}{c}{RA
    (J2000)}&\multicolumn{1}{c}{Dec}&\multicolumn{1}{c}{Type$^b$}&\multicolumn{1}{c}{$N$(HI)}&\multicolumn{1}{c}{$L$(\Ha)}\\
\multicolumn{1}{c}{}&\multicolumn{1}{c}{}&\multicolumn{1}{c}{h:m:s}&\multicolumn{1}{c}{d:m:s}&\multicolumn{1}{c}{}&\multicolumn{1}{c}{$10^{21} cm^{-2}$}&\multicolumn{1}{c}{\ergs}\\[1.5ex] \hline

\endhead
\hline
\multicolumn{3}{l}{{Continued on the Next Page\ldots}}\\
\endfoot
\endlastfoot
MCELS-L1&\nodata&04:44:59.7&-69:03:22&2&1.75&35.12\\
MCELS-L2&\nodata&04:45:04.8&-69:04:25&2&1.90&35.27\\
MCELS-L3&DEM L1&04:47:19.5&-69:18:27&2&1.89&36.41\\
MCELS-L4&DEM L2&04:48:53.7&-69:09:38&1&3.24&36.51\\
MCELS-L5&DEM L6&04:49:06.8&-69:20:26&1&2.71&36.53\\
MCELS-L6&DEM L3&04:49:07.4&-68:24:17&1&2.34&36.55\\
MCELS-L7&DEM L4&04:49:11.3&-69:16:08&0&3.00&36.35\\
MCELS-L8&DEM L4&04:49:18.4&-69:13:54&1&3.57&35.91\\
MCELS-L9&DEM L4&04:49:18.7&-69:13:20&1&3.52&35.70\\
MCELS-L10&DEM L5&04:49:24.4&-69:07:37&3&3.24&35.90\\
MCELS-L11&DEM L4&04:49:25.2&-69:15:24&1&3.29&35.85$^c$\\
MCELS-L12&DEM L4&04:49:25.9&-69:12:00&3&3.08&37.08\\
MCELS-L13&DEM L4&04:49:30.7&-69:13:35&3&3.49&36.66$^c$\\
MCELS-L14&DEM L4&04:49:44.4&-69:12:55&1&3.54&36.02\\
MCELS-L15&DEM L6&04:49:49.2&-69:20:03&3&3.31&37.46\\
MCELS-L16&DEM L4&04:49:51.0&-69:11:44&3&2.97&37.36$^d$\\
MCELS-L17&DEM L4&04:50:02.4&-69:13:23&1&2.94&35.80$^d$\\
MCELS-L18&DEM L4&04:50:05.9&-69:12:05&1&2.98&35.81$^c$\\
MCELS-L19&DEM L7&04:50:08.0&-67:42:10&0&1.35&37.32\\
MCELS-L20&DEM L6&04:50:37.8&-69:25:28&3&2.75&37.64\\
MCELS-L21&\nodata&04:51:32.2&-68:24:24&3&1.53&35.84\\
MCELS-L22&DEM L8a&04:51:44.2&-66:55:14&1&3.16&36.29\\
MCELS-L23&DEM L10a&04:51:47.5&-69:23:10&1&1.73&36.71\\
MCELS-L24&DEM L10a&04:51:49.0&-69:24:25&3&2.00&37.16\\
MCELS-L25&DEM L10a&04:51:52.8&-69:23:29&1&2.22&37.16\\
MCELS-L26&DEM L17&04:51:54.0&-70:47:01&0&1.45&36.63\\
MCELS-L27&DEM L8b&04:52:07.4&-66:55:31&3&3.15&37.41\\
MCELS-L28&DEM L8c&04:52:11.7&-66:54:29&1&3.87&36.63\\
MCELS-L29&DEM L10b&04:52:13.0&-69:20:16&2&3.04&37.69\\
MCELS-L30&BSDL139&04:52:17.5&-70:36:29&0&1.51&35.56\\
MCELS-L31&DEM L9&04:52:19.4&-68:24:39&0&1.59&36.30\\
MCELS-L32&\nodata&04:52:23.3&-66:55:16&1&3.02&35.62\\
MCELS-L33&DEM L10b&04:52:26.5&-69:21:45&1&2.99&36.99\\
MCELS-L34&DEM L11&04:52:34.4&-67:17:23&1&1.62&37.29\\
MCELS-L35&\nodata&04:52:35.0&-66:55:41&1&2.45&35.82\\
MCELS-L36&DEM L15&04:52:46.5&-69:12:54&1&2.73&37.12\\
MCELS-L37&DEM L14&04:52:47.1&-68:54:40&0&1.90&36.37\\
MCELS-L38&DEM L10b&04:52:59.5&-69:23:30&3&2.18&37.20\\
MCELS-L39&DEM L13&04:53:06.3&-68:02:38&2&1.57&37.60\\
MCELS-L40&DEM L15&04:53:07.1&-69:14:11&3&3.21&36.64\\
MCELS-L41&DEM L12&04:53:21.0&-66:56:19&4&1.72&37.16\\
MCELS-L42&DEM L21&04:53:26.0&-70:36:00&3&1.07&36.85\\
MCELS-L43&DEM L20&04:53:30.6&-67:23:22&1&1.34&36.23\\
MCELS-L44&DEM L18&04:53:33.3&-67:03:30&3&1.96&36.26\\
MCELS-L45&DEM L19&04:53:35.5&-67:14:09&3&1.17&36.79\\
MCELS-L46&DEM L16&04:53:38.8&-68:49:00&0&1.28&36.67\\
MCELS-L47&DEM L22&04:53:57.5&-69:10:28&1&2.80&36.64\\
MCELS-L48&DEM L25&04:53:58.9&-70:00:53&2&0.63&37.29\\
MCELS-L49&DEM L22&04:54:03.3&-69:12:08&2&2.56&37.64\\
MCELS-L50&DEM L27&04:54:11.0&-66:54:01&0&1.51&36.64\\
MCELS-L51&DEM L29a&04:54:12.3&-66:45:45&0&2.09&35.34\\
MCELS-L52&DEM L26&04:54:13.4&-68:21:52&1&1.76&37.08\\
MCELS-L53&DEM L23&04:54:24.4&-69:29:43&0&3.10&35.87\\
MCELS-L54&DEM L22&04:54:25.2&-69:10:57&1&2.22&36.87\\
MCELS-L55&DEM L22&04:54:27.6&-69:09:37&1&2.32&37.19\\
MCELS-L56&DEM L29b&04:54:27.7&-66:44:35&2&1.49&35.36\\
MCELS-L57&DEM L30&04:55:03.0&-67:15:53&0&1.74&36.44\\
MCELS-L58&DEM L31&04:55:10.9&-67:11:32&3&1.11&37.77\\
MCELS-L59&DEM L22&04:55:17.7&-69:11:45&2&2.05&37.26\\
MCELS-L60&DEM L24&04:55:25.6&-69:16:06&1&0.99&36.06\\
MCELS-L61&DEM L32&04:55:34.5&-68:25:39&1&2.56&36.58\\
MCELS-L62&DEM L33&04:55:40.8&-68:38:43&0&1.70&37.08\\
MCELS-L63&DEM L35&04:55:58.4&-65:57:50&2&1.78&36.73\\
MCELS-L64&DEM L36&04:56:33.5&-69:29:04&0&1.07&37.55\\
MCELS-L65&DEM L34&04:56:34.3&-66:26:52&2&2.99&39.01\\
MCELS-L66&DEM L37&04:56:58.4&-69:12:45&1&1.91&36.61\\
MCELS-L67&DEM L38&04:57:05.8&-68:44:58&2&1.87&37.36\\
MCELS-L68&DEM L40&04:57:32.7&-67:39:10&0&1.64&36.36\\
MCELS-L69&\nodata&04:57:44.0&-66:15:31&1&4.68&36.11\\
MCELS-L70&DEM L41&04:58:10.3&-66:21:34&1&4.28&37.47\\
MCELS-L71&DEM L39&04:58:29.7&-68:27:02&2&1.12&38.12\\
MCELS-L72&DEM L42&04:58:48.4&-66:11:38&1&2.84&37.38\\
MCELS-L73&N12A&04:58:57.3&-66:13:58&1&3.11&35.98\\
MCELS-L74&DEM L43+48&04:58:58.8&-65:40:28&1&0.93&38.25\\
MCELS-L75&DEM L44&04:59:12.0&-66:16:03&1&3.07&36.24\\
MCELS-L76&DEM L50&04:59:45.5&-70:09:45&4&1.28&37.37\\
MCELS-L77&DEM L45&04:59:57.2&-67:56:48&3&0.98&37.28\\
MCELS-L78&DEM L51&05:00:00.3&-70:03:30&1&0.93&36.05\\
MCELS-L79&DEM L46&05:00:06.6&-66:15:39&1&2.23&36.91\\
MCELS-L80&DEM L47&05:00:10.3&-66:05:37&1&3.29&36.60\\
MCELS-L81&DEM L49&05:00:52.5&-66:23:16&0&2.15&35.94\\
MCELS-L82&DEM L52&05:01:48.0&-68:12:41&0&1.63&36.00\\
MCELS-L83&DEM L55&05:01:50.5&-70:38:37&0&0.62&37.03\\
MCELS-L84&DEM L53&05:01:52.5&-70:04:48&0&0.63&36.27\\
MCELS-L85&DEM L54&05:02:07.0&-69:34:02&0&0.72&35.53\\
MCELS-L86&BSDL453&05:02:09.2&-66:40:18&2&1.29&36.30\\
MCELS-L87&DEM L60&05:02:12.7&-69:03:38&0&2.54&36.46\\
MCELS-L88&DEM L57&05:02:48.6&-67:00:01&0&1.82&36.56\\
MCELS-L89&DEM L61&05:03:13.2&-65:57:24&0&2.38&36.88\\
MCELS-L90&DEM L58&05:03:15.6&-68:27:07&0&1.37&36.20\\
MCELS-L91&DEM L56&05:03:22.8&-66:42:02&0&1.34&36.42\\
MCELS-L92&DEM L59&05:04:20.1&-67:18:35&2&2.27&37.45\\
MCELS-L93&DEM L63&05:04:23.5&-70:44:07&1&1.37&37.18\\
MCELS-L94&DEM L62&05:04:25.2&-69:03:33&0&1.86&37.23\\
MCELS-L95&DEM L64a&05:04:31.4&-70:54:06&1&1.78&36.22\\
MCELS-L96&DEM L64b&05:04:38.9&-70:54:43&1&2.03&36.75\\
MCELS-L97&DEM L68&05:04:39.3&-70:10:32&1&1.34&35.93\\
MCELS-L98&DEM L68&05:04:47.9&-70:05:39&1&2.16&35.25\\
MCELS-L99&DEM L65&05:04:48.8&-67:33:18&2&1.94&36.84\\
MCELS-L100&DEM L67&05:04:55.1&-70:07:41&1&1.91&36.38\\
MCELS-L101&DEM L66&05:05:00.2&-68:03:41&3&1.58&37.56\\
MCELS-L102&DEM L68&05:05:06.7&-70:06:24&3&1.29&37.51$^d$\\
MCELS-L103&DEM L70&05:05:16.7&-68:05:41&3&1.56&36.94\\
MCELS-L104&DEM L69&05:05:16.9&-66:55:15&1&2.01&36.68\\
MCELS-L105&DEM L71&05:05:41.0&-67:52:49&0&1.63&35.58\\
MCELS-L106&DEM L72&05:06:05.0&-65:41:29&3&1.55&36.46\\
MCELS-L107&DEM L73&05:06:12.1&-68:07:28&1&2.33&37.48\\
MCELS-L108&DEM L76&05:06:45.3&-68:26:34&3&1.52&37.51\\
MCELS-L109&DEM L74&05:06:46.3&-68:09:54&1&2.77&36.69\\
MCELS-L110&DEM L75&05:06:47.4&-70:44:44&0&0.93&37.09\\
MCELS-L111&DEM L77&05:06:50.1&-66:54:56&0&1.24&36.65\\
MCELS-L112&DEM L78&05:06:59.9&-67:56:46&0&1.17&35.28\\
MCELS-L113&DEM L80&05:07:19.1&-70:27:02&0&0.93&37.42\\
MCELS-L114&DEM L79&05:07:21.5&-68:32:05&1&2.12&36.76\\
MCELS-L115&DEM L81&05:07:37.5&-71:10:32&0&0.91&36.10\\
MCELS-L116&DEM L83&05:07:39.3&-71:01:30&0&0.94&35.76\\
MCELS-L117&DEM L84&05:08:42.6&-68:45:52&4&1.83&37.86\\
MCELS-L118&DEM L85&05:09:24.0&-68:45:44&1&3.14&36.76\\
MCELS-L119&DEM L86&05:09:34.3&-68:53:45&3&2.55&38.13\\
MCELS-L120&DEM L91&05:09:38.1&-71:26:22&0&1.05&36.29\\
MCELS-L121&DEM L89&05:09:39.5&-67:55:14&3&1.86&37.45\\
MCELS-L122&DEM L88&05:09:53.0&-68:29:11&1&2.37&37.10\\
MCELS-L123&DEM L90&05:10:23.0&-67:09:27&0&2.06&36.32\\
MCELS-L124&DEM L92&05:10:30.3&-69:25:56&0&1.38&36.56\\
MCELS-L125&DEM L93&05:10:43.5&-67:04:50&1&2.03&36.05\\
MCELS-L126&DEM L95&05:10:54.3&-69:03:09&0&2.33&35.57\\
MCELS-L127&DEM L94&05:10:59.4&-67:07:34&1&1.66&36.18\\
MCELS-L128&DEM L96&05:11:23.6&-69:03:54&1&2.71&35.76\\
MCELS-L129&DEM L97&05:12:07.7&-67:06:54&0&2.88&36.96\\
MCELS-L130&DEM L100&05:12:08.2&-70:28:43&1&2.17&35.14\\
MCELS-L131&DEM L100&05:12:15.0&-70:28:05&1&2.03&35.91\\
MCELS-L132&DEM L100&05:12:21.9&-70:27:33&1&1.87&35.33\\
MCELS-L133&DEM L98&05:12:24.5&-67:15:40&0&2.41&36.37\\
MCELS-L134&DEM L101&05:12:28.4&-70:24:52&1&2.50&35.70\\
MCELS-L135&DEM L101&05:12:30.2&-70:24:22&3&2.61&36.52\\
MCELS-L136&DEM L101&05:12:30.4&-70:25:22&1&2.22&34.99\\
MCELS-L137&DEM L102&05:12:44.2&-70:22:02&2&2.22&36.14\\
MCELS-L138&DEM L99&05:12:48.7&-67:02:18&1&2.31&36.29\\
MCELS-L139&DEM L103&05:13:07.1&-69:01:56&0&2.27&36.64\\
MCELS-L140&DEM L104&05:13:10.4&-69:22:35&1&2.81&37.26\\
MCELS-L141&N30D&05:13:16.3&-67:28:24&1&2.94&36.17\\
MCELS-L142&DEM L111&05:13:18.7&-71:22:05&0&1.25&35.98\\
MCELS-L143&DEM L104&05:13:19.0&-69:21:26&1&2.76&37.45\\
MCELS-L144&DEM L104&05:13:21.4&-69:22:38&1&2.60&36.68\\
MCELS-L145&DEM L109&05:13:25.4&-69:10:55&0&2.15&36.29\\
MCELS-L146&DEM L108&05:13:34.9&-69:17:06&2&2.48&37.76\\
MCELS-L147&DEM L104&05:13:40.5&-69:22:30&2&1.60&37.19\\
MCELS-L148&DEM L105&05:13:43.4&-67:22:43&2&1.66&37.53\\
MCELS-L149&DEM L104&05:13:48.5&-69:23:15&1&1.77&36.81\\
MCELS-L150&DEM L106&05:13:53.1&-67:27:05&2&2.93&37.56\\
MCELS-L151&DEM L110&05:14:01.8&-69:31:55&2&1.30&37.37\\
MCELS-L152&DEM L107+115&05:14:06.0&-67:09:02&3&1.75&37.52\\
MCELS-L153&DEM L114&05:14:08.2&-70:07:43&0&1.68&35.54\\
MCELS-L154&DEM L112&05:14:34.4&-67:34:03&0&2.52&36.78\\
MCELS-L155&DEM L119&05:14:53.1&-71:37:09&3&1.21&36.83\\
MCELS-L156&DEM L113&05:14:53.8&-69:25:58&0&1.60&37.44\\
MCELS-L157&DEM L116&05:15:07.2&-66:28:06&1&1.07&36.29\\
MCELS-L158&DEM L120&05:15:20.5&-69:43:30&0&1.36&36.04\\
MCELS-L159&DEM L117&05:15:34.1&-67:20:35&0&1.77&35.73\\
MCELS-L160&DEM L118&05:15:45.7&-66:42:49&0&1.36&36.17\\
MCELS-L161&DEM L124&05:16:53.1&-69:53:01&0&2.02&36.21\\
MCELS-L162&DEM L121&05:16:53.4&-67:19:51&1&2.20&36.90\\
MCELS-L163&\nodata&05:17:02.9&-66:00:55&2&1.07&35.84\\
MCELS-L164&DEM L122&05:17:06.2&-68:52:35&0&3.39&35.05\\
MCELS-L165&DEM L131&05:17:39.1&-71:15:32&2&1.72&37.35\\
MCELS-L166&DEM L130&05:17:44.5&-69:24:00&0&2.34&36.96\\
MCELS-L167&DEM L128&05:17:46.0&-68:47:04&1&3.54&36.13\\
MCELS-L168&DEM L129&05:17:48.0&-67:54:05&2&1.83&36.19\\
MCELS-L169&DEM L127&05:17:49.0&-67:20:58&2&1.28&37.01\\
MCELS-L170&DEM L125&05:17:51.0&-66:01:17&1&1.45&37.11\\
MCELS-L171&DEM L123+132&05:17:59.7&-69:11:01&2&2.23&38.54\\
MCELS-L172&DEM L126&05:18:16.6&-65:57:41&0&1.67&36.49\\
MCELS-L173&DEM L134&05:18:27.4&-69:39:55&1&2.81&36.85\\
MCELS-L174&\nodata&05:18:40.6&-67:05:21&3&1.29&36.32\\
MCELS-L175&DEM L134&05:18:43.7&-69:39:15&1&2.14&36.83\\
MCELS-L176&DEM L134+133&05:19:16.4&-69:38:43&2&1.47&38.09\\
MCELS-L177&DEM L134&05:19:35.8&-69:38:47&3&2.25&36.32\\
MCELS-L178&DEM L135&05:19:48.3&-65:53:53&0&0.91&37.38\\
MCELS-L179&DEM L139&05:20:19.4&-66:28:54&0&1.40&35.37\\
MCELS-L180&DEM L136&05:20:22.1&-66:53:42&2&1.61&36.89\\
MCELS-L181&DEM L141&05:20:30.6&-68:01:02&1&3.59&36.14\\
MCELS-L182&DEM L138&05:20:34.0&-66:46:38&2&2.02&37.03\\
MCELS-L183&\nodata&05:20:40.3&-66:48:46&0&2.40&35.10\\
MCELS-L184&DEM L137&05:20:48.7&-65:27:25&3&0.77&37.64\\
MCELS-L185&DEM L146&05:20:54.5&-71:43:17&0&1.99&35.66\\
MCELS-L186&\nodata&05:21:04.9&-71:41:43&0&1.95&35.02\\
MCELS-L187&\nodata&05:21:17.4&-66:47:14&0&2.31&35.37\\
MCELS-L188&DEM L145&05:21:19.3&-69:40:50&1&2.42&36.71\\
MCELS-L189&DEM L147&05:21:26.3&-69:56:50&0&1.81&35.91\\
MCELS-L190&DEM L143&05:21:26.9&-68:52:04&0&1.87&36.86\\
MCELS-L191&DEM L144&05:21:31.1&-68:10:48&2&3.94&35.94\\
MCELS-L192&DEM L149&05:21:34.6&-69:40:27&1&3.01&37.15\\
MCELS-L193&DEM L150&05:21:36.6&-67:46:34&1&2.90&36.24\\
MCELS-L194&DEM L140&05:21:37.1&-67:54:49&1&3.06&37.46\\
MCELS-L195&DEM L148&05:21:37.3&-69:59:37&0&2.10&36.40\\
MCELS-L196&DEM L151a&05:21:46.8&-67:53:42&0&4.44&36.80\\
MCELS-L197&DEM L153&05:21:49.7&-69:41:09&1&3.15&35.85\\
MCELS-L198&DEM L151&05:21:49.9&-67:51:46&0&3.01&37.39\\
MCELS-L199&DEM L142&05:21:53.9&-65:43:49&0&0.97&37.18\\
MCELS-L200&DEM L154&05:22:00.9&-65:58:22&2&0.92&37.57\\
MCELS-L201&DEM L152&05:22:06.9&-67:56:46&3&3.97&38.19\\
MCELS-L202&DEM L164&05:22:13.1&-71:26:10&2&1.25&37.59\\
MCELS-L203&DEM L157&05:22:16.3&-68:38:58&1&1.68&36.61\\
MCELS-L204&DEM L158&05:22:19.4&-68:04:26&1&4.89&36.64\\
MCELS-L205&DEM L155&05:22:25.5&-65:44:46&0&1.24&37.32$^d$\\
MCELS-L206&DEM L156&05:22:26.9&-67:53:41&1&4.23&37.39\\
MCELS-L207&DEM L165&05:22:27.8&-71:35:51&1&1.88&37.57\\
MCELS-L208&DEM L158&05:22:29.9&-68:04:52&1&4.45&35.93\\
MCELS-L209&DEM L163&05:22:32.3&-70:08:45&2&1.10&36.50\\
MCELS-L210&DEM L159&05:22:36.2&-68:08:32&1&4.68&36.56\\
MCELS-L211&\nodata&05:22:37.0&-66:38:52&2&2.70&35.46\\
MCELS-L212&DEM L160&05:22:45.2&-68:04:00&1&4.51&37.89\\
MCELS-L213&DEM L161&05:22:48.0&-66:41:06&1&3.15&36.38\\
MCELS-L214&DEM L168&05:22:52.7&-69:50:58&0&2.20&36.40\\
MCELS-L215&DEM L155a&05:22:53.6&-65:43:03&1&1.54&36.88\\
MCELS-L216&DEM L166a&05:23:06.0&-68:00:17&1&3.74&36.28\\
MCELS-L217&DEM L162&05:23:06.8&-66:22:32&1&1.97&36.52\\
MCELS-L218&DEM L166b&05:23:12.5&-68:00:18&1&4.23&36.55\\
MCELS-L219&DEM L167&05:23:16.5&-67:56:14&1&3.53&36.65\\
MCELS-L220&DEM L170&05:23:27.4&-68:12:25&0&2.55&36.10\\
MCELS-L221&DEM L171&05:23:33.5&-69:38:51&0&2.04&36.60\\
MCELS-L222&DEM L169&05:23:39.5&-68:00:43&1&3.51&36.26\\
MCELS-L223&DEM L172&05:23:40.6&-69:37:01&2&1.00&36.91\\
MCELS-L224&DEM L173&05:23:50.8&-69:41:26&0&2.66&36.26\\
MCELS-L225&DEM L176a&05:24:03.0&-68:56:21&0&2.47&35.34\\
MCELS-L226&DEM L173&05:24:04.1&-69:40:18&1&3.19&36.08\\
MCELS-L227&DEM L173&05:24:07.0&-69:38:43&1&2.18&36.05\\
MCELS-L228&DEM L176b&05:24:09.4&-68:55:53&0&2.46&35.24\\
MCELS-L229&DEM L174&05:24:12.3&-68:30:09&1&2.76&36.31\\
MCELS-L230&N132B&05:24:17.5&-69:38:56&1&2.09&35.95\\
MCELS-L231&DEM L174&05:24:22.7&-68:31:34&3&2.21&37.56$^d$\\
MCELS-L232&DEM L175&05:24:24.1&-66:14:29&2&2.69&37.30\\
MCELS-L233&DEM L194&05:24:28.2&-71:38:36&0&2.16&36.00\\
MCELS-L234&DEM L178&05:24:34.2&-69:27:09&0&1.15&36.92\\
MCELS-L235&DEM L177&05:24:36.9&-69:06:55&0&1.76&38.32\\
MCELS-L236&DEM L182+184&05:24:38.5&-66:57:13&0&1.51&37.43\\
MCELS-L237&DEM L179&05:24:39.2&-68:28:47&1&3.24&36.26\\
MCELS-L238&DEM L180&05:24:58.2&-68:28:41&2&3.01&37.23\\
MCELS-L239&DEM L175a&05:24:58.5&-66:26:03&3&3.83&37.17\\
MCELS-L240&DEM L186&05:25:02.4&-69:38:34&3&2.03&36.26\\
MCELS-L241&DEM L188&05:25:06.1&-71:27:49&2&1.58&35.98\\
MCELS-L242&DEM L181&05:25:21.7&-66:02:46&2&2.56&37.17\\
MCELS-L243&DEM L187&05:25:25.4&-69:26:10&0&1.00&36.91\\
MCELS-L244&DEM L183&05:25:27.4&-66:21:52&1&3.98&36.07\\
MCELS-L245&DEM L185&05:25:54.5&-65:55:53&0&2.34&35.94\\
MCELS-L246&DEM L190&05:26:01.3&-66:04:57&4&3.91&37.19\\
MCELS-L247&DEM L189&05:26:04.1&-66:15:47&0&4.41&37.67\\
MCELS-L248&DEM L193&05:26:10.0&-67:10:37&3&1.43&37.31\\
MCELS-L249&DEM L195&05:26:12.7&-66:21:47&0&1.75&37.48\\
MCELS-L250&DEM L192+201&05:26:13.1&-67:29:56&1&1.90&38.52\\
MCELS-L251&DEM L196&05:26:13.9&-67:37:13&1&2.27&36.38\\
MCELS-L252&DEM L196&05:26:20.7&-67:37:55&1&1.86&37.60\\
MCELS-L253&DEM L189&05:26:23.2&-66:14:31&2&4.40&36.28\\
MCELS-L254&DEM L197&05:26:28.6&-69:18:58&3&0.78&37.38\\
MCELS-L255&DEM L196&05:26:28.6&-67:42:01&2&2.07&35.98\\
MCELS-L256&DEM L198&05:26:32.2&-69:02:04&0&1.20&37.77\\
MCELS-L257&DEM L196&05:26:37.9&-67:43:28&1&2.58&36.00\\
MCELS-L258&DEM L199&05:26:42.5&-68:49:34&3&1.18&38.36\\
MCELS-L259&DEM L202&05:27:10.3&-71:32:06&2&1.99&37.57\\
MCELS-L260&DEM L208&05:27:17.6&-70:34:46&3&1.01&37.73\\
MCELS-L261&DEM L206&05:27:26.8&-71:23:28&2&3.09&36.01\\
MCELS-L262&\nodata&05:27:31.3&-67:27:31&0&2.18&37.51\\
MCELS-L263&DEM L203&05:27:31.4&-68:27:00&0&1.68&38.38\\
MCELS-L264&DEM L207&05:27:41.6&-71:24:46&1&2.47&36.26\\
MCELS-L265&DEM L204&05:27:54.2&-65:50:10&4&0.65&36.49\\
MCELS-L266&DEM L209&05:28:02.2&-69:21:36&0&0.87&37.63\\
MCELS-L267&DEM L205&05:28:06.2&-67:26:50&2&2.56&37.59\\
MCELS-L268&BSDL1844&05:28:14.6&-67:23:57&1&2.50&36.67\\
MCELS-L269&DEM L210&05:28:19.2&-69:01:21&0&1.52&37.96\\
MCELS-L270&DEM L213&05:28:43.1&-70:20:20&0&1.44&36.44\\
MCELS-L271&DEM L211&05:28:54.3&-67:43:24&0&1.58&36.57\\
MCELS-L272&DEM L214&05:29:17.4&-66:57:21&0&0.69&36.95\\
MCELS-L273&DEM L215&05:29:33.4&-69:48:48&0&1.23&37.26\\
MCELS-L274&\nodata&05:29:52.2&-71:04:33&1&2.17&36.56\\
MCELS-L275&DEM L216&05:30:09.4&-69:45:09&0&0.91&36.18\\
MCELS-L276&DEM L218&05:30:33.9&-70:07:59&0&1.41&37.02\\
MCELS-L277&DEM L221&05:30:34.0&-71:01:21&1&2.02&38.43\\
MCELS-L278&N206B&05:30:51.2&-71:08:01&1&3.48&36.83\\
MCELS-L279&DEM L222a&05:30:58.1&-67:20:28&0&1.77&35.56\\
MCELS-L280&DEM L224&05:31:03.0&-69:19:19&0&0.93&38.37\\
MCELS-L281&DEM L222b&05:31:11.6&-67:22:47&0&1.47&35.91\\
MCELS-L282&DEM L223&05:31:17.7&-67:26:54&0&1.08&36.59\\
MCELS-L283&DEM L225&05:31:49.2&-67:21:35&0&1.43&35.47\\
MCELS-L284&DEM L226&05:32:01.7&-68:40:24&1&2.43&37.21\\
MCELS-L285&DEM L228a&05:32:05.4&-66:24:46&1&1.84&36.91\\
MCELS-L286&BSDL2224&05:32:10.6&-68:39:04&1&3.63&35.68\\
MCELS-L287&DEM L227&05:32:10.9&-68:28:23&1&3.77&38.07$^d$\\
MCELS-L288&DEM L228b&05:32:13.6&-66:23:26&1&1.42&36.89\\
MCELS-L289&DEM L219+229&05:32:19.6&-67:40:57&4&2.04&38.29\\
MCELS-L290&BSDL2247&05:32:30.7&-68:40:13&1&3.26&36.15\\
MCELS-L291&DEM L228&05:32:32.9&-66:24:20&2&0.87&37.98$^d$\\
MCELS-L292&N148A&05:32:44.0&-68:24:28&1&0.00&33.31\\
MCELS-L293&DEM L230&05:32:53.5&-67:49:09&0&1.27&35.59\\
MCELS-L294&DEM L232&05:33:03.9&-68:56:15&0&2.05&38.37\\
MCELS-L295&DEM L231&05:33:10.6&-67:42:45&3&2.74&37.24\\
MCELS-L296&DEM L234&05:33:40.0&-67:31:03&0&1.93&38.01\\
MCELS-L297&DEM L233&05:33:43.9&-68:46:01&1&2.68&36.56\\
MCELS-L298&DEM L238&05:34:17.7&-70:33:35&4&1.29&36.02\\
MCELS-L299&DEM L236&05:34:25.9&-67:05:46&0&0.76&36.17\\
MCELS-L300&DEM L235&05:34:31.1&-66:07:46&1&1.19&37.24\\
MCELS-L301&DEM L258&05:34:37.5&-68:12:37&0&3.22&37.04\\
MCELS-L302&DEM L239&05:34:40.0&-66:14:19&2&0.82&37.47\\
MCELS-L303&DEM L237&05:34:40.4&-65:57:10&2&1.83&37.00\\
MCELS-L304&DEM L242&05:34:51.0&-69:31:23&1&2.70&36.64\\
MCELS-L305&DEM L240&05:34:53.1&-67:21:24&3&0.81&36.61\\
MCELS-L306&DEM L246&05:35:04.8&-69:43:25&2&2.58&38.51\\
MCELS-L307&DEM L241&05:35:15.5&-67:34:04&3&2.28&38.37\\
MCELS-L308&DEM L245&05:35:22.4&-67:42:08&0&1.96&37.13\\
MCELS-L309&DEM L244&05:35:28.2&-66:38:43&0&0.52&37.42\\
MCELS-L310&DEM L243&05:35:33.6&-66:02:26&1&1.56&37.66\\
MCELS-L311&N59C&05:35:39.1&-67:37:07&1&3.11&36.83\\
MCELS-L312&DEM L249&05:36:08.3&-70:38:56&0&2.66&35.99\\
MCELS-L313&DEM L248&05:36:09.8&-69:31:54&0&3.27&37.91\\
MCELS-L314&DEM L247&05:36:28.0&-66:02:12&0&1.37&36.29\\
MCELS-L315&DEM L259&05:36:30.8&-69:49:06&1&3.75&36.72\\
MCELS-L316&DEM L250&05:36:39.8&-67:26:51&1&2.37&37.22\\
MCELS-L317&DEM L251&05:36:40.2&-66:26:31&0&2.15&36.86\\
MCELS-L318&DEM L252&05:37:02.7&-66:21:18&1&2.08&37.62\\
MCELS-L319&DEM L255&05:37:03.0&-66:39:46&1&1.14&37.62\\
MCELS-L320&DEM L260&05:37:16.7&-69:46:00&1&5.09&36.24\\
MCELS-L321&DEM L253&05:37:18.0&-66:17:58&1&2.45&37.10\\
MCELS-L322&DEM L256&05:37:28.9&-66:27:51&0&2.34&36.20\\
MCELS-L323&DEM L261&05:37:43.0&-69:22:03&3&3.10&38.24\\
MCELS-L324&DEM L262&05:37:50.7&-69:39:20&0&4.32&37.65\\
MCELS-L325&DEM L254&05:38:12.5&-66:18:16&1&1.52&36.98\\
MCELS-L326&DEM L265&05:38:18.6&-70:41:15&2&2.82&37.37\\
MCELS-L327&DEM L264&05:38:21.6&-66:35:25&0&1.28&36.43\\
MCELS-L328&DEM L263&05:38:36.0&-69:05:11&3&4.24&39.66\\
MCELS-L329&DEM L266&05:38:48.2&-70:04:28&0&5.47&35.43\\
MCELS-L330&DEM L269&05:38:57.6&-69:29:55&0&4.73&38.27\\
MCELS-L331&DEM L267&05:39:14.1&-70:12:39&0&5.51&36.43\\
MCELS-L332&N158D&05:39:16.4&-69:33:17&1&4.88&36.70\\
MCELS-L333&DEM L269&05:39:33.1&-69:25:25&2&3.46&38.15\\
MCELS-L334&\nodata&05:39:34.3&-69:39:23&1&5.37&37.53\\
MCELS-L335&DEM L269&05:39:37.4&-69:28:07&1&4.65&36.89\\
MCELS-L336&\nodata&05:39:46.1&-69:38:49&1&4.29&37.82\\
MCELS-L337&DEM L284&05:39:47.9&-69:37:08&1&4.96&38.56$^d$\\
MCELS-L338&DEM L270&05:39:48.0&-66:08:52&2&1.23&36.97\\
MCELS-L339&DEM L274&05:39:51.1&-71:09:22&1&5.05&35.59\\
MCELS-L340&DEM L274&05:39:53.8&-71:09:43&1&5.15&36.24\\
MCELS-L341&DEM L274&05:39:58.0&-71:10:16&1&5.20&36.30\\
MCELS-L342&DEM L272&05:40:03.4&-69:49:14&1&6.54&36.02\\
MCELS-L343&DEM L271&05:40:06.8&-69:45:28&3&6.33&38.27\\
MCELS-L344&DEM L276&05:40:09.5&-71:11:05&1&5.06&36.77\\
MCELS-L345&DEM L278&05:40:10.1&-71:12:26&1&5.46&36.02\\
MCELS-L346&DEM L275&05:40:12.3&-69:55:01&1&5.59&36.19\\
MCELS-L347&DEM L273&05:40:13.5&-68:59:26&1&4.99&36.62\\
MCELS-L348&DEM L277&05:40:22.9&-69:53:16&1&5.76&35.82\\
MCELS-L349&\nodata&05:40:23.4&-69:40:17&0&4.64&37.16\\
MCELS-L350&DEM L279&05:40:26.9&-69:50:24&0&8.43&36.02\\
MCELS-L351&DEM L281&05:40:42.0&-70:02:31&1&4.74&36.99\\
MCELS-L352&\nodata&05:40:43.6&-69:47:19&2&7.14&36.12\\
MCELS-L353&\nodata&05:40:48.3&-69:49:33&2&7.29&36.04\\
MCELS-L354&DEM L280&05:40:48.5&-70:10:06&1&5.33&36.06\\
MCELS-L355&DEM L283&05:40:50.8&-69:46:08&1&6.01&35.85\\
MCELS-L356&DEM L287&05:40:53.4&-71:12:05&3&5.34&36.21\\
MCELS-L357&DEM L290&05:40:57.3&-70:54:41&1&4.39&35.79\\
MCELS-L358&DEM L282&05:41:03.0&-69:55:24&0&4.35&36.91\\
MCELS-L359&DEM L288&05:41:07.1&-71:13:43&0&4.90&35.79\\
MCELS-L360&DEM L291&05:41:12.1&-70:29:45&0&4.71&35.80\\
MCELS-L361&DEM L289&05:41:25.6&-71:15:48&1&4.40&35.86\\
MCELS-L362&DEM L292&05:41:25.7&-71:17:00&1&4.08&35.70\\
MCELS-L363&DEM L285&05:41:28.4&-69:46:36&0&4.47&37.05\\
MCELS-L364&DEM L295&05:41:34.5&-70:01:19&0&4.16&36.35\\
MCELS-L365&DEM L294&05:41:35.1&-70:35:13&3&5.14&36.64\\
MCELS-L366&DEM L286&05:41:35.8&-66:59:16&0&0.81&36.26\\
MCELS-L367&DEM L293&05:41:38.1&-71:19:49&2&4.13&37.78\\
MCELS-L368&DEM L297&05:42:10.0&-68:58:17&0&5.58&36.57\\
MCELS-L369&DEM L298&05:42:17.2&-69:05:45&1&5.00&37.70\\
MCELS-L370&DEM L296&05:42:25.9&-66:39:50&0&0.89&36.55\\
MCELS-L371&DEM L299&05:42:55.3&-68:56:53&4&5.10&37.30\\
MCELS-L372&DEM L300&05:43:04.6&-69:45:57&1&5.35&37.48\\
MCELS-L373&DEM L301&05:43:17.5&-67:50:48&2&1.77&37.84\\
MCELS-L374&DEM L303&05:43:46.7&-67:27:12&1&2.73&36.35\\
MCELS-L375&DEM L306&05:44:15.4&-66:23:40&0&1.22&37.71\\
MCELS-L376&DEM L305&05:44:22.2&-67:27:35&0&2.48&35.83\\
MCELS-L377&DEM L307&05:44:24.2&-69:22:43&1&4.80&36.78\\
MCELS-L378&DEM L310&05:44:55.5&-69:27:44&0&4.49&38.53$^d$\\
MCELS-L379&DEM L309&05:45:08.9&-67:08:53&2&1.20&37.57\\
MCELS-L380&DEM L311&05:45:22.2&-69:46:21&1&4.01&37.04\\
MCELS-L381&DEM L308&05:45:25.0&-67:17:44&0&1.20&37.79\\
MCELS-L382&DEM L309&05:45:43.0&-67:09:49&1&1.67&36.50\\
MCELS-L383&DEM L312&05:46:05.8&-69:33:28&0&5.34&36.84\\
MCELS-L384&DEM L313&05:46:27.3&-69:35:22&1&5.39&35.86\\
MCELS-L385&DEM L314&05:46:33.9&-69:34:22&1&5.27&36.28\\
MCELS-L386&DEM L315&05:46:38.8&-67:10:39&3&0.86&37.40\\
MCELS-L387&DEM L317&05:47:04.0&-70:09:04&0&3.99&36.56\\
MCELS-L388&DEM L316&05:47:06.3&-69:42:33&4&6.24&36.85\\
MCELS-L389&DEM L318&05:47:56.6&-69:52:10&1&3.78&35.44\\
MCELS-L390&DEM L320&05:48:01.9&-69:53:51&1&3.69&36.32\\
MCELS-L391&DEM L319&05:48:02.9&-69:53:02&1&3.76&35.63\\
MCELS-L392&DEM L321&05:48:11.3&-69:52:44&1&3.83&35.35\\
MCELS-L393&DEM L322&05:48:15.1&-70:02:03&1&4.19&36.78\\
MCELS-L394&DEM L324&05:48:44.6&-69:50:38&1&2.91&35.91\\
MCELS-L395&DEM L325&05:48:58.0&-69:59:43&1&3.72&36.59\\
MCELS-L396&DEM L323+326&05:49:24.0&-70:06:20&2&3.03&38.17\\
MCELS-L397&DEM L327&05:49:27.3&-69:19:34&0&2.84&36.84\\
MCELS-L398&DEM L328&05:51:29.3&-68:13:15&0&1.88&37.60\\
MCELS-L399&DEM L329&05:51:41.9&-69:55:49&0&1.43&35.40\\
MCELS-L400&N75A&05:55:42.2&-68:09:47&1&2.08&35.76\\
MCELS-L401&\nodata&05:55:54.4&-68:13:54&1&1.47&37.43\\
\hline
\label{tab:LMCObjs}
\end{longtable}
{\footnotesize
  \noindent $^a$Identifiers in column 2 are from \citet{DEM} (DEM),
 \citet{Bica1999} (BSDL), or \citet{Henize1956} (N). \\ $^b$Optical
  depth classifications in column 5 are: (0) indeterminate, (1)
  optically thick, (2) blister, (3) optically thin, and (4) shocked
  nebulae.  \\ $^c$Local \Ha\ backgrounds could not be unambiguously
  determined for $L$ measurements of these objects due to a high DIG
  luminosity, even though structure is seen in \SII/\OIII.  Therefore
  the background was set to the surface brightness of the outermost
  area of the LMC observed by the MCELS survey, as discussed in the
  text.\\ $^d$Object includes separately catalogued substructure in
  the line of sight.  Photometry for the substructure
is not included in the photometry of the larger region.}

\clearpage
\renewcommand*\thetable{6}
\LongTables
\begin{longtable}{lcccccc}
\caption{REVISED - MCELS SMC HII Region Catalog - REVISED}\\
\hline \hline \\[-2ex]
\multicolumn{1}{c}{Object ID}&\multicolumn{1}{c}{Other ID$^a$}&\multicolumn{1}{c}{RA (J2000)}&\multicolumn{1}{c}{Dec}&\multicolumn{1}{c}{Type$^b$}&\multicolumn{1}{c}{$N$(HI)}&\multicolumn{1}{c}{$L$(\Ha)}\\
\multicolumn{1}{c}{}&\multicolumn{1}{c}{}&\multicolumn{1}{c}{h:m:s}&\multicolumn{1}{c}{d:m:s}&\multicolumn{1}{c}{}&\multicolumn{1}{c}{$10^{21}
  cm^{-2}$}&\multicolumn{1}{c}{\ergs}\\[1.5ex] \hline
\endfirsthead

\hline \hline \\[-2ex]
\multicolumn{1}{c}{Object ID}&\multicolumn{1}{c}{Other ID$^a$}&\multicolumn{1}{c}{RA (J2000)}&\multicolumn{1}{c}{Dec}&\multicolumn{1}{c}{Type$^b$}&\multicolumn{1}{c}{$N$(HI)}&\multicolumn{1}{c}{$L$(\Ha)}\\
\multicolumn{1}{c}{}&\multicolumn{1}{c}{}&\multicolumn{1}{c}{h:m:s}&\multicolumn{1}{c}{d:m:s}&\multicolumn{1}{c}{}&\multicolumn{1}{c}{$10^{21}
  cm^{-2}$}&\multicolumn{1}{c}{\ergs}\\[1.5ex] \hline
\endhead

\hline
\multicolumn{3}{l}{{Continued on the Next Page\ldots}}\\
\endfoot
\endlastfoot
MCELS-S1&DEM S1&00:31:41.1&-73:47:39&3&2.38&36.30\\
MCELS-S2&DEM S2&00:36:59.0&-72:59:42&1&2.66&36.67\\
MCELS-S3&\nodata&00:39:59.6&-73:33:30&1&2.88&35.88\\
MCELS-S4&DEM S5&00:41:02.6&-73:36:20&1&3.88&36.21\\
MCELS-S5&\nodata&00:41:39.6&-73:24:26&2&4.94&34.76\\
MCELS-S6&\nodata&00:42:10.2&-73:14:52&2&5.77&36.34\\
MCELS-S7&DEM S6&00:42:15.5&-72:59:38&2&5.36&35.84\\
MCELS-S8&DEM S7&00:42:27.3&-73:43:55&1&3.11&35.95\\
MCELS-S9&\nodata&00:42:55.3&-74:28:52&1&1.93&36.11\\
MCELS-S10&\nodata&00:43:06.7&-73:20:30&2&6.82&35.85\\
MCELS-S11&DEM S8&00:43:08.0&-72:35:54&2&1.86&34.54\\
MCELS-S12&\nodata&00:43:18.7&-73:13:53&3&6.48&34.57\\
MCELS-S13&\nodata&00:43:34.9&-73:15:45&1&7.63&34.67\\
MCELS-S14&DEM S9&00:43:35.8&-73:02:28&2&5.36&36.34\\
MCELS-S15&\nodata&00:43:44.9&-73:08:53&1&6.30&36.70\\
MCELS-S16&\nodata&00:43:47.3&-73:15:55&2&8.11&34.34\\
MCELS-S17&DEM S10&00:43:50.8&-73:28:29&1&6.12&36.96\\
MCELS-S18&\nodata&00:44:54.2&-72:56.00&1&4.13&35.07\\
MCELS-S19&N10&00:44:54.4&-73:10:24&1&8.43&36.13\\
MCELS-S20&DEM S15+18&00:45:20.6&-73:03:39&1&8.60&37.44\\
MCELS-S21&DEM S12&00:45:21.0&-73:59:13&3&2.89&36.29\\
MCELS-S22&N13B&00:45:21.4&-73:22:28&1&8.82&36.71\\
MCELS-S23&DEM S16&00:45:24.4&-73:23:06&1&8.43&36.65\\
MCELS-S24&DEM S14+19&00:45:25.9&-73:15:12&1&9.54&37.11\\
MCELS-S25&DEM S17&00:45:32.3&-73:12:30&1&9.16&36.48\\
MCELS-S26&DEM S20&00:46:09.6&-73:05:59&1&10.01&37.16\\
MCELS-S27&DEM S28&00:46:11.0&-73:25:37&1&7.43&35.67\\
MCELS-S28&DEM S21&00:46:19.8&-73:23:33&1&8.35&36.59\\
MCELS-S29&DEM S29&00:46:24.3&-73:26:22&1&7.80&37.61$^e$\\
MCELS-S30&DEM S22&00:46:24.8&-73:12:28&1&8.78&36.55\\
MCELS-S31&DEM S23&00:46:31.5&-73:06:16&1&10.86&37.40\\
MCELS-S32&\nodata&00:46:38.5&-72:54:40&1&6.61&34.89\\
MCELS-S33&DEM S24&00:46:41.0&-73:21:37&1&9.64&36.61\\
MCELS-S34&DEM S25&00:46:43.0&-73:31:48&1&6.44&37.14\\
MCELS-S35&\nodata&00:47:00.9&-73:18:04&1&10.16&36.76\\
MCELS-S36&\nodata&00:47:07.5&-73:14:11&3&9.46&36.24\\
MCELS-S37&DEM S31&00:47:29.1&-73:05:08&2&13.93&36.82\\
MCELS-S38&DEM S30&00:47:30.3&-73:22:17&1&8.90&36.66\\
MCELS-S39&DEM S32&00:47:43.9&-73:08:22&1&11.73&37.74\\
MCELS-S40&DEM S35&00:47:46.4&-73:17:31&1&10.46&36.26\\
MCELS-S41&DEM S34&00:47:53.7&-73:17:36&2&10.00&34.89\\
MCELS-S42&DEM S36&00:47:57.4&-73:17:38&1&9.96&36.48\\
MCELS-S43&DEM S33&00:48:03.3&-73:35:12&1&4.94&36.14\\
MCELS-S44&DEM S37&00:48:03.9&-73:16:23&1&9.98&37.33\\
MCELS-S45&DEM S38&00:48:09.1&-73:14:09&1&10.73&36.80\\
MCELS-S46&\nodata&00:48:15.8&-73:11:17&2&10.61&35.57\\
MCELS-S47&DEM S39&00:48:18.1&-73:10:19&1&10.98&35.8\\
MCELS-S48&DEM S42&00:48:18.9&-73:19:43&1&10.02&36.84\\
MCELS-S49&DEM S41&00:48:21.6&-73:32:52&1&4.93&36.68\\
MCELS-S50&DEM S40&00:48:24.0&-73:05:50&0&13.27&36.41\\
MCELS-S51&DEM S43&00:48:26.7&-73:15:16&1&9.78&37.12\\
MCELS-S52&\nodata&00:48:28.3&-72:15:58&1&2.87&36.73\\
MCELS-S53&\nodata&00:48:32.6&-72:52:58&2&9.57&35.78\\
MCELS-S54&DEM S51&00:48:56.3&-73:03:55&1&10.80&37.51$^e$\\
MCELS-S55&\nodata&00:48:56.6&-73:11:40&1&10.64&35.59\\
MCELS-S56&DEM S45&00:49:01.8&-73:08:24&1&12.06&37.43\\
MCELS-S57&DEM S44&00:49:12.0&-73:28:19&1&6.72&37.26$^e$\\
MCELS-S58&DEM S47&00:49:14.6&-72:52:45&1&9.83&36.90\\
MCELS-S59&DEM S46e&00:49:29.4&-72:47:44&1&7.01&36.45$^e$\\
MCELS-S60&DEM S49&00:49:36.5&-73:15:43&1&8.15&36.77\\
MCELS-S61&\nodata&00:49:37.5&-73:25:30&2&7.20&35.47\\
MCELS-S62&DEM S48&00:49:39.4&-72:48:47&2&8.57&36.26\\
MCELS-S63&DEM S50&00:49:45.0&-73:10:31&2&9.64&35.81\\
MCELS-S64&DEM S47&00:49:47.0&-72:56:34&1&10.64&36.80\\
MCELS-S65&\nodata&00:49:52.3&-73:24:11&3&7.66&36.08\\
MCELS-S66&\nodata&00:49:52.7&-73:25:39&2&7.69&36.51\\
MCELS-S67&DEM S46+55&00:49:57.2&-72:44:57&2&4.94&37.95\\
MCELS-S68&DEM S57&00:50:00.6&-72:32:43&1&4.20&36.43\\
MCELS-S69&DEM S56&00:50:25.7&-72:35:36&4&3.86&37.21\\
MCELS-S70&\nodata&00:50:30.2&-73:31:36&1&6.74&36.57\\
MCELS-S71&DEM S54&00:50:33.9&-72:53:26&1&10.46&37.84$^e$\\
MCELS-S72&DEM S52&00:50:34.6&-73:20:11&1&8.04&36.49\\
MCELS-S73&DEM S53&00:50:48.8&-73:24:22&3&7.20&35.76\\
MCELS-S74&N41&00:51:01.7&-72:52:52&1&9.91&36.24\\
MCELS-S75&\nodata&00:51:06.3&-73:31:36&0&6.87&35.04\\
MCELS-S76&\nodata&00:51:14.1&-73:31:36&1&6.91&34.92\\
MCELS-S77&DEM S59&00:51:19.1&-73:30:15&1&6.71&36.54\\
MCELS-S78&DEM S63&00:51:34.7&-72:41:22&1&5.39&37.14\\
MCELS-S79&\nodata&00:51:40.6&-73:31:51&1&7.56&34.62\\
MCELS-S80&DEM S62&00:51:47.8&-72:50:47&1&9.58&36.15\\
MCELS-S81&DEM S65&00:51:58.3&-72:16:31&1&5.18&36.70\\
MCELS-S82&DEM S60&00:52:01.7&-73:13:19&3&8.80&36.64\\
MCELS-S83&\nodata&00:52:04.9&-72:44:58&1&6.72&37.07\\
MCELS-S84&DEM S63&00:52:05.6&-72:39:36&1&7.02&36.68\\
MCELS-S85&DEM S67&00:52:15.1&-71:50:18&1&1.50&36.56\\
MCELS-S86&DEM S71&00:52:18.2&-73:27:07&1&8.16&35.92\\
MCELS-S87&DEM S66&00:52:25.4&-72:08:59&1&2.82&37.50$^e$\\
MCELS-S88&DEM S72&00:52:37.2&-73:26:11&1&8.21&36.21\\
MCELS-S89&\nodata&00:52:39.5&-72:55:30&1&7.05&35.17\\
MCELS-S90&DEM S69&00:53:01.5&-72:53:42&1&6.32&36.62\\
MCELS-S91&DEM S74&00:53:23.6&-73:12:08&1&9.13&36.96\\
MCELS-S92&\nodata&00:53:25.5&-72:28:30&1&4.79&35.36\\
MCELS-S93&N52A&00:53:40.6&-72:39:38&2&7.52&35.82\\
MCELS-S94&N52B&00:53:43.1&-72:39:21&2&7.48&35.82\\
MCELS-S95&DEM S76&00:54:01.5&-72:22:23&1&3.54&36.60\\
MCELS-S96&DEM S78&00:54:11.2&-73:17:17&3&8.34&36.19\\
MCELS-S97&\nodata&00:54:15.7&-73:32:17&1&5.56&35.49\\
MCELS-S98&DEM S80&00:54:23.0&-72:42:49&2&7.10&37.78\\
MCELS-S99&DEM S79&00:54:29.2&-71:57:58&3&1.86&36.02\\
MCELS-S100&\nodata&00:54:56.9&-73:19:14&4&5.57&35.04\\
MCELS-S101&\nodata&00:55:02.4&-72:55:59&1&6.37&35.94\\
MCELS-S102&\nodata&00:55:14.7&-72:26:33&1&5.58&35.65\\
MCELS-S103&\nodata&00:55:34.4&-72:29:12&1&5.30&35.37\\
MCELS-S104&DEM S83b&00:55:34.4&-72:17:12&3&5.63&35.94\\
MCELS-S105&DEM S83a&00:55:44.1&-72:16:02&2&5.71&35.87\\
MCELS-S106&\nodata&00:55:48.9&-72:38:14&1&7.61&35.39\\
MCELS-S107&DEM S81&00:55:57.4&-73:23:17&1&3.87&36.86\\
MCELS-S108&DEM S84&00:56:02.2&-72:15:44&2&6.42&35.66\\
MCELS-S109&\nodata&00:56:03.5&-72:27:11&3&5.63&36.11\\
MCELS-S110&DEM S85&00:56:16.4&-72:17:25&1&6.59&36.56\\
MCELS-S111&DEM S88&00:56:18.6&-72:47:24&3&8.53&36.13\\
MCELS-S112&DEM S90&00:56:46.9&-72:03:22&1&3.23&36.70\\
MCELS-S113&DEM S89&00:56:48.1&-72:47:47&2&9.21&35.47\\
MCELS-S114&\nodata&00:57:02.5&-72:21:51&2&7.22&35.80\\
MCELS-S115&DEM S91&00:57:10.3&-73:34:17&1&3.06&36.19\\
MCELS-S116&\nodata&00:57:18.3&-71:54:39&1&3.28&35.31\\
MCELS-S117&\nodata&00:57:38.4&-72:24:40&1&8.04&35.42\\
MCELS-S118&DEM S93&00:57:56.7&-72:39:21&0&7.86&36.47\\
MCELS-S119&DEM S94&00:58:16.6&-72:38:53&1&7.75&36.57\\
MCELS-S120&SNR B0056-72.5$^c$&00:58:17.0&-72:17:53&4&5.66&36.45\\
MCELS-S121&DEM S95&00:58:20.1&-72:40:10&1&7.50&36.68\\
MCELS-S122&DEM S96&00:58:28.1&-71:44:40&1&3.18&36.11\\
MCELS-S123&DEM S97&00:58:30.6&-71:31:07&1&1.84&36.66$^e$\\
MCELS-S124&DEM S98&00:58:36.6&-72:14:08&3&6.80&36.37\\
MCELS-S125&DEM S99&00:58:57.3&-72:14:36&3&5.85&36.18\\
MCELS-S126&\nodata&00:59:05.9&-71:45:19&2&3.66&34.51\\
MCELS-S127&DEM S100&00:59:15.0&-72:24:15&3&7.43&36.18\\
MCELS-S128&DEM S102&00:59:18.8&-72:17:31&1&4.69&36.91\\
MCELS-S129&DEM S105&00:59:42.1&-71:43:43&0&3.86&36.44\\
MCELS-S130&N66,DEM S103&00:59:42.6&-72:12:05&1&3.91&38.81\\
MCELS-S131&DEM S107&01:00:15.2&-71:48:25&2&4.47&36.28\\
MCELS-S132&DEM S109&01:00:58.6&-71:35:27&3&4.22&36.11\\
MCELS-S133&DEM S108&01:01:17.7&-71:30:59&1&3.31&37.28\\
MCELS-S134&DEM S113&01:01:30.3&-71:47:45&2&5.00&35.64\\
MCELS-S135&DEM S112&01:01:30.7&-71:51:08&1&4.25&36.47\\
MCELS-S136&DEM S111&01:01:43.2&-71:56:16&0&4.53&36.42\\
MCELS-S137&\nodata&01:01:53.6&-72:06:10&1&4.37&36.67\\
MCELS-S138&\nodata&01:01:59.9&-71:54:54&2&5.41&35.63\\
MCELS-S139&DEM S115&01:02:16.6&-71:51:25&1&5.05&36.23\\
MCELS-S140&DEM S116&01:02:28.6&-71:56:34&2&5.62&35.41\\
MCELS-S141&DEM S118&01:02:41.1&-72:24:41&1&5.22&36.74\\
MCELS-S142&DEM S117a&01:02:43.5&-71:53:34&1&5.69&35.56\\
MCELS-S143&DEM S117b&01:02:47.8&-71:53:18&3&5.71&36.41\\
MCELS-S144&DEM S119+120&01:03:01.2&-72:05:41&1&6.24&36.49\\
MCELS-S145&DEM S121&01:03:03.1&-71:53:30&3&5.54&36.29\\
MCELS-S146&SNR B0101-72.6$^d$&01:03:17.4&-72:09:43&1&5.75&35.29\\
MCELS-S147&DEM S123&01:03:25.0&-72:03:45&1&6.04&37.90$^e$\\
MCELS-S148&DEM S123&01:03:48.6&-72:03:56&3&5.99&36.18\\
MCELS-S149&DEM S122&01:03:58.7&-72:40:53&1&5.16&35.86\\
MCELS-S150&\nodata&01:04:08.4&-72:02:07&1&5.62&36.51\\
MCELS-S151&\nodata&01:04:15.0&-72:24:00&2&4.76&34.72\\
MCELS-S152&\nodata&01:04:22.0&-71:56:46&2&5.73&35.77\\
MCELS-S153&N78&01:05:03.1&-71:59:22&2&6.52&36.35\\
MCELS-S154&N78A&01:05:04.4&-71:58:58&3&6.37&36.03\\
MCELS-S155&DEM S125&01:05:07.7&-72:12:14&1&5.14&37.34\\
MCELS-S156&DEM S129&01:05:09.5&-72:48:05&1&5.31&37.05\\
MCELS-S157&DEM S127&01:05:12.1&-71:58:25&2&6.28&36.41\\
MCELS-S158&DEM S126&01:05:12.8&-72:00:38&1&6.00&37.51\\
MCELS-S159&DEM S128&01:05:23.5&-72:08:42&3&5.19&36.29\\
MCELS-S160&DEM S130&01:05:41.3&-72:03:48&3&5.96&36.25\\
MCELS-S161&DEM S134&01:05:52.1&-72:29:56&1&4.60&37.21\\
MCELS-S162&\nodata&01:05:55.5&-72:19:47&2&4.51&35.16\\
MCELS-S163&DEM S131&01:06:18.1&-72:05:24&4&5.96&36.53\\
MCELS-S164&DEM S132&01:06:25.0&-71:57:37&1&6.56&37.21\\
MCELS-S165&\nodata&01:06:41.5&-73:10:02&2&5.76&35.67\\
MCELS-S166&DEM S134&01:06:56.1&-72:33:06&1&5.87&36.69$^e$\\
MCELS-S167&\nodata&01:07:27.0&-73:33:13&1&2.84&35.65\\
MCELS-S168&DEM S133&01:07:34.7&-72:51:20&1&5.32&36.91\\
MCELS-S169&DEM S135&01:08:09.9&-71:59:50&1&6.17&37.48\\
MCELS-S170&DEM S136&01:09:05.0&-71:51:37&1&4.48&36.58\\
MCELS-S171&DEM S138&01:09:17.0&-73:10:59&1&4.83&37.39\\
MCELS-S172&\nodata&01:09:27.3&-72:01:28&3&4.83&36.10\\
MCELS-S173&\nodata&01:09:41.5&-73:18:16&1&4.15&35.23\\
MCELS-S174&\nodata&01:09:50.4&-72:30:50&1&4.00&36.42\\
MCELS-S175&\nodata&01:10:44.1&-72:21:25&2&3.97&35.87\\
MCELS-S176&DEM S140&01:10:49.7&-72:43:17&1&4.64&37.41\\
MCELS-S177&\nodata&01:11:05.1&-72:13:48&3&4.23&36.03\\
MCELS-S178&DEM S142&01:11:20.5&-72:09:50&1&4.22&36.69\\
MCELS-S179&DEM S141&01:11:35.4&-72:21:57&1&4.03&37.06\\
MCELS-S180&DEM S150&01:11:41.9&-73:13:26&1&5.34&36.29\\
MCELS-S181&\nodata&01:11:43.5&-73:17:52&1&4.46&36.46$^e$\\
MCELS-S182&DEM S144&01:11:53.5&-72:44:19&3&4.95&36.24\\
MCELS-S183&DEM S143&01:12:14.0&-72:15:28&2&4.18&35.80\\
MCELS-S184&DEM S147&01:13:42.1&-73:17:50&1&6.16&37.68\\
MCELS-S185&DEM S148&01:13:51.6&-73:15:46&2&7.35&35.94\\
MCELS-S186&DEM S147&01:14:00.3&-73:17:16&1&6.82&36.73\\
MCELS-S187&DEM S149&01:14:16.9&-73:15:53&3&7.10&36.35\\
MCELS-S188&DEM S150&01:14:22.1&-73:14:34&1&6.99&36.57\\
MCELS-S189&N84&01:14:27.7&-73:12:51&1&6.80&36.86$^e$\\
MCELS-S190&DEM S151&01:14:38.1&-73:16:05&1&6.63&36.53\\
MCELS-S191&DEM S151&01:14:41.7&-73:18:06&1&6.13&37.25\\
MCELS-S192&N84b&01:14:47.2&-73:19:48&3&5.86&36.03\\
MCELS-S193&DEM S152&01:14:55.7&-73:20:10&0&5.78&36.16\\
MCELS-S194&DEM S154&01:15:04.6&-72:19:33&2&3.38&35.88\\
MCELS-S195&DEM S152&01:15:04.7&-73:19:10&1&5.71&36.98\\
MCELS-S196&DEM S153&01:15:12.9&-72:56:32&1&4.39&35.79\\
MCELS-S197&DEM S156&01:16:13.6&-73:10:06&1&6.47&36.87\\
MCELS-S198&DEM S157&01:16:14.7&-73:25:42&1&4.04&38.27$^e$\\
MCELS-S199&DEM S158&01:16:49.3&-73:09:06&2&6.36&36.17\\
MCELS-S200&DEM S159&01:17:05.7&-73:12:24&1&5.04&36.34\\
MCELS-S201&\nodata&01:19:35.6&-73:05:48&3&4.55&35.95\\
MCELS-S202&N87&01:21:06.7&-73:15:04&0&3.45&35.91\\
MCELS-S203&\nodata&01:21:13.3&-73:06:15&2&4.20&36.26\\
MCELS-S204&DEM S160&01:23:13.1&-73:22:28&2&3.11&36.55\\
MCELS-S205&DEM S160&01:23:39.0&-73:24:03&2&2.75&35.56\\
MCELS-S206&\nodata&01:24:02.4&-73:17:53&2&2.87&35.23\\
MCELS-S207&DEM S161n&01:24:14.6&-73:09:31&1&4.64&36.88\\
MCELS-S208&DEM S161&01:24:45.4&-73:09:34&1&4.18&37.07$^e$\\
MCELS-S209&DEM S162&01:24:48.9&-73:27:34&1&2.71&36.68\\
MCELS-S210&DEM S163&01:25:04.0&-73:16:41&1&2.91&36.93\\
MCELS-S211&DEM S164&01:25:53.9&-73:22:41&1&2.51&37.37\\
MCELS-S212&DEM S165&01:27:03.3&-73:08:32&1&2.92&36.95\\
MCELS-S213&DEM S167&01:29:09.9&-73:24:50&3&1.51&37.99$^e$\\
MCELS-S214&DEM S166&01:29:26.8&-73:32:38&1&2.02&37.54\\
\hline
\label{tab:SMCObjs}
\end{longtable}

\end{document}